\documentclass[11pt,a4paper]{article}
\pdfoutput=1
\usepackage{a4wide}
\usepackage{amsmath}
\usepackage{amssymb}
\usepackage{amsfonts}
\usepackage{cite}
\usepackage[table]{xcolor}
\usepackage{soul}
\usepackage{adjustbox}
\usepackage{multirow}
\usepackage{pdflscape}
\usepackage{gensymb}
\usepackage{graphicx}
\graphicspath{{figures/}}
\usepackage{diagbox}
\usepackage{pifont}
\usepackage{bigstrut}
\usepackage[small,bf]{caption}
\usepackage{enumerate}
\usepackage{tabulary}
\usepackage{float}
\usepackage{arydshln}
\usepackage{enumitem}
\usepackage{mathtools}
\usepackage[utf8]{inputenc}
\usepackage[T1]{fontenc}
\usepackage{footnote}
\usepackage{footmisc}
\usepackage{ctable}
\newcommand{\Fig}{Fig.}
\newcommand{\Figs}{Figs.}

\def\1{\mathbf{1}}
\def\3{\mathbf{3}}
\def\2{\mathbf{2}}

\DeclareMathOperator{\diag}{diag}

\DeclareMathOperator{\im}{Im}
\DeclareMathOperator{\re}{Re}

\DeclareMathOperator{\sign}{sgn}
\DeclarePairedDelimiter{\abs}{\lvert}{\rvert}
\numberwithin{equation}{section}
\makeatletter
\g@addto@macro\bfseries{\boldmath}
\makeatother
\definecolor{model11}{HTML}{76d9fe}
\definecolor{model12}{HTML}{4cb2fe}
\definecolor{model13}{HTML}{3c8dfd}
\definecolor{model14}{HTML}{203bf0}
\definecolor{model15}{HTML}{2600bd}
\definecolor{model16}{HTML}{fe76d9}
\definecolor{model17}{HTML}{fe4cb2}
\definecolor{model18}{HTML}{fd3c8d}
\definecolor{model19}{HTML}{fc2a4e}
\definecolor{model110}{HTML}{e31c1a}
\definecolor{model111}{HTML}{b12600}
\definecolor{model21}{HTML}{ffeda0}
\definecolor{model22}{HTML}{fed976}
\definecolor{model23}{HTML}{feb24c}
\definecolor{model24}{HTML}{fd8d3c}
\definecolor{model25}{HTML}{fc4e2a}
\definecolor{model26}{HTML}{e31a1c}
\definecolor{model27}{HTML}{b10026}
\newcounter{savefootnote}
\newcounter{symfootnote}
\newcommand{\symfootnote}[1]{%
   \setcounter{savefootnote}{\value{footnote}}%
   \setcounter{footnote}{\value{symfootnote}}%
   \ifnum\value{footnote}>8\setcounter{footnote}{0}\fi%
   \let\oldthefootnote=\thefootnote%
   \renewcommand{\thefootnote}{\fnsymbol{footnote}}%
   \footnote{#1}%
   \let\thefootnote=\oldthefootnote%
   \setcounter{symfootnote}{\value{footnote}}%
   \setcounter{footnote}{\value{savefootnote}}%
}

\begin{document}
%=============

\begin{titlepage}

\vspace*{-15mm}

\begin{flushright}
IFT-UAM/CSIC-20-48
\end{flushright}
\vspace*{0.8cm}

\begin{center}

{\bf\LARGE{Testing Lepton Flavor Models at ESSnuSB}}\\[10mm]

Mattias Blennow,$^{a,b}$\symfootnote{On leave of absence}$^{,c,}$\footnote{E-mail: \texttt{emb@kth.se}} 
Monojit Ghosh,$^{a,c,}$\footnote{E-mail: \texttt{manojit@kth.se}} 
Tommy Ohlsson,$^{a,c,d,}$\footnote{E-mail: \texttt{tohlsson@kth.se}} 
and
Arsenii Titov$^{\,e,}$\footnote{E-mail: \texttt{arsenii.titov@pd.infn.it}} \\
\vspace{8mm}
$^{a}$\,{\it Department of Physics, School of Engineering Sciences,\\ KTH Royal Institute of Technology, AlbaNova University Center,\\ Roslagstullsbacken 21, SE--106 91 Stockholm, Sweden} \\
\vspace{2mm}
$^{b}$\,{\it Departamento de F\'isica Te\'orica and Instituto de F\'{\i}sica Te\'orica, IFT-UAM/CSIC,\\
Universidad Aut\'onoma de Madrid, Cantoblanco, 28049, Madrid, Spain} \\
\vspace{2mm}
$^{c}$\,{\it The Oskar Klein Centre for Cosmoparticle Physics, AlbaNova University Center, Roslagstullsbacken 21, SE--106 91 Stockholm, Sweden} \\
\vspace{2mm}
$^{d}$\,{\it University of Iceland, Science Institute, Dunhaga 3, IS--107 Reykjavik, Iceland} \\
\vspace{2mm}
$^{e}$\,{\it Dipartimento di Fisica e Astronomia ``G. Galilei'', Universit{\`a} degli Studi di Padova \\
and INFN, Sezione di Padova, Via Francesco Marzolo 8, I--35131 Padova, Italy}
\end{center}
\vspace{8mm}

\begin{abstract}
\noindent 
We review and investigate lepton flavor models, stemming from discrete non-Abelian flavor symmetries, described by one or two free model parameters. First, we confront eleven one- and seven two-parameter models with current results on leptonic mixing angles from global fits to neutrino oscillation data. We find that five of the one- and five of the two-parameter models survive the confrontation test at $3\sigma$. Second, we investigate how these ten one- and two-parameter lepton flavor models may be discriminated at the proposed ESSnuSB experiment in Sweden. We show that the three one-parameter models that predict $\sin\delta_{\rm CP}=0$ can be distinguished from those two that predict $|\sin\delta_{\rm CP}|=1$ by at least $7\sigma$. Finally, we find that three of the five one-parameter models can be excluded by at least $5\sigma$ and two of the one-parameter as well as at most two of the five two-parameter models can be excluded by at least $3\sigma$ with ESSnuSB if the true values of the leptonic mixing parameters remain close to the present best-fit values. 
\end{abstract}
\end{titlepage}

\setcounter{footnote}{0}

\tableofcontents

%===============
\section{Introduction}
\label{sec:intro}
%===============
The flavor puzzle, which consists in understanding 
the number of fermion generations as well as
the patterns of masses and mixing fermions obey, 
remains one of the most important problems in particle physics. 
It has no description in the Standard Model (SM)
and thus calls for physics beyond the SM. 
In particular, this new physics is supposed to provide 
a mechanism for neutrino mass generation 
and, ideally, reveal an organizing principle 
behind the quark and lepton mixing patterns 
emerged from the data, if such a principle exists. 

The tremendous experimental progress made in neutrino physics 
in the last two decades allowed us to establish that 
two leptonic mixing angles are large, while the third one 
is relatively small, but non-zero~\cite{Tanabashi:2018oca}. 
Such a mixing pattern appears to be very different from that 
in the quark sector, where all three mixing angles are small.
In the attempts to quantitatively describe the peculiar structure 
of leptonic mixing, flavor symmetries have gained a significant interest. 
In particular, models based on discrete non-Abelian symmetries, 
which naturally allow for rotation in flavor space by fixed large angles, 
have been extensively studied over the past years 
(see Refs.~\cite{Altarelli:2010gt,Ishimori:2010au,King:2013eh,King:2014nza,Feruglio:2015jfa,Petcov:2017ggy,Xing:2019vks,Feruglio:2019ktm} for reviews).

The main features of the discrete symmetry approach 
to lepton flavor include
(i)~specific predictions for some of the leptonic mixing angles 
and the Dirac CP-violating (CPV) phase, and/or 
(ii)~existence of algebraic relations between some of 
the mixing parameters. 
The latter are commonly referred to as 
lepton (or neutrino) mixing sum rules (see, e.g., Refs.~\cite{King:2005bj,Antusch:2005kw,Ge:2011ih,Ge:2011qn,Hernandez:2012ra,Hernandez:2012sk,Marzocca:2013cr,Petcov:2014laa,Girardi:2014faa,Girardi:2015vha,Girardi:2015rwa,Petcov:2018snn,Delgadillo:2018tza,Everett:2019idp}).
Furthermore, if a discrete flavor symmetry is combined with 
the so-called generalized CP symmetry~\cite{Feruglio:2012cw,Holthausen:2012dk,Chen:2014tpa}, predictions for 
the Majorana CPV phases can also be obtained~(see, e.g., Refs.~\cite{Feruglio:2012cw,Feruglio:2013hia,Ding:2013bpa,Girardi:2013sza,Hagedorn:2014wha,Li:2015jxa,DiIura:2015kfa,Ballett:2015wia,Turner:2015uta,Girardi:2016zwz,Lu:2016jit,Penedo:2017vtf}).
These phases are present in the Pontecorvo--Maki--Nakagawa--Sakata (PMNS) leptonic mixing matrix $U_\mathrm{PMNS}$
if massive neutrinos are Majorana fermions~\cite{Bilenky:1980cx}. 
However, they do not affect flavor neutrino oscillations~\cite{Bilenky:1980cx,Langacker:1986jv}.

The predictions and sum rules for the leptonic mixing angles 
and the Dirac CPV phase can be tested 
at current and, most importantly, future
neutrino oscillation experiments~\cite{Antusch:2007rk,Hanlon:2013ska,Ballett:2013wya,Ballett:2014dua,Ballett:2016yod,Chatterjee:2017xkb,Chatterjee:2017ilf,Agarwalla:2017wct,Pasquini:2018udd,Chakraborty:2018dew,Nath:2018xkz}. 
Neutrino physics is entering a precision era, 
which is crucial for probing different flavor models. 
At present, the leptonic mixing angle $\theta_{13}$ is the 
best-measured quantity among the leptonic mixing parameters. Measurements of this angle with high precision have been performed by the reactor neutrino experiments Daya Bay~\cite{An:2012eh,Adey:2018zwh}, Double Chooz~\cite{Abe:2011fz,Abe:2014bwa}, and RENO~\cite{Ahn:2012nd,Bak:2018ydk}.
The NO$\nu$A~\cite{Ayres:2004js} and T2K~\cite{Abe:2011ks}
long-baseline (LBL) neutrino oscillation 
experiments 
provide measurements of the leptonic mixing angle $\theta_{23}$ as well as the first hints of leptonic CP violation~\cite{Abe:2018wpn,Acero:2019ksn,Abe:2019vii} in terms of the leptonic Dirac CPV phase $\delta_{\rm CP}$. 
However, the status of the latter still basically remains unknown.
Nevertheless, T2K has recently measured and reported the best-fit value of $\delta_{\rm CP}$ to be $-1.89 \approx -108^\circ$ with the $3\sigma$ confidence interval as $[-3.41,-0.03] \approx [-195^\circ,-2^\circ]$, excluding 46~\% of the total parameter space for normal neutrino mass ordering (NO) \cite{Abe:2019vii}. This result also shows that both values of leptonic CP conservation for $\delta_{\rm CP}$, i.e., $0$ and $180^\circ$, are ruled out at 95~\% C.L.

The ESSnuSB experiment~\cite{Baussan:2013zcy,Wildner:2015yaa} 
is a future LBL facility proposed to be built in Sweden that will significantly improve the precision on $\delta_{\rm CP}$. The novel feature of ESSnuSB is the measurement of $\delta_{\rm CP}$ at the second oscillation maximum, where the measurement of $\delta_{\rm CP}$ is much less sensitive to systematic errors compared to the measurement at the first oscillation maximum. Therefore, it has the capability of measuring $\delta_{\rm CP}$ with excellent precision even with low statistics. As we will discuss, this would be of paramount importance for discriminating among various flavor models. The capability of ESSnuSB in measuring the unknown neutrino oscillation parameters within the standard three-flavor oscillation scenario has recently been studied in Refs.~\cite{Chakraborty:2017ccm,Ghosh:2019sfi,Blennow:2019bvl}. Other future LBL facilities, which are also capable of measuring $\theta_{23}$ and $\delta_{\rm CP}$ with very good precision, are DUNE~\cite{Acciarri:2015uup,Abi:2020evt} and T2HK~\cite{Abe:2015zbg} 
that are expected to become operational in the next decade.
In addition, the proposed medium-baseline JUNO experiment~\cite{An:2015jdp,Djurcic:2015vqa} 
will improve the precision on the leptonic mixing angle~$\theta_{12}$.

In the present work, we consider several well-motivated 
lepton flavor models, which lead to different mixing patterns. 
Our classification of these patterns is based on the number 
of free parameters they depend upon. 
First, we comment on fully-fixed mixing patterns, 
which do not contain any free parameter, and thus, 
all mixing angles (and sometimes $\delta_\mathrm{CP}$) 
are predicted to have certain values. 
For small (in terms of the number of elements) discrete non-Abelian groups, 
such as $A_4$, $S_4$, and $A_5$, some of the angles 
(typically, but not universally, $\theta_{13}$) turn out to be 
many sigmas away from their measured values. 
Next, we examine scenarios for which $U_\mathrm{PMNS}$ depends on 
one real continuous parameter $\theta$. 
Such mixing patterns arise from breaking of a flavor symmetry group $G_f$
combined with a generalized CP symmetry to a $Z_2 \times \mathrm{CP}$ 
residual symmetry~\cite{Feruglio:2012cw}. 
As illustrative examples, we consider the patterns derived from 
$G_f = A_4$~\cite{Ding:2013bpa}, $S_4$~\cite{Feruglio:2012cw}, 
and $A_5$~\cite{Li:2015jxa,DiIura:2015kfa,Ballett:2015wia} 
combined with the CP symmetry. 
Further, we explore the mixing patterns 
obtained from breaking the same flavor symmetries, 
but with no CP symmetry, to a $Z_2$ residual symmetry 
in either the charged lepton or neutrino sector~\cite{Girardi:2015rwa}. 
They are characterized by two real continuous parameters~---~an angle $\theta$ and a phase $\phi$.
Finally, we briefly discuss cases when the leptonic mixing matrix depends 
on three free parameters (either two angles and one phase or three angles). 

Concentrating on the one- and two-parameter scenarios, 
we first confront their predictions with the current 
global neutrino oscillation data~\cite{Esteban:2018azc,NuFiTv41} 
(see Refs.~\cite{Capozzi:2018ubv,deSalas:2018bym,deSalas:2017kay} for alternative global analyses)
to single out those which are well compatible with the data. 
For the selected models, we investigate in detail the potential 
of ESSnuSB to discriminate among them. 

This article is organized as follows. 
In Section~\ref{sec:models}, we review lepton mixing patterns 
derived from discrete non-Abelian flavor symmetries 
and classify them according to the number of free parameters 
they depend upon, whereas in Section~\ref{sec:confronting}, we confront the predictions of the
one- and two-parameter scenarios with
global neutrino oscillation data. 
Then, in Section~\ref{sec:essnusb}, we describe the ESSnuSB experimental setup, whereas in Section~\ref{sec:stat}, we present the details of the simulation method used and the statistical analysis performed.
Next, in Section~\ref{sec:results}, we present and discuss the results of this analysis. 
Finally, in Section~\ref{sec:sc}, we present a summary of our work and draw our conclusions.

%===============
\section{Lepton Mixing Patterns from Residual Symmetries}
\label{sec:models}
%===============

In the discrete symmetry approach to lepton flavor, 
it is assumed that at energies higher than a certain scale $\Lambda$,
there exists a flavor symmetry described 
by a discrete non-Abelian group~$G_f$. 
The group needs to be non-Abelian, since only in this case, 
it has multi-dimensional (in particular, three-dimensional) 
irreducible representations to which 
three lepton $\mathrm{SU(2)_L}$ doublets 
can be assigned. This in turn allows for 
predictions of the leptonic mixing matrix
(see, e.g., Ref.~\cite{Feruglio:2019ktm} for more details). 
At energies below $\Lambda$, 
the flavor symmetry must be completely broken 
to account for three distinct charged lepton and neutrino masses.
However, the charged lepton and neutrino mass matrices, 
$M_e$ and $M_\nu$, may be separately invariant under 
non-trivial Abelian subgroups $G_e$ and $G_\nu$ of $G_f$.
These so-called residual symmetries constrain $M_e$ and $M_\nu$, 
and hence, the form of the unitary matrices $U_e$ and $U_\nu$, 
which diagonalize the mass matrices as follows
%%%%%
\begin{align}
U_e^\dagger M_e M_e^\dagger U_e &= \diag\left(m_e^2,m_\mu^2,m_\tau^2\right)\,, \\
U_\nu^T M_\nu U_\nu &= \diag\left(m_1,m_2,m_3\right)\,,
\end{align}
%%%%%
%
where $m_i$, $i=1,2,3$, are three neutrino masses.%
\footnote{For definiteness, we concentrate on the case of Majorana neutrinos. The case of Dirac neutrinos is analogous to that of charged leptons (see Ref.~\cite{Girardi:2015rwa} for details).}
Thus, the form of the leptonic mixing matrix given by
%%%%%
\begin{equation}
U_\mathrm{PMNS} = U_e^\dagger U_\nu
\end{equation}
%%%%%
%
is also constrained. In the so-called standard parametrization~\cite{Tanabashi:2018oca}, 
$U_\mathrm{PMNS}$ is expressed 
in terms of the three leptonic mixing angles, $\theta_{12}$, $\theta_{13}$, and $\theta_{23}$, the Dirac, $\delta_\mathrm{CP}$, and two Majorana, 
$\alpha_{21}$ and $\alpha_{31}$, CPV phases:
%%%%%
\begin{equation}
U_\mathrm{PMNS} = \begin{pmatrix}
 c_{12} c_{13} & s_{12} c_{13} & s_{13} e^{-i \delta_\mathrm{CP}} \\
 -s_{12} c_{23} - c_{12} s_{23} s_{13} e^{i \delta_\mathrm{CP}}
 & c_{12} c_{23} - s_{12} s_{23} s_{13} e^{i \delta_\mathrm{CP}}
 & s_{23} c_{13} \\
 s_{12} s_{23} - c_{12} c_{23} s_{13} e^{i \delta_\mathrm{CP}} &
 - c_{12} s_{23} - s_{12} c_{23} s_{13} e^{i \delta_\mathrm{CP}}
 & c_{23} c_{13} 
\end{pmatrix}
P\,,
\end{equation}
%%%%%
%
where $c_{ij} \equiv \cos\theta_{ij}$, $s_{ij} \equiv \sin\theta_{ij}$, 
$P \equiv \diag\left(1,e^{i\alpha_{21}/2},e^{i\alpha_{31}/2}\right)$ 
and $\theta_{ij} \in [0,\pi/2] = [0,90^\circ]$ whilst
$\delta$, $\alpha_{21}$, $\alpha_{31} \in (-\pi,\pi] = (-180^\circ,180^\circ]$. 
The diagonal matrix $P$ has a physical meaning only if 
massive neutrinos are of Majorana nature.

It can further be shown that depending on $G_e$ and $G_\nu$ 
the leptonic mixing matrix is either completely fixed (up to permutations of rows and columns and external phases) 
or predicted to depend on a number of free parameters. 
In the following subsections, we consider lepton mixing patterns that arise from particular $G_e$ and $G_\nu$. 
We classify them according to the number of free parameters 
entering the predicted form of $U_\mathrm{PMNS}$.

%-----------------------------------------------------------
\subsection{Fully-fixed mixing patterns}
\label{sec:0params}
%-----------------------------------------------------------

If $G_e = Z_k$, $k>2$ or $Z_m \times Z_n$, $m,n \geq 2$ 
and $G_\nu = Z_2 \times Z_2$, which is the maximal symmetry 
of the neutrino Majorana mass matrix,%
\footnote{If the smallest neutrino mass is zero, this symmetry is enhanced.}
$U_\mathrm{PMNS}$ is completely predicted 
(up to permutations of rows and columns and external phases). 
In this case, putting it into the standard parametrization~\cite{Tanabashi:2018oca},
one can obtain the values of the leptonic mixing parameters 
$\sin^2\theta_{12}$, $\sin^2\theta_{13}$, and $\sin^2\theta_{23}$, 
as well as the Dirac CPV phase $\delta_\mathrm{CP}$ 
(if $\theta_{13} \neq 0$).

Well-known examples include bimaximal (BM)~\cite{Vissani:1997pa,Barger:1998ta,Baltz:1998ey}, 
tri-bimaximal (TBM)~\cite{Harrison:2002er,Harrison:2002kp,Xing:2002sw}, and golden ratio (GR)~\cite{Datta:2003qg,Kajiyama:2007gx} mixing. 
All of them are characterized by maximal mixing 
$\sin^2\theta_{23} = 1/2$ and zero mixing angle $\theta_{13} = 0$. 
The difference is in the predicted value of the mixing angle $\theta_{12}$. Namely, $\sin^2\theta_{12} = 1/2$ for BM, 
$\sin^2\theta_{12}=1/3$ for TBM,
and $\sin^2\theta_{12} = 1/(2+\varphi) \approx 0.276$ for GR mixing, 
$\varphi \equiv (1+\sqrt{5})/2$ being the golden ratio. 
TBM mixing can be realized by breaking $G_f = S_4$ to 
$G_e = Z_3^T$ and $G_\nu = Z_2^S \times Z_2^U$~\cite{Lam:2008rs}, 
where $S$, $T$, and $U$ are $S_4$ generators. 
It can also arise from $G_f = A_4$ 
generated by $S$ and $T$,
in that case the $U$ symmetry arises accidentally~\cite{Altarelli:2005yp,Altarelli:2005yx}. 
Analogously, BM mixing can be obtained 
from $G_f = S_4$ broken to different subgroups~\cite{Altarelli:2009gn}, 
while GR mixing can stem from $G_f = A_5$~\cite{Everett:2008et}. 
Note that the groups $A_4$, $S_4$, and $A_5$ have a relatively small 
number of elements, namely, 12, 24, and 60, respectively.
The aforementioned mixing patterns were very appealing before 2012, 
when the value of $\theta_{13}$ was compatible with zero.
Currently, all of them are ruled out. 

A complete classification of all possible mixing patterns 
fully-fixed by the residual symmetries $G_e$ and $G_\nu$ 
has been performed in Ref.~\cite{Fonseca:2014koa}. 
From 17 sporadic cases and one infinite series of mixing matrices 
found therein, only the latter can lead to phenomenologically viable 
values of the leptonic mixing angles. Namely, this infinite series 
is given (up to permutations of rows and columns) by
\footnote{The matrix $\abs{U_\mathrm{PMNS}}^2$ is defined as 
the matrix with the elements $\abs{\left(U_\mathrm{PMNS}\right)_{\ell i}}^2$, 
$\ell = e,\mu,\tau$ and $i=1,2,3$.}
%%%%%
\begin{equation}
\abs{U_\mathrm{PMNS}}^2 = \frac{1}{3}\begin{pmatrix}
1+\re\sigma & 1 & 1 - \re\sigma \\
1+\re\left(\omega\sigma\right) & 1 & 1-\re\left(\omega\sigma\right) \\
1+\re\left(\omega^2\sigma\right) & 1 & 1-\re\left(\omega^2\sigma\right)
\end{pmatrix}\,,
\end{equation}
%%%%%
%
where $\sigma = e^{2\pi i p/n}$  and $\omega = e^{2\pi i/3}$ 
are roots of unity and $p$ and $n$ are coprime. 
Thus, for all viable mixing patterns, we have the following prediction
%%%%%
\begin{equation}
\sin^2\theta_{12} = \frac{1}{3\left(1-\sin^2\theta_{13}\right)} > \frac{1}{3}\,.
\label{eq:TM2}
\end{equation}
%%%%%
Furthermore, $\sin\delta_\mathrm{CP} = 0$, i.e., no Dirac CP violation 
is predicted. 
A given choice of $\sigma$ defines the flavor symmetry group $G_f$. 
It turns out that the smallest viable group leading to $n=9,18$ is 
$(Z_{18} \times Z_6) \rtimes S_3$, which has 648 elements. 
The next ``minimal'' group implying $n=11,22,33,66$ is 
$\Delta(6\times22^2)$, which has order 2904. 
Such groups are much more complex (less natural) 
than those originally proposed for description 
of the observed lepton mixing pattern.

%-----------------------------------------------------------
\subsection{Models with one free parameter}
\label{1p_models}
%-----------------------------------------------------------

A discrete flavor symmetry can be consistently combined 
with a generalized CP symmetry~\cite{Feruglio:2012cw,Holthausen:2012dk,Chen:2014tpa}. 
The latter is given by a CP transformation, 
which can act non-trivially in flavor space. 
This action is represented by a unitary symmetric matrix $X$. 
The full symmetry group, in this case $G_\mathrm{CP}$, is given by 
a semi-direct product $G_\mathrm{CP} = G_f \rtimes \text{CP}$~\cite{Feruglio:2012cw}. 
Breaking this symmetry to $G_e = Z_k$, $k > 2$ 
or $Z_m\times Z_n$, $m,n\geq2$ and $G_\nu = Z_2 \times \text{CP}$ 
leads to $U_\mathrm{PMNS}$ that 
is defined up to a rotation $R_{ij}(\theta)$ in the neutrino sector. 
Here, $ij$ refers to the plane of rotation, e.g.,
%%%%%
\begin{equation}
R_{13}\left(\theta\right) = \begin{pmatrix}
\cos\theta & 0 & \sin\theta \\
0 & 1 & 0 \\
-\sin\theta & 0 & \cos\theta
\end{pmatrix}\,,
\end{equation}
%%%%%
%
where $\theta$ is a free real continuous parameter
and its fundamental interval is $[0,\pi) = [0,180^\circ)$~\cite{Feruglio:2012cw}.
The advantage of this approach is that the Majorana CPV phases 
are also predicted (contrary to the case without CP).

In what follows, we consider the lepton mixing patterns 
originating from $S_4 \rtimes \text{CP}$~\cite{Feruglio:2012cw} 
and $A_5 \rtimes \text{CP}$~\cite{Li:2015jxa,DiIura:2015kfa,Ballett:2015wia} broken 
to the residual symmetries as described above. 
Note that the results obtained from $A_4 \rtimes \text{CP}$~\cite{Ding:2013bpa}
are contained in those for $S_4 \rtimes \text{CP}$~\cite{Feruglio:2012cw}.
Thus, we do not need to consider the $A_4 \rtimes \text{CP}$ case 
separately. 

In Ref.~\cite{Feruglio:2012cw}, it has been shown that fixing $G_e = Z_3$, 
there are five inequivalent choices of the $Z_2 \times \text{CP}$ 
transformations leaving the neutrino sector invariant 
(due to different $Z_2$ subgroups of $S_4$ 
and CP transformations $X$ compatible with them). 
Four of them, denoted Cases~I, II, IV, and V, lead to the results 
summarized in Table~\ref{tab:FHZ1}. 
%%%%%%%%%%%%%%%%%%%%
\begin{table}[t]
\renewcommand{\arraystretch}{1.5}
\centering
\begin{tabular}{|c|c|c|c|c|} 
\hline
Case & I & II & IV & V\\
\hline
\hline
$\sin^2 \theta_{13}$ & $\frac{2}{3} \sin^2 \theta$ & $\frac{2}{3} \sin^2 \theta$ & $\frac{1}{3} \sin^2 \theta$& $\frac{1}{3} \sin^2 \theta$\\
$\sin^2 \theta_{12}$ & $\frac{1}{2+\cos 2 \theta}$ & $ \frac{1}{2+\cos 2 \theta}$ & $\frac{\cos^2 \theta}{2+\cos^2 \theta}$ & $\frac{\cos^2 \theta}{2+\cos^2 \theta}$\\
$\sin^2 \theta_{23}$ & $\frac{1}{2}$ & $\frac{1}{2} \, \left( 1-\frac{\sqrt{3} \sin 2 \theta}{2 +\cos 2\theta}\right)$ & $\frac{1}{2}$ & $\frac{1}{2} \, \left( 1- \frac{2 \sqrt{6} \sin 2 \theta}{5+\cos 2\theta} \right)$\\
$|\sin \delta_{\rm CP}|$ & $1$ & 0 & $1$ &  0\\
\hline
\end{tabular}
\caption{Predictions for the leptonic mixing parameters in terms of the free parameter $\theta$ 
in Cases~I, II, IV, and V originating from $G_\mathrm{CP} = S_4 \rtimes \text{CP}$ broken to $G_e = Z_3$ and $G_\nu = Z_2 \times \text{CP}$ 
derived in Ref.~\cite{Feruglio:2012cw}. 
}
\label{tab:FHZ1}
\end{table}
%%%%%%%%%%%%%%%%%%%%
%
Case~III 
has been found to be phenomenologically not viable 
and we do not present it here. 
Cases~I and II realize the so-called trimaximal (TM) mixing pattern 2 (TM$_2$)~\cite{Grimus:2008tt}, for which 
$\abs{\left(U_\mathrm{PMNS}\right)_{\ell 2}}^2 = 1/3$, $\ell = e,\mu,\tau$,
while Cases~IV and V give rise to TM$_1$~\cite{Albright:2008rp}, 
characterized by 
$\abs{\left(U_\mathrm{PMNS}\right)_{e1}}^2 = 2/3$ and
$\abs{\left(U_\mathrm{PMNS}\right)_{\mu1}}^2 = 
\abs{\left(U_\mathrm{PMNS}\right)_{\tau1}}^2 = 1/6$. 
In particular, Cases~I and II  lead to Eq.~\eqref{eq:TM2},% 
\footnote{Note that now all mixing parameters are functions of $\theta$.}
i.e., they predict $\sin^2\theta_{12} \geq 1/3$, while for 
Cases~IV and V
%%%%%
\begin{equation}
\sin^2\theta_{12} = \frac{1-3\sin^2\theta_{13}}{3\left(1-\sin^2\theta_{13}\right)} < \frac{1}{3}\,.
\label{eq:TM1}
\end{equation}
%%%%%
%
Furthermore, the following sum rule for $\cos\delta_\mathrm{CP}$ holds 
in Cases~I and II
%%%%%
\begin{equation}
\cos\delta_\mathrm{CP} = \frac{\left(1-2\sin^2\theta_{13}\right) \cot2\theta_{23}}{\sin\theta_{13} \sqrt{2 - 3\sin^2\theta_{13}}}\,.
\label{eq:cosdeltaTM2}
\end{equation}
%%%%%
%

In Case~I, $\sin^2\theta_{23}$ is predicted to be $1/2$, and thus, $\cos\delta_\mathrm{CP} = 0$. To establish for which values of $\theta$ 
the Dirac CPV phase $\delta_\mathrm{CP} = -90^\circ$ and for which it is $90^\circ$, 
one needs to look at the rephasing invariant $J_\mathrm{CP}$~\cite{Jarlskog:1985ht} 
that controls the magnitude of CPV effects in neutrino oscillations~\cite{Krastev:1988yu}. In terms of the elements of $U_\mathrm{PMNS}$, it can be chosen as
%%%%%
\begin{align}
J_\mathrm{CP} &= \im\left[\left(U_\mathrm{PMNS}\right)_{e1} \left(U_\mathrm{PMNS}\right)_{e3}^\ast \left(U_\mathrm{PMNS}\right)_{\tau1}^\ast \left(U_\mathrm{PMNS}\right)_{\tau3}\right] \nonumber \\ 
&= \frac{1}{8} \sin2\theta_{12} \sin2\theta_{23} \sin2\theta_{13} \cos\theta_{13} \sin\delta_\mathrm{CP}\,,
\label{eq:JCP}
\end{align}
%%%%%
%
where in the second line we have reported its expression in 
the standard parametrization of the leptonic mixing matrix. 
Instead, in the parametrization corresponding to Case~I, it reads
%%%%%
\begin{equation}
J_\mathrm{CP} = - \frac{\sin2\theta}{6\sqrt{3}}\,.
\label{eq:JCPS4I}
\end{equation}
%%%%%
%
Equating \eqref{eq:JCP} and \eqref{eq:JCPS4I}, we find
%%%%%
\begin{equation}
\sin\delta_\mathrm{CP} = -\sign\left(\sin2\theta\right), \quad \text{i.e.,} \quad
\delta_\mathrm{CP} = \begin{cases}
-90^\circ\,, \quad \theta \in \left(0,90^\circ\right)\,, \\
+90^\circ\,, \quad \theta \in \left(90^\circ,180^\circ\right)\,.
\end{cases}
\label{eq:deltaCPS4I}
\end{equation}
%%%%%
%

In Case~II, we have $J_\mathrm{CP} = 0$, whilst substituting 
the expressions for $\sin^2\theta_{ij}$ from Table~\ref{tab:FHZ1} in Eq.~\eqref{eq:cosdeltaTM2}, we obtain
%%%%%
\begin{equation}
\cos\delta_\mathrm{CP} = \sign\left[\sin2\theta\left(1+2\cos2\theta\right)\right]\,, 
\end{equation}
%%%%%
%
i.e.,
%%%%%
\begin{equation}
\delta_\mathrm{CP} = \begin{cases}
\phantom{18}0\phantom{^\circ}\,, \quad \theta \in \left(0,60^\circ\right) \cup \left(90^\circ,120^\circ\right)\,, \\
180^\circ\,, \quad \theta \in \left(60^\circ,90^\circ\right) \cup \left(120^\circ,180^\circ\right)\,.
\end{cases}
\end{equation}
%%%%%
%
 
Then, Cases~IV and V lead to a different sum rule for $\cos\delta_\mathrm{CP}$, 
which reads 
%%%%%
\begin{equation}
\cos\delta_\mathrm{CP} = -\frac{\left(1-5\sin^2\theta_{13}\right) \cot2\theta_{23}}{2\sqrt{2} \sin\theta_{13} \sqrt{1 - 3\sin^2\theta_{13}}}\,.
\label{eq:cosdeltaTM1}
\end{equation}
%%%%%
%
In Case~IV, $\sin^2\theta_{23} = 1/2$, and hence, $\cos\delta_\mathrm{CP} = 0$.
Furthermore, the $J_\mathrm{CP}$ invariant takes the following form
%%%%%
\begin{equation}
J_\mathrm{CP} = -\frac{\sin2\theta}{6\sqrt{6}}\,,
\end{equation}
%%%%%
%
resulting in the same predictions for $\delta_\mathrm{CP}$ as in Eq.~\eqref{eq:deltaCPS4I}. In Case~V, $J_\mathrm{CP} = 0$, whereas
%%%%%
\begin{equation}
\cos\delta_\mathrm{CP} = -\sign\left[\sin2\theta\left(1+5\cos2\theta\right)\right]\,, 
\end{equation}
%%%%%
%
and
%%%%%
\begin{equation}
\delta_\mathrm{CP} = \begin{cases}
\phantom{18}0\phantom{^\circ}\,, \quad \theta \in \left(50.8^\circ,90^\circ\right) \cup \left(129.2^\circ,180^\circ\right)\,, \\
180^\circ\,, \quad \theta \in \left(0,50.8^\circ\right) \cup \left(90^\circ,129.2^\circ\right)\,,
\end{cases}
\end{equation}
%%%%%
%
where $50.8^\circ$ and $129.2^\circ$ correspond to 
$\arccos(-1/5)/2$ and $\pi - \arccos(-1/5)/2$, respectively.

For $G_e = Z_4$, only one viable case has been found. 
It leads to the predictions shown in Table~\ref{tab:FHZ2}.
%%%%%%%%%%%%%%%%%%%%
\begin{table}[t]
\renewcommand{\arraystretch}{1.5}
\centering
\begin{tabular}{|c|c|c|} 
\hline
Case & VI-a& VI-b\\
\hline
\hline
$\sin^2 \theta_{13}$ & \multicolumn{2}{c|}{$\frac{1}{4} \left( \sqrt{2} \cos \theta + \sin \theta \right)^2$}\\
$\sin^2 \theta_{12}$ & \multicolumn{2}{c|}{$\frac{2}{5 - \cos 2 \theta -2  \sqrt{2} \sin 2 \theta}$}\\
$ \sin^2 \theta_{23}$ & $\frac{4 \sin^2 \theta}{5 - \cos 2 \theta - 2 \sqrt{2} \sin 2 \theta}$ & $1-\frac{4 \sin^2 \theta}{5 - \cos 2 \theta - 2 \sqrt{2} \sin 2 \theta}$\\
$\sin \delta_{\rm CP}$ &  \multicolumn{2}{c|}{$0$}\\
\hline
 \end{tabular}
\caption{Predictions for the leptonic mixing parameters in terms of the free parameter $\theta$ in the case originating from $G_\mathrm{CP} = S_4 \rtimes \text{CP}$ broken to 
$G_e=Z_4$ (or $Z_2 \times Z_2$) and $G_\nu = Z_2 \times \mathrm{CP}$
derived in Ref.~\cite{Feruglio:2012cw}. 
}
\label{tab:FHZ2}
\end{table}
%%%%%%%%%%%%%%%%%%%%
%
The existence of two solutions, marked as Cases~VI-a and VI-b,
arises from the freedom to exchange the second and the third rows 
of the leptonic mixing matrix. 
For both solutions, we have
%%%%%
\begin{equation}
\sin^2\theta_{12} = \frac{1}{4\left(1-\sin^2\theta_{13}\right)} > \frac{1}{4}\,,
\end{equation}
%%%%%
%
and in addition, we find that
%%%%%
\begin{align}
\cos\delta_\mathrm{CP} &= \frac{\left(1-2\sin^2\theta_{13}\right) \cos2\theta_{23} 
- \left(1-\sin^2\theta_{13}\right) \sin^2\theta_{23}}
{\sin2\theta_{23} \sin\theta_{13} \sqrt{3 - 4\sin^2\theta_{13}}} 
\quad \text{in} \quad \text{Case VI-a}\,,
\label{eq:cosdeltaBM1} \\
\cos\delta_\mathrm{CP} &= \frac{\left(1-2\sin^2\theta_{13}\right) \cos2\theta_{23} 
+ \left(1-\sin^2\theta_{13}\right) \cos^2\theta_{23}}
{\sin2\theta_{23} \sin\theta_{13} \sqrt{3 - 4\sin^2\theta_{13}}} 
\quad \text{in} \quad \text{Case VI-b}\,.
\label{eq:cosdeltaBM2}
\end{align}
%%%%%
%
Note that these equations are consistent with the fact that under exchange 
of the second and the third rows of $U_\mathrm{PMNS}$, $\theta_{12}$ and $\theta_{13}$ remain unchanged, while $\theta_{23} \to \pi/2 - \theta_{23}$ 
and $\delta_\mathrm{CP} \to \delta_\mathrm{CP} + \pi$.
The $J_\mathrm{CP}$ invariant vanishes, which means that 
$\abs{\cos\delta_\mathrm{CP}} = 1$. Substituting the expressions for $\sin^2\theta_{ij}$ given in Table~\ref{tab:FHZ2} in Eqs.~\eqref{eq:cosdeltaBM1} 
and \eqref{eq:cosdeltaBM2},
we find
%%%%%
\begin{equation}
\cos\delta_\mathrm{CP} = \pm\sign\left[\left(1+3\cos2\theta\right)\left(\sqrt{2}\sin\theta-\cos\theta\right)\right]\,,
\end{equation}
%%%%%
%
where ``$+$'' corresponds to Case~VI-a and ``$-$'' to Case~VI-b. 
This in turn implies that
%%%%%
\begin{equation}
\delta_\mathrm{CP} = \begin{cases}
0~[180^\circ]\,, \quad \theta \in \left(35.3^\circ,54.7^\circ\right) \cup \left(125.3^\circ,180^\circ\right), \\
180^\circ~[0]\,, \quad \theta \in \left(0,35.3^\circ\right) \cup \left(54.7^\circ,125.3^\circ\right), 
\end{cases}
\quad \text{in} \quad \text{Case VI-a [VI-b]}\,.
\end{equation}
%%%%%
%
Here, $35.3^\circ$, $54.7^\circ$, and $125.3^\circ$ correspond to 
$\arctan(1/\sqrt2)$, $\arccos(-1/3)/2$, and $\pi-\arccos(-1/3)/2$, respectively.
Finally, $G_e = Z_2 \times Z_2$ leads to the same results as $G_e = Z_4$.

Similarly, lepton mixing patterns originating from 
$G_\mathrm{CP} = A_5 \rtimes \text{CP}$ broken to 
$G_\nu = Z_2 \times \text{CP}$ in the neutrino sector 
have been derived in Refs.~\cite{Li:2015jxa,DiIura:2015kfa,Ballett:2015wia}. 
The results for $G_e = Z_3$ (Case~V) and 
$G_e = Z_2 \times Z_2$ (Case~VII)
are presented in Table~\ref{tab:LD1} and those for
$G_e = Z_5$ (Cases~II, III, and IV) in Table~\ref{tab:LD2}. 
%%%%%%%%%%%%%%%%%%%%
\begin{table}[t]
\renewcommand{\arraystretch}{1.5}
\centering
\begin{tabular}{|c|c|c|c|} 
\hline
Case & V & VII-a & VII-b\\
\hline
\hline
$\sin^2 \theta_{13}$ & $\frac{1 - \sin2\theta}{3}$ & 
\multicolumn{2}{c|}{$\frac{(\cos\theta-\varphi\sin\theta)^2}{4\varphi^{2}}$} \\
$\sin^2 \theta_{12}$ & $\frac{1}{2+\sin{2\theta}}$ & 
\multicolumn{2}{c|}{$\frac{(\varphi\cos\theta+ \sin\theta)^2}{4\varphi^2-(\cos\theta- \varphi\sin\theta)^2}$} \\
$ \sin^2 \theta_{23}$ & $\frac{1}{2}$ & 
$\frac{(\varphi^2\cos\theta-\sin\theta)^2}{4\varphi^2-(\cos\theta-\varphi\sin\theta)^2}$ & 
$\frac{\varphi^{2}(\cos\theta+\varphi\sin\theta)^2}{4\varphi^{2}-(\cos\theta- \varphi\sin\theta)^2}$ \\
$|\sin \delta_{\rm CP}|$ & 1 &  \multicolumn{2}{c|}{0} \\
\hline
\end{tabular}
\caption{Predictions for the leptonic mixing parameters in terms of the free parameter $\theta$ 
in Case~V (VII) originating from $G_\mathrm{CP} = A_5 \rtimes \text{CP}$ broken to 
$G_e = Z_3$ ($Z_2 \times Z_2$) and $G_\nu = Z_2 \times \text{CP}$ 
derived in Ref.~\cite{Li:2015jxa}. 
The quantity $\varphi = (1 + \sqrt5)/2$ is the golden ratio.
}
\label{tab:LD1}
\end{table}
%%%%%%%%%%%%%%%%%%%%
%
%
%
%
%%%%%%%%%%%%%%%%%%%%
\begin{table}[t]
\renewcommand{\arraystretch}{1.5}
\centering
\begin{tabular}{|c|c|c|c|} 
\hline
Case & II & III & IV \\
\hline
\hline
$\sin^2 \theta_{13}$ & $\frac{3-\varphi}{5}\sin^{2}\theta$ & 
$\frac{\varphi}{\sqrt{5}}\sin^{2}\theta$ & $\frac{\varphi}{\sqrt{5}}\sin^{2}\theta$ \\
$\sin^2 \theta_{12}$ & $\frac{2\cos^{2}\theta}{3+2\varphi+\cos2\theta}$ & 
$\frac{4-2\varphi}{5-2\varphi+\cos{2\theta}}$ & $\frac{4-2\varphi}{5-2\varphi+\cos{2\theta}}$ \\
$ \sin^2 \theta_{23}$ & $\frac{1}{2}$ & 
$\frac{1}{2}-\frac{\sqrt{3-\varphi}\sin2\theta}{3\varphi-2+\varphi\cos2\theta}$ & 
$\frac{1}{2}$ \\
$|\sin \delta_{\rm CP}|$ & 1 &  0 & 1 \\
\hline
\end{tabular}
\caption{Predictions for the leptonic mixing parameters in terms of the free parameter $\theta$ 
in Cases~II, III, and IV originating from $G_\mathrm{CP} = A_5 \rtimes \text{CP}$ broken to 
$G_e = Z_5$ and $G_\nu = Z_2 \times \text{CP}$ derived in Ref.~\cite{Li:2015jxa}. 
The quantity $\varphi = (1 + \sqrt5)/2$ is the golden ratio.
}
\label{tab:LD2}
\end{table}
%%%%%%%%%%%%%%%%%%%%
%
The case numbering follows Ref.~\cite{Li:2015jxa}. 
Several comments are in order. 
First, Cases~I, VI, and VIII have been found to be not viable, 
so we do not present them here.
Secondly, since $\theta$ is a free parameter and the corresponding rotation 
can be performed either clockwise or counterclockwise, 
we can replace $\theta \to \pi/4 \pm \theta$. Then, the results for Case~V 
in Table~\ref{tab:LD1} match those for Case~I in Table~\ref{tab:FHZ1}. 
Thus, these two cases are identical (cf.~Ref.~\cite{Ballett:2015wia}). 
Further, two solutions in Case~VII arise from the exchange 
of the second and the third rows of the leptonic mixing matrix. 
Finally, the predictions in all cases except for Case~V involve 
the golden ratio $\varphi$ characteristic for the $A_5$ group.

To establish the values of $\theta$ leading to $\delta_\mathrm{CP}=0~(-90^\circ)$ and those giving rise to $\delta_\mathrm{CP}=180^\circ~(90^\circ)$ in Cases~VII and III 
(Cases~II and IV), one needs to perform an analysis similar to that we have 
described for $G_\mathrm{CP} = S_4 \rtimes \text{CP}$. In this work, we do not present 
the corresponding analytical expressions for the sake of brevity. 
However, we have established such a correspondence and used it 
in our main analyses in Sections~\ref{sec:confronting} and \ref{sec:results}.

It is worth noting that all cases in Tables~\ref{tab:FHZ1}--\ref{tab:LD2} 
predict either CP conservation or maximal CP violation. 
In addition, the cases predicting the latter also lead to 
$\sin^2\theta_{23} = 1/2$. 
Finally, for these cases, $\sin^2\theta_{12}$ and $\sin^2\theta_{13}$ 
remain invariant under $\theta \to \pi - \theta$, whereas $\delta_\mathrm{CP}$ 
changes from $-90^\circ$ to $90^\circ$. This fact will manifest itself 
in Section~\ref{sec:confronting}, where we fit the models to global 
neutrino oscillation data.

%-----------------------------------------------------------
\subsection{Models with two free parameters}
\label{sec:2params}
%-----------------------------------------------------------

Now, if we relax the assumption of generalized CP invariance 
and break the flavor symmetry group $G_f$ to either
%%%%%
\begin{enumerate}[label=(\Alph*)]
\item $G_e = Z_2$ and $G_{\nu} = Z_k$, $k > 2$ or $Z_m \times Z_n$, $m,n \geq 2$,%
\footnote{Note that here we consider both Majorana and Dirac neutrinos. 
In the latter case, $G_\nu$ can be different from $Z_2$ or $Z_2\times Z_2$.}
\end{enumerate}
or
\begin{enumerate}[label=(\Alph*),resume]
\item $G_e = Z_k$, $k > 2$ or $Z_m \times Z_n$, $m,n \geq 2$ and $G_{\nu} = Z_2$,
\end{enumerate}
%%%%%
% 
the leptonic mixing matrix $U_\mathrm{PMNS}$ 
is defined up to a complex rotation 
$U_{ij}(\theta^{l}_{ij},\delta^{l}_{ij})$
in either the charged lepton ($l=e$) or neutrino ($l=\nu$) sector 
and thus contains two free real continuous parameters. 
Here, $ij$ refers to the plane in which the rotation is imposed, e.g., 
%%%%%
\begin{equation}
U_{13}\left(\theta^{l}_{13},\delta^{l}_{13}\right) = \begin{pmatrix}
\cos\theta^{l}_{13} & 0 & \sin\theta^l_{13} e^{-i\delta^l_{13}} \\
0 & 1 & 0 \\
-\sin\theta^l_{13} e^{i\delta^l_{13}} & 0 & \cos\theta^{l}_{13}
\end{pmatrix}\,.
\end{equation}
%%%%%
%
All possible cases of this type have been considered in Ref.~\cite{Girardi:2015rwa}, where they have been classified according to the plane 
in which the free rotation is performed. 
Below, we summarize the expressions 
for $\sin^2\theta_{ij}$, $\cos\delta_\mathrm{CP}$, and $J_\mathrm{CP}$
in terms of the two free parameters. 
We will use them further in our statistical analysis.

 \textbf{Case A1.}
In this case, a fixed part of the leptonic mixing matrix 
parametrized in terms of the angles $\theta^\circ_{ij}$ 
defined by the choice of residual symmetries 
$G_e, G_\nu \subset G_f$, 
is corrected from the left 
by $U_{12}(\theta^{e}_{12},\delta^{e}_{12})$. 
The leptonic mixing parameters and $J_\mathrm{CP}$ are given by
%%%%%
\begin{align}
\sin^2 \theta_{13} & = \cos^2 \theta \sin^2 \theta^{\circ}_{13} + \cos^2 \theta^{\circ}_{13} \sin^2 \theta \sin^2 \theta^{\circ}_{23} 
+\frac{1}{2} \sin 2 \theta \sin 2 \theta^{\circ}_{13} \sin \theta^{\circ}_{23} \cos \phi  \,, 
\label{eq:ss13A1}\\
\sin^2 \theta_{23} & = \frac{\sin^2 \theta^{\circ}_{13} - \sin^2 \theta_{13} + \cos^2 \theta^{\circ}_{13} \sin^2 \theta^{\circ}_{23}}{1 - \sin^2 \theta_{13}} \,, 
\label{eq:ss23A1}\\
\sin^2 \theta_{12} & = \frac{\cos^2 \theta^{\circ}_{23} \sin^2 \theta}{1 - \sin^2 \theta_{13}}  \,,
\label{eq:ss12A1} \\
\cos \delta_{\rm CP} &= \frac{\cos^2 \theta_{13} (\sin^2 \theta^{\circ}_{23} - \cos^2 \theta_{12}) + \cos^2 \theta^{\circ}_{13} \cos^2 \theta^{\circ}_{23} (\cos^2 \theta_{12} - \sin^2 \theta_{12} \sin^2 \theta_{13})}{\sin 2 \theta_{12} \sin \theta_{13} |\cos \theta^{\circ}_{13} \cos \theta^{\circ}_{23}| (\cos^2 \theta_{13} - \cos^2 \theta^{\circ}_{13} \cos^2 \theta^{\circ}_{23})^{\frac{1}{2}}} \,,
\label{eq:cosdeltaA1} \\
J_\mathrm{CP} &= -\frac{1}{8} \sin2\theta^{\circ}_{13} \sin2\theta^{\circ}_{23} \cos\theta^{\circ}_{23} \sin2\theta \sin\phi\,,
\label{eq:JCPA1}
\end{align}
%%%%%
%
where $\theta \in [0,\pi) = [0,180^\circ)$ and $\phi \in [0,2\pi) = [0,360^\circ)$ are free parameters related to $\theta^e_{12}$ and $\delta^e_{12}$ (see Appendix~B of Ref.~\cite{Girardi:2015rwa} for details). 
From Eqs.~\eqref{eq:ss23A1} and \eqref{eq:cosdeltaA1}, we see that 
in this case there are two relations not explicitly involving the free parameters: 
(i)~between $\sin^2\theta_{23}$ and $\sin^2\theta_{13}$ and 
(ii)~among $\cos\delta_\mathrm{CP}$, $\theta_{12}$, and $\theta_{13}$. 
This can be expected, since four observables ($\sin^2\theta_{ij}$ and $\delta_\mathrm{CP}$) have been expressed in terms of two parameters 
($\theta$ and $\phi$). 
To obtain the value of $\delta_\mathrm{CP}$ for given $(\theta,\phi)$, 
the expression for $J_\mathrm{CP}$ in Eq.~\eqref{eq:JCPA1} has to be compared 
to its expression in the standard parametrization given in Eq.~\eqref{eq:JCP}. 
We note that $\sin^2\theta_{ij}$ in Eqs.~\eqref{eq:ss13A1}--\eqref{eq:ss12A1}, 
and hence, $\cos\delta_\mathrm{CP}$ in Eq.~\eqref{eq:cosdeltaA1}, 
depend on $\theta$ and $\phi$ via $\sin^2\theta$ and the product $\sin2\theta\cos\phi$. There are two transformations that leave them invariant:
%%%%%
\begin{equation}
\begin{cases}
\theta \to \pi - \theta \\
\phi \to \pi - \phi
\end{cases} 
\qquad \text{and} \qquad
\phi \to 2\pi-\phi\,.
\end{equation}
%%%%%
%
Under both of these transformations, $J_\mathrm{CP}$ changes sign, and as a consequence, also 
$\delta_\mathrm{CP} \to -\delta_\mathrm{CP}$. Thus, for given $(\theta,\phi)$, 
we have four solutions that lead to the same values of $\sin^2\theta_{ij}$, 
of which two~---~$(\theta,\phi)$ and $(\pi-\theta,\pi+\phi)$~---~give $\delta_\mathrm{CP}$ and the other two~---~$(\pi-\theta,\pi-\phi)$ and $(\theta,2\pi-\phi)$~---~yield $-\delta_\mathrm{CP}$. 
This observation applies to all two-parameter cases considered below.
 
 \textbf{Case A2.}
In this case, the corresponding free rotation matrix is 
$U_{13}(\theta^{e}_{13},\delta^{e}_{13})$. 
The leptonic mixing angles, $\cos\delta_\mathrm{CP}$, 
and $J_\mathrm{CP}$ are defined by
%%%%%
\begin{align}
\sin^2 \theta_{13} & = \sin^2 \theta \cos^2 \theta^{\circ}_{23} \,, 
\label{eq:ss13A2}\\
\sin^2 \theta_{23} & = \frac{\sin^2 \theta^{\circ}_{23}}{1 - \sin^2 \theta_{13}}\,, 
\label{eq:ss23A2}\\
\sin^2 \theta_{12} & = \frac{\cos^2 \theta \sin^2 \theta^{\circ}_{12} + \cos^2 \theta^{\circ}_{12} \sin^2 \theta \sin^2 \theta^{\circ}_{23} 
-\frac{1}{2} \sin 2 \theta \sin 2 \theta^{\circ}_{12} \sin \theta^{\circ}_{23} \cos \phi }{1 - \sin^2 \theta_{13}} \,,
\label{eq:ss12A2} \\
\cos \delta_\mathrm{CP} &= \frac{\cos^2 \theta_{13} (\cos^2 \theta_{12} -\cos^2 \theta^{\circ}_{12} \cos^2 \theta^{\circ}_{23}) - \sin^2 \theta^{\circ}_{23} (\cos^2 \theta_{12} - \sin^2 \theta_{12} \sin^2 \theta_{13})}{\sin 2 \theta_{12} \sin \theta_{13} |\sin \theta^{\circ}_{23}| (\cos^2 \theta_{13} - \sin^2 \theta^{\circ}_{23})^{\frac{1}{2}}} \,,
\label{eq:cosdeltaA2} \\
J_\mathrm{CP} &= \frac{1}{8} \sin2\theta^{\circ}_{12} \sin2\theta^{\circ}_{23} \cos\theta^{\circ}_{23} \sin2\theta \sin\phi\,.
\label{eq:JCPA2}
\end{align}
%%%%%
%
The free parameters $\theta$ and $\phi$ are related to 
$\theta^e_{13}$ and $\delta^e_{13}$. As in the previous case, 
we find sum rules for $\sin^2\theta_{23}$ and $\cos\delta_\mathrm{CP}$.

 \textbf{Case A3.} The correction to a fixed part of $U_\mathrm{PMNS}$ 
due to $U_{23}(\theta^{e}_{23},\delta^{e}_{23})$ leads to 
%%%%%
\begin{align}
\sin^2 \theta_{13} &= \sin^2 \theta^{\circ}_{13}\,, 
\label{eq:ss13A3} \\
\sin^2 \theta_{23} &= \sin^2 \theta \,, 
\label{eq:ss23A3} \\
\sin^2 \theta_{12} &= \sin^2 \theta^{\circ}_{12}\,, 
\label{eq:ss12A3} \\
\cos \delta_{\rm CP} &= \pm \cos \phi\,,
\label{eq:cosdeltaA3} \\
\sin\delta_\mathrm{CP} &= \mp \sin\phi\,.
\label{eq:sindeltaA3}
\end{align}
%%%%%
%
Therefore, while the mixing angles $\theta_{12}$ and $\theta_{13}$ are predicted to have 
certain fixed values, the mixing angle $\theta_{23}$ and the Dirac CPV phase $\delta_\mathrm{CP}$ remain 
unconstrained in this case. 

 \textbf{Case B1.} 
For pattern B, the leptonic mixing matrix is defined up to a free complex rotation from the right. 
If this rotation is $U_{13}(\theta^\nu_{13},\delta^\nu_{13})$, one has
%%%%%
\begin{align}
\sin^2 \theta_{13} & = \cos^2 \theta^{\circ}_{12} \sin^2 \theta \,, 
\label{eq:ss13B1}\\
\sin^2 \theta_{23} & = \frac{\cos^2 \theta^{\circ}_{23} \sin^2 \theta \sin^2 \theta^{\circ}_{12} 
+ \cos^2 \theta \sin^2 \theta^{\circ}_{23} 
- \frac{1}{2} \sin 2 \theta \sin 2 \theta^{\circ}_{23} \sin \theta^{\circ}_{12} \cos \phi }{1 - \sin^2 \theta_{13}} \,, 
\label{eq:ss23B1}\\
\sin^2 \theta_{12} & = \frac{\sin^2 \theta^{\circ}_{12}}{1 - \sin^2 \theta_{13}} \,,
\label{eq:ss12B1} \\
\cos \delta_{\rm CP} &= \frac{\cos^2 \theta_{13} (\cos^2 \theta_{23} - \cos^2 \theta^{\circ}_{12} \cos^2 \theta^{\circ}_{23}) - \sin^2 \theta^{\circ}_{12} (\cos^2 \theta_{23} - \sin^2 \theta_{13} \sin^2 \theta_{23})}
{\sin 2 \theta_{23} \sin \theta_{13} |\sin \theta^{\circ}_{12}| (\cos^2 \theta_{13} - \sin^2 \theta^{\circ}_{12})^{\frac{1}{2}}} \,,
\label{eq:cosdeltaB1} \\
J_\mathrm{CP} &= -\frac{1}{8} \sin2\theta^{\circ}_{23} \sin2\theta^{\circ}_{12} \cos\theta^{\circ}_{12} \sin2\theta \sin\phi\,,
\label{eq:JCPB1}
\end{align}
%%%%%
%
where $\theta$ and $\phi$ are related to $\theta^\nu_{13}$ and $\delta^\nu_{13}$ 
and the angles $\theta^\circ_{ij }$ are fixed once the residual symmetries 
$G_e$ and $G_\nu$ originating from breaking $G_f$ are specified. 
This case is characterized by the sum rules for $\sin^2\theta_{12}$ and $\cos\delta_\mathrm{CP}$, i.e., Eqs.~\eqref{eq:ss12B1} and \eqref{eq:cosdeltaB1}, respectively.

 \textbf{Case B2.} 
In the case of $U_{23}(\theta^{\nu}_{23}, \delta^{\nu}_{23})$, 
the leptonic mixing parameters and $J_\mathrm{CP}$ are given by
%%%%%
\begin{align}
\sin^2 \theta_{13} & = \cos^2 \theta^{\circ}_{13} \sin^2 \theta^{\circ}_{12} \sin^2 \theta + \sin^2 \theta^{\circ}_{13} \cos^2 \theta 
+ \frac{1}{2} \sin 2 \theta \sin 2 \theta^{\circ}_{13} \sin \theta^{\circ}_{12} \cos \phi \,, 
\label{eq:ss13B2}\\
\sin^2 \theta_{23} & = \frac{\cos^2 \theta^{\circ}_{12} \sin^2 \theta}{1 - \sin^2 \theta_{13}} \,, 
\label{eq:ss23B2}\\
\sin^2 \theta_{12} & = \frac{\cos^2 \theta_{13} - \cos^2 \theta^{\circ}_{12} \cos^2 \theta^{\circ}_{13}  }{1 - \sin^2 \theta_{13}} \,,
\label{eq:ss12B2} \\
\cos \delta_{\rm CP} &= 
\dfrac{\cos^2 \theta_{13} (\sin^2 \theta^{\circ}_{12} - \cos^2 \theta_{23}) + \cos^2 \theta^{\circ}_{12} \cos^2 \theta^{\circ}_{13} ( \cos^2 \theta_{23} - \sin^2 \theta_{13} \sin^2 \theta_{23} )}
{ \sin 2 \theta_{23} \sin \theta_{13} | \cos \theta^{\circ}_{12} \cos \theta^{\circ}_{13}| (\cos^2 \theta_{13} - \cos^2 \theta^{\circ}_{12} \cos^2 \theta^{\circ}_{13} )^{\frac{1}{2}}} \,,
\label{eq:cosdeltaB2} \\
J_\mathrm{CP} &= \frac{1}{8} \sin2\theta^{\circ}_{13} \sin2\theta^{\circ}_{12} \cos\theta^{\circ}_{12} \sin2\theta \sin\phi\,.
\label{eq:JCPB2}
\end{align}
%%%%%
%
Here, $\theta$ and $\phi$ are free parameters arising from $\theta^\nu_{23}$ and $\delta^\nu_{23}$. 
Analogously to the previous case, $\sin^2\theta_{12}$ is related to 
$\sin^2\theta_{13}$ via Eq.~\eqref{eq:ss12B2}, while $\cos\delta_\mathrm{CP}$ 
is expressed in terms of $\theta_{13}$ and $\theta_{23}$ 
through Eq.~\eqref{eq:cosdeltaB2}. 

 \textbf{Case B3.} 
For $U_{12}(\theta^\nu_{12},\delta^\nu_{12})$, 
the following simple relations hold
%%%%%
\begin{align}
\sin^2 \theta_{13} &= \sin^2 \theta^{\circ}_{13}\,,
\label{eq:ss13B3} \\
\sin^2 \theta_{23} &= \sin^2 \theta^{\circ}_{23}\,,
\label{eq:ss23B3} \\
\sin^2 \theta_{12} &= \sin^2 \theta\,,
\label{eq:ss12B3} \\
\cos \delta_{\rm CP} &= \pm \cos \phi\,,
\label{eq:cosdeltaB3} \\
\sin\delta_\mathrm{CP} &= \pm \sin\phi\,,
\label{eq:sindeltaB3}
\end{align}
%%%%%
%
i.e., while the mixing angles $\theta_{13}$ and $\theta_{23}$ are predicted to have 
certain fixed values, the mixing angle $\theta_{12}$ and the Dirac CPV phase $\delta_\mathrm{CP}$ remain unconstrained. 

It is worth noting that Cases A1, A2, B1, and B2 lead to non-trivial predictions for the Dirac CPV phase $\delta_\mathrm{CP}$.
The study of these cases for $G_f = A_4$, $S_4$, and $A_5$ 
and all their Abelian subgroups, 
which can play the role of the residual symmetries $G_e$ and $G_\nu$, 
has been performed in Refs.~\cite{Girardi:2015rwa,Petcov:2018snn} 
in light of global neutrino oscillation data. 
For $G_f = S_4$, only two viable cases have been found~\cite{Petcov:2018snn}. 
We display the corresponding residual symmetries and 
the values of the parameters $\sin^2\theta^\circ_{ij}$ 
fixed by them in Table~\ref{tab:S4}. 
%%%%%%%%%%%%%%%%%%%%%%%%%%%%%%%%%%%%%
\begin{table}[t]
\centering
\renewcommand*{\arraystretch}{1.5}
\begin{tabular}{|cc|c|ccc|cc|}
\hline
$G_e$ & $G_\nu$ & Case & 
$\sin^2\theta^{\circ}_{12}$ & $\sin^2\theta^\circ_{13}$& $\sin^2\theta^\circ_{23}$ \\
\hline
\hline
\multirow{2}{*}{$Z_3$} & \multirow{2}{*}{$Z_2$} & B1 & 
$1/3$ & $-$ & $1/2$ \\ 
& & B2 ($S_4$) & $1/6$ & $1/5$ & $-$ \\
\hline
\end{tabular}
\caption{Values of the fixed parameters $\sin^2\theta^\circ_{ij}$ 
for viable cases originating from $G_f=S_4$ broken 
to the residual symmetries $G_e$ and $G_\nu$, 
as derived in Ref.~\cite{Girardi:2015rwa} (see also Ref.~\cite{Petcov:2018snn}).
The entries marked with ``$-$'' are not relevant. 
}
\label{tab:S4}
\end{table}
%%%%%%%%%%%%%%%%%%%%%%%%%%%%%%%%%%%%%
%
For $G_f = A_5$, there are six viable cases~\cite{Petcov:2018snn}, 
which we summarize in Table~\ref{tab:A5}.
%%%%%%%%%%%%%%%%%%%%%%%%%%%%%%%%%%%%%
\begin{table}[t]
\centering
\renewcommand*{\arraystretch}{1.5}
\begin{tabular}{|cc|c|ccc|ccc|}
\hline
$G_e$ & $G_\nu$ & Case & $\sin^2\theta^{\circ}_{12}$ & $\sin^2\theta^\circ_{13}$ & $\sin^2\theta^\circ_{23}$ \\ 
\hline
\hline
\multirow{2}{*}{$Z_2$} & \multirow{2}{*}{$Z_3$} & A1 ($A_5$) & 
$-$ & $0.226$ & $0.436$ \\ 
& & A2 ($A_5$) & 
$0.226$ & $-$ & $0.436$ \\ 
\hline
$Z_3$ & $Z_2$ & B1 & 
$1/3$ & $-$ & $1/2$ \\
\hdashline
$Z_5$ & $Z_2$ & B1 ($A_5$) & 
$0.276$ & $-$ & $1/2$ \\ 
\hdashline
\multirow{2}{*}{$Z_2 \times Z_2$} & \multirow{2}{*}{$Z_2$} & B2 ($A_5$) & 
$0.095$ & $0.276$ & $-$ \\
& & B2 ($A_5$ II) & 
$1/4$ & $0.127$ & $-$ \\ 
\hline
\end{tabular}
\caption{Values of the fixed parameters $\sin^2\theta^\circ_{ij}$ 
for viable cases originating from $G_f=A_5$ broken 
to the residual symmetries $G_e$ and $G_\nu$, 
as derived in Ref.~\cite{Girardi:2015rwa} (see also Ref.~\cite{Petcov:2018snn}).
The entries marked with ``$-$'' are not relevant.
}
\label{tab:A5}
\end{table}
%%%%%%%%%%%%%%%%%%%%%%%%%%%%%%%%%%%%%
%
Irrational values of $\sin^2 \theta^{\circ}_{ij}$ quoted therein
can be expressed in terms of the golden ratio $\varphi$ as follows~\cite{Girardi:2015rwa}:
$2/(4 \varphi^2 - \varphi) \approx 0.226$, 
$\varphi/(6\varphi - 6) \approx 0.436$,
$1/(2 + \varphi) \approx 0.276$,
$1/(4 \varphi^2) \approx 0.095$, 
and $1/(3 + 3 \varphi) \approx 0.127$. 
Finally, we notice that Case~B1 common to both $S_4$ and $A_5$ 
can also be realized from $A_4$. 
In fact, it is the only viable case arising from $A_4$ 
broken to non-trivial $G_e$ and $G_\nu$~\cite{Girardi:2015rwa}.

%-----------------------------------------------------------
\subsection{Models with three free parameters}
%-----------------------------------------------------------

Here, we briefly discuss scenarios for which 
$U_\mathrm{PMNS}$ depends on three free parameters. 
For instance, this happens if $G_f$ is broken to 
$G_e = Z_2$ and $G_\nu = Z_2$. 
In this case both $U_e$ and $U_\nu$ are defined up to 
complex rotations. Thus, naively, we have four free parameters~---~two 
angles and two phases. However, as demonstrated in Ref.~\cite{Girardi:2015rwa}, the number of free parameters reduces to three~---~two angles 
and one phase~---~after an appropriate rearrangement. 
Since four observables are now expressed in terms of three parameters, 
one relation between the observables can be expected. 
Indeed, depending on the planes in which the two complex rotations 
are imposed, either $\cos\delta_\mathrm{CP}$ is determined 
by the mixing angles or one relation among the latter is found.

Another example of a three-parameter setup is given by 
$G_e = Z_2$ and $G_\nu = Z_2 \times \text{CP}$. 
In this case, the free rotation in the neutrino sector is real. 
Such a breaking pattern has been investigated in Ref.~\cite{Penedo:2017vtf} 
for $G_\mathrm{CP} = S_4 \rtimes \text{CP}$ 
and in Ref.~\cite{Turner:2015uta} for $G_\mathrm{CP} = A_5 \rtimes \text{CP}$.
Furthermore, if $G_e$ is larger than $Z_2$ and a single 
residual CP transformation is preserved in the neutrino sector, $U_\mathrm{PMNS}$ depends 
on three free angles~\cite{Lu:2016jit}.

Finally, there exists a possibility that $G_f$ is fully broken 
in the charged lepton (neutrino) sector, 
while the form of the matrix $U_\nu$ ($U_e$) is fully determined 
by a residual $G_\nu$ ($G_e$) symmetry (which is larger than $Z_2$).
In such a case, the unitary matrix $U_e$ ($U_\nu$) is unconstrained, 
unless its form is motivated by some additional arguments.
A well-studied example is given by 
$U_e^\dagger = U_{12}(\theta^e_{12},\delta^e_{12})
U_{23}(\theta^e_{23},\delta^e_{23})$ and 
$U_\nu$ being the BM, TBM, or GR mixing matrix.
In this case, $U_e$ provides necessary corrections to $U_\nu$, 
in particular, generating a non-zero $\theta_{13}$ and shifting 
$\theta_{12}$ from its leading order value $\theta^\nu_{12}$.%
\footnote{We abuse the notation $\theta^\nu_{12}$, 
which here denotes the value of $\theta_{12}$ for BM, TBM, or GR mixing, 
and not a free parameter as in Subsection~\ref{sec:2params}.}
Such scenario leads to the following sum rule~\cite{Petcov:2014laa}
%%%%%%%%%%
\begin{equation}
\cos\delta_\mathrm{CP} = \frac{\tan\theta_{23}}{\sin2\theta_{12}\, \sin\theta_{13}} 
\left[\cos2\theta^\nu_{12} + \left(\sin^2\theta_{12} - \cos^2\theta^\nu_{12}\right)
\left(1 - \cot^2\theta_{23}\,\sin^2\theta_{13}\right)\right]\,. 
\label{eq:cosdelta}
\end{equation}
%%%%%%%%%%
It has been studied in detail in Refs.~\cite{Girardi:2014faa,Ballett:2014dua} 
and the effects of renormalization group evolution 
of the leptonic mixing parameters on its predictions 
have been further investigated in Refs.~\cite{Zhang:2016djh,Gehrlein:2016fms}.
In Ref.~\cite{Agarwalla:2017wct}, this sum rule 
has been thoroughly analyzed in the context of DUNE and T2HK. 
Alternative sum rules arising from the charged lepton corrections 
have been derived in Ref.~\cite{Girardi:2015vha}. 

In the next section, we will confront eleven 
one-parameter mixing patterns 
(see Tables~\ref{tab:FHZ1}--\ref{tab:LD2}) and seven two-parameter 
ones (see Tables~\ref{tab:S4}--\ref{tab:A5}) 
with the current global neutrino oscillation data. 
Note that we will not consider the two-parameter Cases~A3 and B3, 
since they do not lead to predictions for $\cos\delta_\mathrm{CP}$ 
[cf.~Eqs.~\eqref{eq:cosdeltaA3} and \eqref{eq:cosdeltaB3}]. 
Finally, we will consider neither fully-fixed mixing patterns 
nor those depending on three free parameters.
While the former to be viable requires 
very large discrete groups (see Subsection~\ref{sec:0params}), 
the predictive power of the latter is less than 
that of models with one or two free parameters.

%===============
\section{Confronting the Flavor Models with Global Neutrino Oscillation Data}
\label{sec:confronting}
%===============

Fits to global neutrino oscillation data have basically been performed by three groups (see Refs.~\cite{Esteban:2018azc,Capozzi:2018ubv,deSalas:2017kay}). In our analysis, we use the results of the so-called NuFIT group announced in July 2019 \cite{NuFiTv41}. In Table~\ref{tab:NuFit}, we list the current best-fit values of the leptonic mixing parameters as well as their corresponding $1\sigma$ errors and $3\sigma$ ranges.
\begin{table}[!t]
\centering
\renewcommand*{\arraystretch}{1.5}
\begin{tabular}{|c|c|c|}
\hline
Parameter & Best-fit value with $1\sigma$ error & $3\sigma$ range \\
\hline
\hline
$\sin^2 \theta_{12}$ & $0.310^{+0.013}_{-0.012}$ & $(0.275,0.350)$ \\
$\sin^2 \theta_{13}$ & $0.02237^{+0.00066}_{-0.00065}$ & $(0.02044,0.02435)$ \\
$\sin^2 \theta_{23}$ & $0.563^{+0.018}_{-0.024}$ & $(0.433,0.609)$ \\
$\delta_{\rm CP}$ & $\left(221^{+39}_{-28}\right)^\circ$ & $(144^\circ,357^\circ)$ \\
$\Delta m^2_{21}$ & $(7.39^{+0.21}_{-0.20}) \cdot 10^{-5} \, {\rm eV}^2$ & $(6.79,8.01) \cdot 10^{-5} \, {\rm eV}^2$ \\
$\Delta m^2_{31}$ & $+(2.528^{+0.029}_{-0.031}) \cdot 10^{-3} \, {\rm eV}^2$ & $+(2.436,2.618) \cdot 10^{-3} \, {\rm eV}^2$ \\
\hline
\end{tabular}
\caption{Current best-fit values of the leptonic mixing parameters with their corresponding $1\sigma$ errors and $3\sigma$ ranges from NuFIT~4.1 [July 2019, NO with atmospheric neutrino oscillation data from Super-Kamiokande~(SK)], see Ref.~\cite{NuFiTv41}.}
\label{tab:NuFit}
\end{table}
In order to confront the one- and two-parameter lepton flavor models discussed in Section~\ref{sec:models} with global neutrino oscillation data and to determine the allowed values of the model parameters, we define the $\chi^2$ function as follows
\begin{equation}
\chi^2(\theta) = \left[ \frac{\sin^2 \theta_{12}(\theta) - \sin^2 \theta_{12}}{\sigma(\sin^2 \theta_{12})} \right]^2 + \left[ \frac{\sin^2 \theta_{13}(\theta) - \sin^2 \theta_{13}}{\sigma(\sin^2 \theta_{13})} \right]^2 + \left[ \frac{\sin^2 \theta_{23}(\theta) - \sin^2 \theta_{23}}{\sigma(\sin^2 \theta_{23})} \right]^2
\label{eq:chisq1p}
\end{equation}
for the one-parameter models and
\begin{align}
\chi^2(\theta,\phi) &= \left[ \frac{\sin^2 \theta_{12}(\theta,\phi) - \sin^2 \theta_{12}}{\sigma(\sin^2 \theta_{12})} \right]^2 + \left[ \frac{\sin^2 \theta_{13}(\theta,\phi) - \sin^2 \theta_{13}}{\sigma(\sin^2 \theta_{13})} \right]^2 \nonumber\\
&+ \left[ \frac{\sin^2 \theta_{23}(\theta,\phi) - \sin^2 \theta_{23}}{\sigma(\sin^2 \theta_{23})} \right]^2
\label{eq:chisq2p}
\end{align}
for the two-parameter models, where the parameter values have been chosen as $\sin^2\theta_{12} = 0.310$, $\sin^2\theta_{13} =  0.02237$, and $\sin^2\theta_{23} = 0.563$ and the errors have been assumed to be $\sigma(\sin^2\theta_{12}) = 0.013$, $\sigma(\sin^2\theta_{13}) = 0.00066$, and $\sigma(\sin^2\theta_{23}) = 0.024$.

In Table~\ref{tab:fits}, we present the results of our fits including the minimal values of the $\chi^2$ function (i.e., $\chi^2_{\min{}}$) for the eleven relevant one-parameter models and the seven relevant two-parameter models and rearrange them according to how well they fit the data of NuFIT~4.1. In addition, we give the best-fit values of the corresponding model parameters. In the case of the one-parameter models, the best-fit parameter and its corresponding $3\sigma$ interval are denoted $\theta_{\rm bf}$ and $\theta_{3\sigma}$, respectively, whereas in the case of the two-parameter models, the best-fit parameters and their corresponding $3\sigma$ intervals are denoted by $\theta_{\rm bf}, \theta_{3\sigma}$ and $\phi_{\rm bf}, \phi_{3\sigma}$, respectively. To calculate the $3 \sigma$ intervals for both one- and two-parameter models, we impose $\chi^2 - \chi^2_{\min{}} = 3^2 = 9$, corresponding to one degree of freedom ({\rm d.o.f.}). In fact, this holds also for the two-parameter models, since presenting the $3\sigma$ interval for one of the two parameters, we minimize $\chi^2$ over the other parameter. 
Note that the model parameters $\theta$ and $\phi$ have different meaning for different models.
\begin{table}%
\vspace{-1.75cm}
\renewcommand*{\arraystretch}{1.5}
\begin{center}
\begin{tabular}{|c|c|c|cc|cc|}
\hline
Model & Case & $\chi^2_{\min}$ & $\theta_{\rm bf}$ & $\theta_{3\sigma}$ & $\phi_{\rm bf}$ & $\phi_{3\sigma}$ \\
\hline
\hline
\cellcolor{model11}1.1 & VII-b ($A_5$) & 5.37 & $17.0^\circ$ & $(16.3^\circ,17.7^\circ)$ & $-$ & $-$ \\
\hline
\cellcolor{model12}1.2 & III ($A_5$) & 5.97 & $169.9^\circ$ & $(169.4^\circ,170.4^\circ)$ & $-$ & $-$ \\
\hline
\cellcolor{model13} & \multirow{2}{*}{IV ($S_4$)} & & $15.0^\circ$ & $(14.3^\circ,15.7^\circ)$ & $-$ & $-$ \\
\cellcolor{model13}\multirow{-2}{*}{1.3} & & \multirow{-2}{*}{7.28} & $165.0^\circ$ & $(164.3^\circ,165.7^\circ)$ & $-$ & $-$ \\
\hline
\cellcolor{model14}{\color{white} 1.4} & II ($S_4$) & 8.91 & $169.5^\circ$ & $(169.0^\circ,170.0^\circ)$ & $-$ & $-$ \\
\hline
\cellcolor{model15} & \multirow{2}{*}{IV ($A_5$)} & & $10.1^\circ$ & $(9.6^\circ,10.6^\circ)$ & $-$ & $-$ \\
\cellcolor{model15}\multirow{-2}{*}{{\color{white} 1.5}} & & \multirow{-2}{*}{11.3} & $169.9^\circ$ & $(169.4^\circ,170.4^\circ)$ & $-$ & $-$ \\
\hline
\cellcolor{model16} & \multirow{2}{*}{I ($S_4$)} & & $10.5^\circ$ & $(10.0^\circ,11.1^\circ)$ & $-$ & $-$ \\
\cellcolor{model16}\multirow{-2}{*}{1.6} & & \multirow{-2}{*}{12.6} & $169.5^\circ$ & $(168.9^\circ,170.0^\circ)$ & $-$ & $-$ \\
\hline
\cellcolor{model17}1.7 & VII-a ($A_5$) & 14.8 & $16.9^\circ$ & $(16.2^\circ,17.6^\circ)$ & $-$ & $-$ \\
\hline
\cellcolor{model18}1.8 & VI-b ($S_4$) & 18.1 & $115.3^\circ$ & $(114.8^\circ,115.8^\circ)$ & $-$ & $-$ \\
\hline
\cellcolor{model19} & \multirow{2}{*}{II ($A_5$)} & & $16.5^\circ$ & $(15.7^\circ,17.3^\circ)$ & $-$ & $-$ \\
\cellcolor{model19}\multirow{-2}{*}{1.9} & & \multirow{-2}{*}{21.8} & $163.5^\circ$ & $(162.7^\circ,164.3^\circ)$ & $-$ & $-$ \\
\hline
\cellcolor{model110}{\color{white} 1.10} & V ($S_4$) & 36.8 & $165.2^\circ$ & $(164.4^\circ,165.9^\circ)$ & $-$ & $-$ \\
\hline
\cellcolor{model111}{\color{white} 1.11} & VI-a ($S_4$) & 53.8 & $115.3^\circ$ & $(114.8^\circ,115.8^\circ)$ & $-$ & $-$ \\
\hline
\hline
\cellcolor{model21} & \multirow{2}{*}{A1 ($A_5$)} & & $47.2^\circ$& $(43.2^\circ,50.9^\circ)$ & $163.2^\circ$ & $(158.0^\circ,180^\circ]$ \\
\cellcolor{model21}\multirow{-2}{*}{2.1} & & \multirow{-2}{*}{0.151} & $132.8^\circ$& $(129.1^\circ,136.8^\circ)$ & $16.8^\circ$ & $[0,22.0^\circ)$ \\
\hline
\cellcolor{model22} & \multirow{2}{*}{B2 ($S_4$)} & & $54.4^\circ$ & $(49.3^\circ,59.7^\circ)$ & $149.7^\circ$ & $(148.0^\circ,154.3^\circ)$ \\
\cellcolor{model22}\multirow{-2}{*}{2.2} & & \multirow{-2}{*}{0.386} & $125.6^\circ$ & $(120.3^\circ,130.7^\circ)$ & $30.3^\circ$ & $(25.7^\circ,32.0^\circ)$ \\
\hline
\cellcolor{model23} & \multirow{2}{*}{B2 ($A_5$)} & & $51.3^\circ$ & $(48.2^\circ,56.0^\circ)$ & $161.4^\circ$ & $(150.4^\circ,180^\circ]$ \\
\cellcolor{model23}\multirow{-2}{*}{2.3} & & \multirow{-2}{*}{2.49} & $128.7^\circ$ & $(124.0^\circ,131.8^\circ)$ & $18.6^\circ$ & $[0,29.6^\circ)$ \\
\hline
\cellcolor{model24} & \multirow{2}{*}{B1 ($A_5$)} & & $10.1^\circ$ & $(9.6^\circ,10.6^\circ)$ & $132.6^\circ$ & $(84.4^\circ,180^\circ]$ \\
\cellcolor{model24}\multirow{-2}{*}{2.4} & & \multirow{-2}{*}{4.40} & $169.9^\circ$ & $(169.4^\circ,170.4^\circ)$ & $47.4^\circ$ & $[0,95.6^\circ)$ \\
\hline
\cellcolor{model25} & \multirow{2}{*}{B1} & & $10.5^\circ$ & $(10.0^\circ,11.0^\circ)$ & $126.4^\circ$ & $(85.1^\circ,180^\circ]$ \\
\cellcolor{model25}\multirow{-2}{*}{2.5} & & \multirow{-2}{*}{5.67} & $169.5^\circ$ & $(169.0^\circ,170.0^\circ)$ & $53.6^\circ$ & $[0,94.9^\circ)$ \\
\hline
\cellcolor{model26} & \multirow{2}{*}{B2 ($A_5$ II)} & & $52.2^\circ$ & $(50.1^\circ,52.9^\circ)$ & $180^\circ$ & $(164.7^\circ,180^\circ]$ \\
\cellcolor{model26}\multirow{-2}{*}{{\color{white} 2.6}} & & \multirow{-2}{*}{14.8} & $127.8^\circ$ & $(127.1^\circ,129.9^\circ)$ & 0 & $[0,15.3^\circ)$ \\
\hline
\cellcolor{model27} & \multirow{2}{*}{A2 ($A_5$)} & & $11.5^\circ$ & $(10.9^\circ,12.0^\circ)$ & $132.4^\circ$ & $(108.6^\circ,180^\circ]$ \\
\cellcolor{model27}\multirow{-2}{*}{{\color{white} 2.7}} & & \multirow{-2}{*}{23.6} & $168.5^\circ$ & $(168.0^\circ,169.1^\circ)$ & $47.6^\circ$ & $[0,71.4^\circ)$ \\
\hline
\end{tabular}
\end{center}
\caption{Fits of the 18 one- and two-parameter models to global neutrino oscillation data from NuFIT~4.1 (July 2019, NO with SK atmospheric data) \cite{NuFiTv41}. See also Table~\ref{tab:NuFit}. Note that there are four degenerate best-fit points $(\theta_{\rm bf},\phi_{\rm bf})$ for each two-parameter model (i.e., Models~2.1--2.7) that lead to the same value of the $\chi^2_{\min{}}$ function. Two best-fit points are presented for each model and the other two can be found performing the replacement rule $\phi_{\rm bf} \to 2\pi - \phi_{\rm bf}$. In addition, the two presented best-fit points for each model are related by the simultaneous replacement of $\theta_{\rm bf} \to \pi - \theta_{\rm bf}$ and $\phi_{\rm bf} \to \pi - \phi_{\rm bf}$.}
\label{tab:fits}
\end{table}
In Table~\ref{tab:fits2}, as an outcome of the minimization of the $\chi^2$ function, we display the best-fit values of the leptonic mixing parameters $\sin^2 \theta_{12}({\rm bf})$, $\sin^2 \theta_{13}({\rm bf})$, and $\sin^2 \theta_{23}({\rm bf})$, which are evaluated at the best-fit values of the model parameters. Note that some of the models have fixed values of some of the leptonic mixing parameters, which stems from the fact that these parameters are independent from the corresponding model parameters. For example, $\sin^2 \theta_{23}({\rm bf}) = 1/2$ for Models~1.3, 1.5, 1.6, and 1.9. Furthermore, we calculate the predicted values of the CPV phase $\delta_{\rm CP}$ for the various one- and two-parameter models at the best-fit values of the corresponding model parameters (given in Table~\ref{tab:fits}), which are also presented in Table~\ref{tab:fits2}. We observe that all one-parameter models have fixed values of $\delta_{\rm CP}$, which are either $0$ and $180^\circ$ (CP conservation) or $-90^\circ$ and $90^\circ$ (maximal CP violation) and these values are, by construction, given by the respective model. For the two-parameter models, it is interesting to note that the negative values of $\delta_{\rm CP}$ for Models~2.3--2.5 and 2.7 are within the $1\sigma$ interval of $(-167\degree,-100\degree)$ \cite{NuFiTv41}.
\begin{table}[!t]
\renewcommand*{\arraystretch}{1.5}
\begin{center}
\begin{tabular}{|c|c|ccc|c|}
\hline
Model & $\chi^2_{\min}$ & $\sin^2 \theta_{12}({\rm bf})$ & $\sin^2 \theta_{13}({\rm bf})$ & $\sin^2 \theta_{23}({\rm bf})$ & $\delta_{\rm CP}$ \\
\hline
\hline
\cellcolor{model11}1.1 & 5.37 & 0.331 & 0.0223 & 0.523 & $180^\circ$ \\
\hline
\cellcolor{model12}1.2 & 5.97 & 0.283 & 0.0223 & 0.593 & $180^\circ$ \\
\hline
\cellcolor{model13}1.3 & 7.28 & 0.318 & 0.0224 & $\tfrac{1}{2}$ & $\{-90^\circ,90^\circ\}$ \\
\hline
\cellcolor{model14}{\color{white} 1.4} & 8.91 & 0.341 & 0.0223 & 0.606 & $180^\circ$ \\
\hline
\cellcolor{model15}{\color{white} 1.5} & 11.3 & 0.283 & 0.0224 & $\tfrac{1}{2}$ & $\{-90^\circ,90^\circ\}$ \\
\hline
\cellcolor{model16}1.6 & 12.6 & 0.341 & 0.0223 & $\tfrac{1}{2}$ & $\{-90^\circ,90^\circ\}$ \\
\hline
\cellcolor{model17}1.7 & 14.8 & 0.330 & 0.0227 & 0.480 & 0 \\
\hline
\cellcolor{model18}1.8 & 18.1 & 0.256 & 0.0224 & 0.582 & 0 \\
\hline
\cellcolor{model19}1.9 & 21.8 & 0.260 & 0.0223 & $\tfrac{1}{2}$ & $\{-90^\circ,90^\circ\}$ \\
\hline
\cellcolor{model110}{\color{white} 1.10} & 36.8 & 0.318 & 0.0219 & 0.707 & 0 \\
\hline
\cellcolor{model111}{\color{white} 1.11} & 53.8 & 0.256 & 0.0226 & 0.418 & $180^\circ$ \\
\hline
\hline
\cellcolor{model21}2.1 & 0.151 & 0.310 & 0.0224 & 0.554 & $\{-41.9^\circ,41.9^\circ\}$ \\
\hline
\cellcolor{model22}2.2 & 0.386 & 0.318 & 0.0224 & 0.563 & $\{74.0^\circ,-74.0^\circ\}$ \\
\hline
\cellcolor{model23}2.3 & 2.49 & 0.330 & 0.0224 & 0.563 & $\{145.2^\circ,-145.2^\circ\}$ \\
\hline
\cellcolor{model24}2.4 & 4.40 & 0.283 & 0.0224 & 0.563 & $\{-132.1^\circ,132.1^\circ\}$ \\
\hline
\cellcolor{model25}2.5 & 5.67 & 0.341 & 0.0223 & 0.563 & $\{-125.7^\circ,125.7^\circ\}$ \\
\hline
\cellcolor{model26}{\color{white} 2.6} & 14.8 & 0.330 & 0.0227 & 0.480 & $0$ \\
\hline
\cellcolor{model27}{\color{white} 2.7} & 23.6 & 0.310 & 0.0224 & 0.446 & $\{138.1^\circ,-138.1^\circ\}$ \\
\hline
\end{tabular}
\caption{Best-fit values of leptonic mixing parameters $\sin^2 \theta_{12}({\rm bf})$, $\sin^2 \theta_{13}({\rm bf})$, and $\sin^2 \theta_{23}({\rm bf})$ for the 18 one- and two-parameter models (including the predicted values of $\delta_{\rm CP}$ for each model), using the best-fit values of the model parameters given in Table~\ref{tab:fits}. Note that for the models that predict $\delta_{\rm CP} \in \{\delta_1,\delta_2\}$, $\delta_1$ corresponds to the first presented best-fit point in Table~\ref{tab:fits} with $\theta_{\rm bf} < 90^\circ$, whereas $\delta_2$ corresponds to the second one with $\theta_{\rm bf} > 90^\circ$.}
\label{tab:fits2}
\end{center}
\end{table}

\begin{figure}%[!t]
\vspace{-2.5cm}
\begin{center}
\includegraphics[scale=0.625]{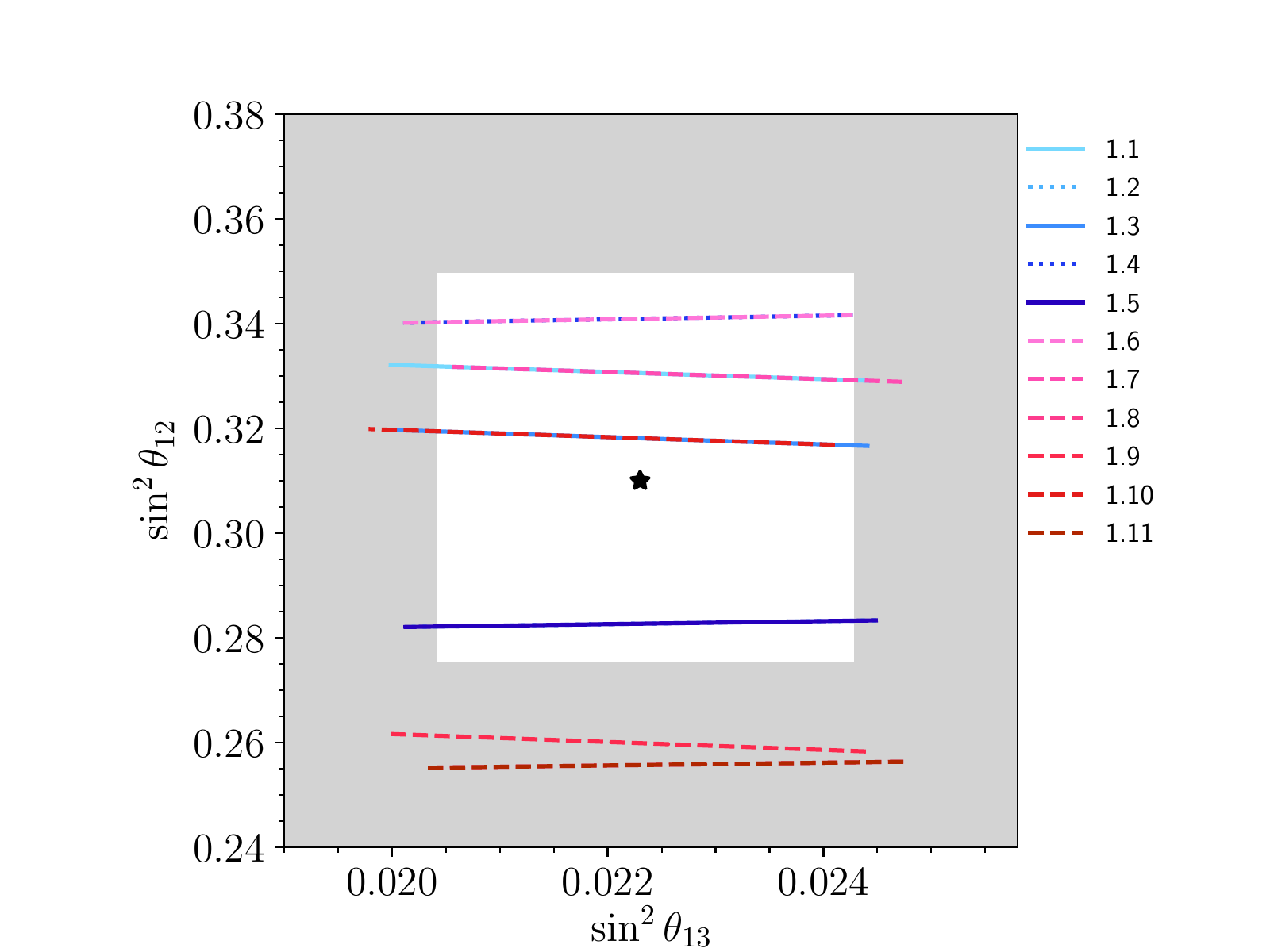} \\
\includegraphics[scale=0.625]{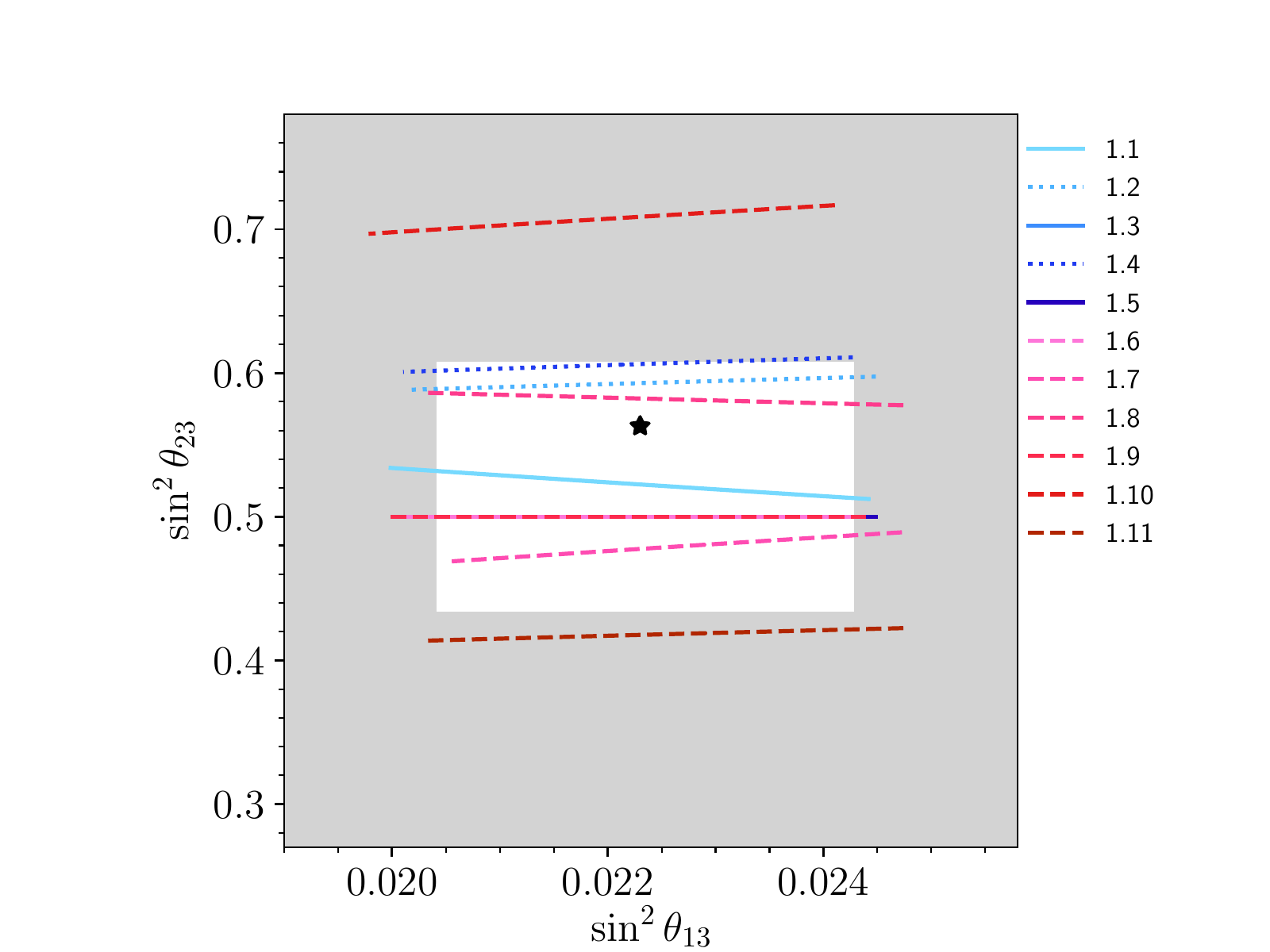} \\
\includegraphics[scale=0.625]{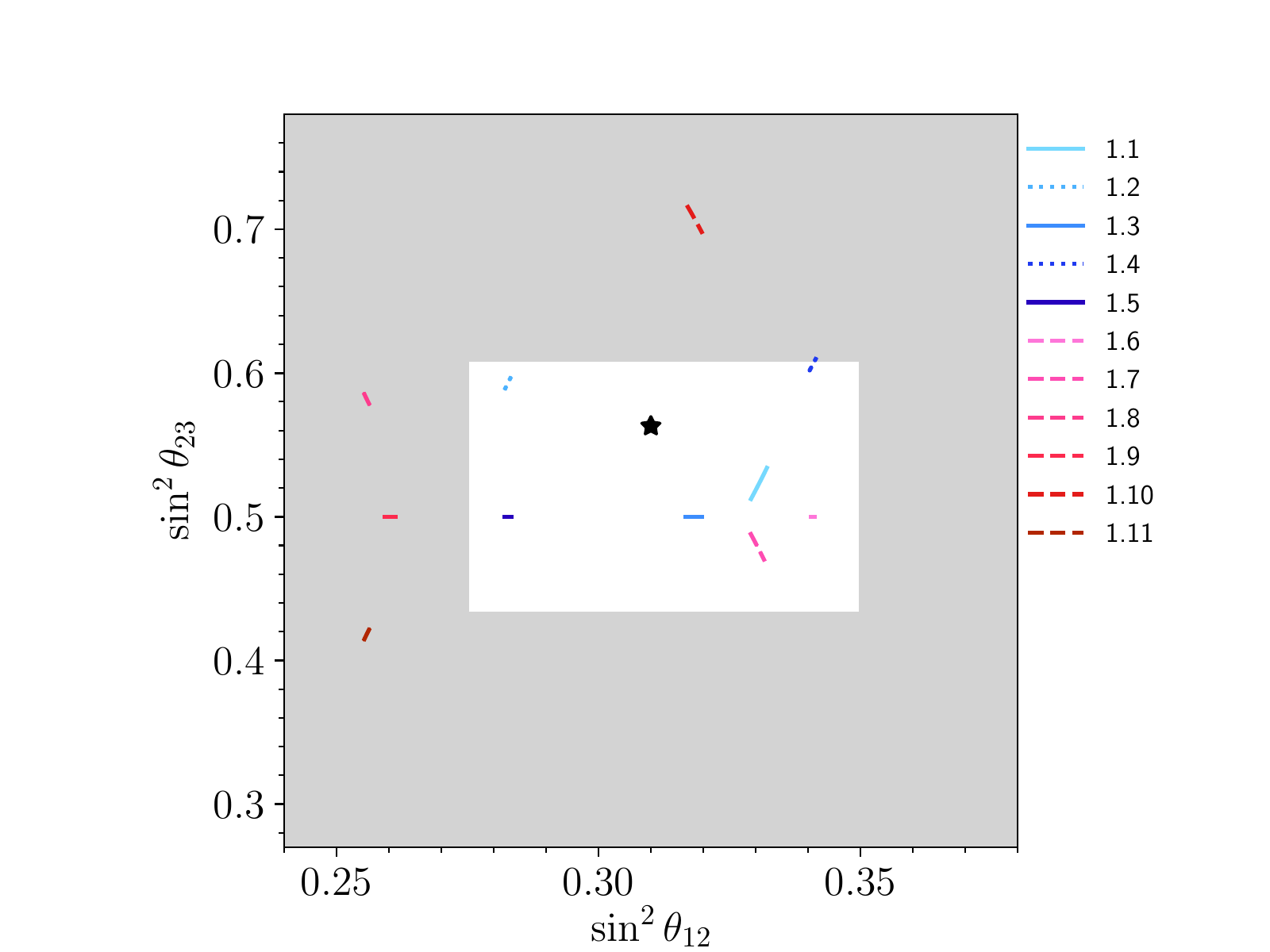} 
\end{center}
\caption{Prediction of leptonic mixing angles for the eleven one-parameter models. The lines show the allowed values of the leptonic mixing angles for the $3\sigma$ intervals $\theta_{3\sigma}$ of the respective model parameters $\theta$ given in Table~\ref{tab:fits}. The white areas show the allowed $3\sigma$ regions of the leptonic mixing angles from fits to global neutrino oscillation data, whereas the gray-shaded areas show the corresponding excluded regions. A star (``$\star$'') indicates the present best-fit values of the leptonic mixing angles from global neutrino oscillation data. {\it Top panel:} $\sin^2 \theta_{13}$ -- $\sin^2 \theta_{12}$. {\it Middle panel:} $\sin^2 \theta_{13}$ -- $\sin^2 \theta_{23}$. {\it Bottom panel:} $\sin^2 \theta_{12}$ -- $\sin^2 \theta_{23}$.}
\label{fig:pred_oneparamod}
\end{figure}

\begin{figure}[t!]
\vspace{-2cm}
\begin{center}
\includegraphics[scale=0.5]{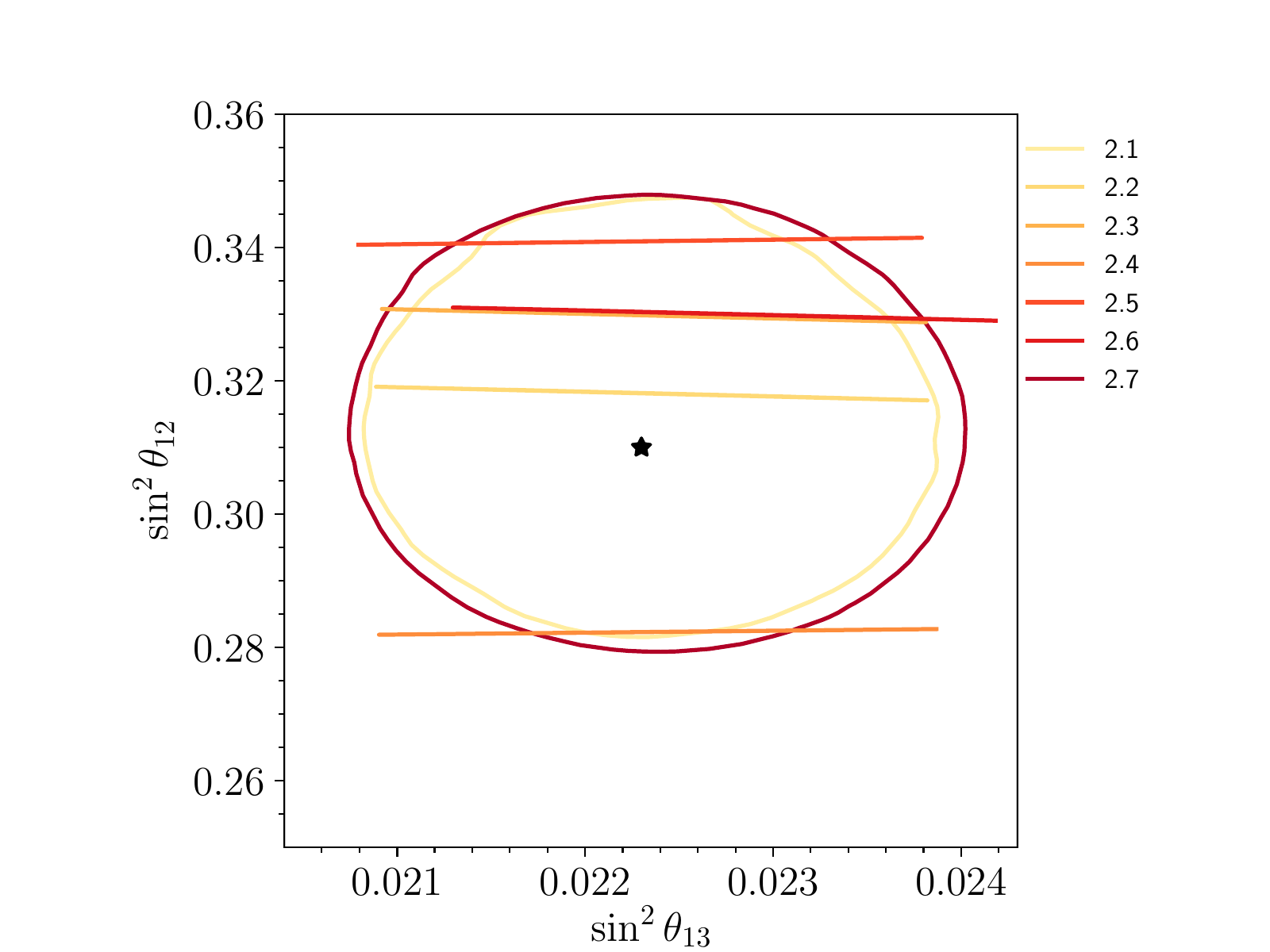}
\hspace{-30pt}
\includegraphics[scale=0.5]{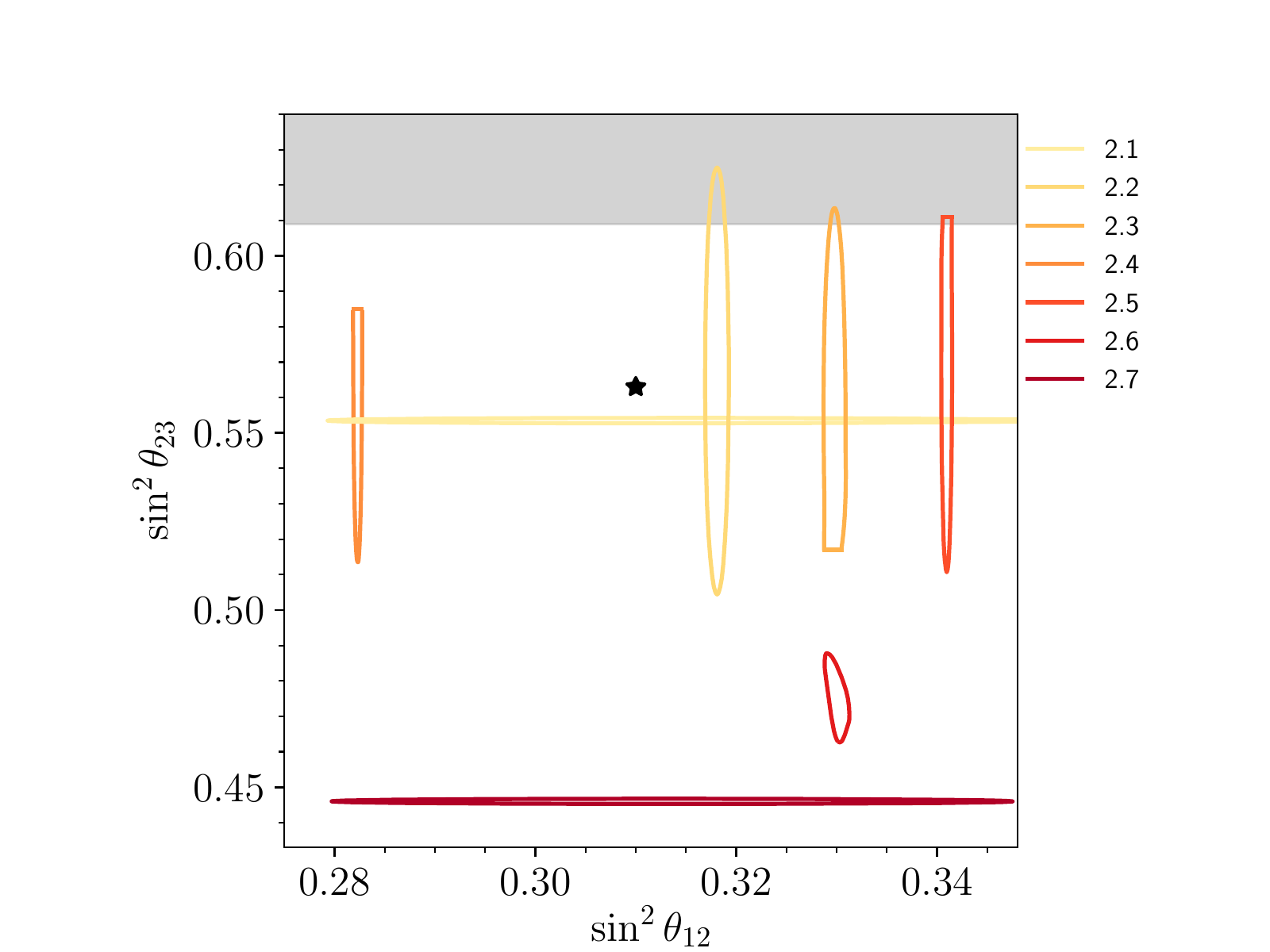} \\
\includegraphics[scale=0.5]{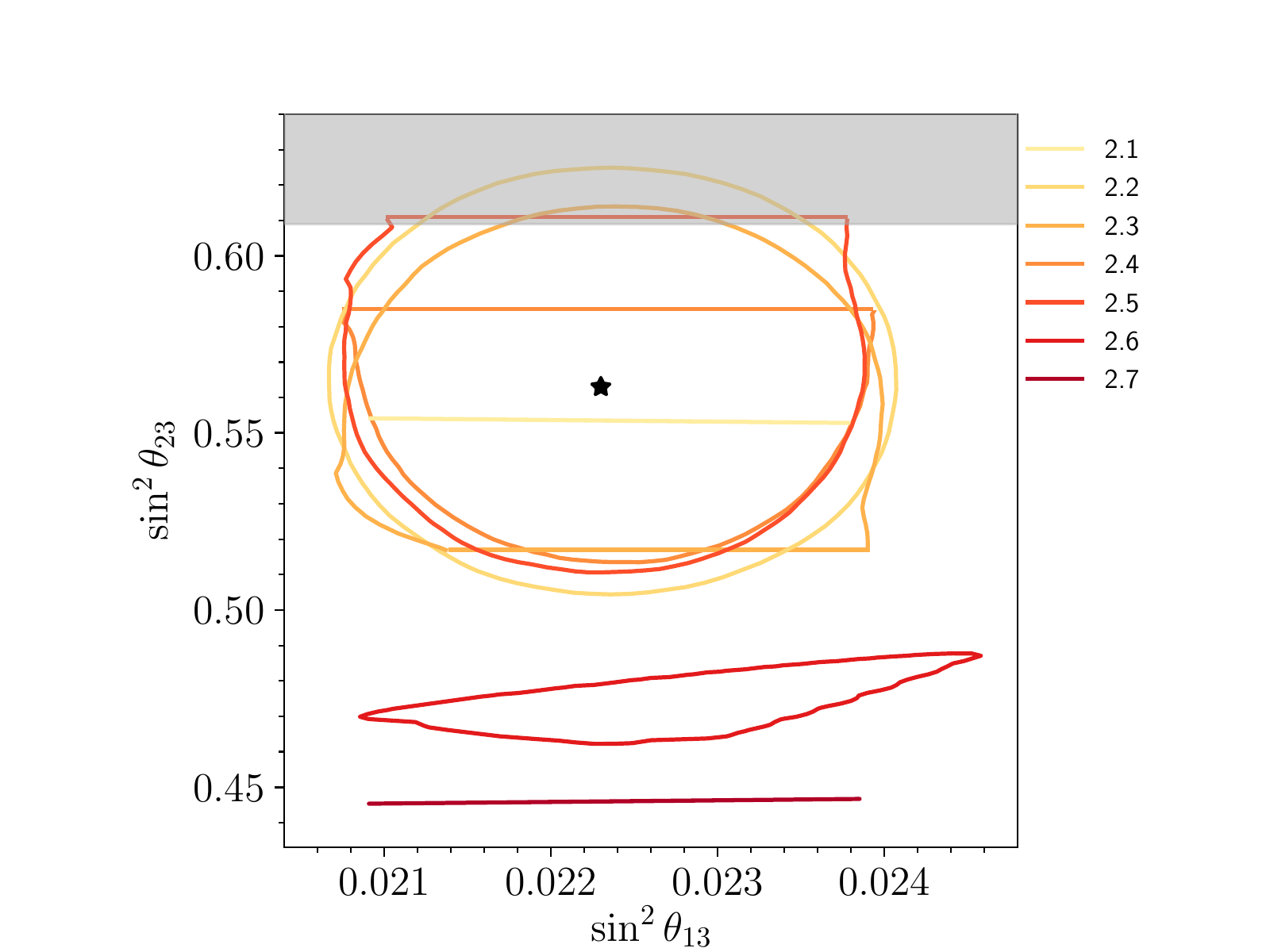}
\hspace{-30pt}
\includegraphics[scale=0.5]{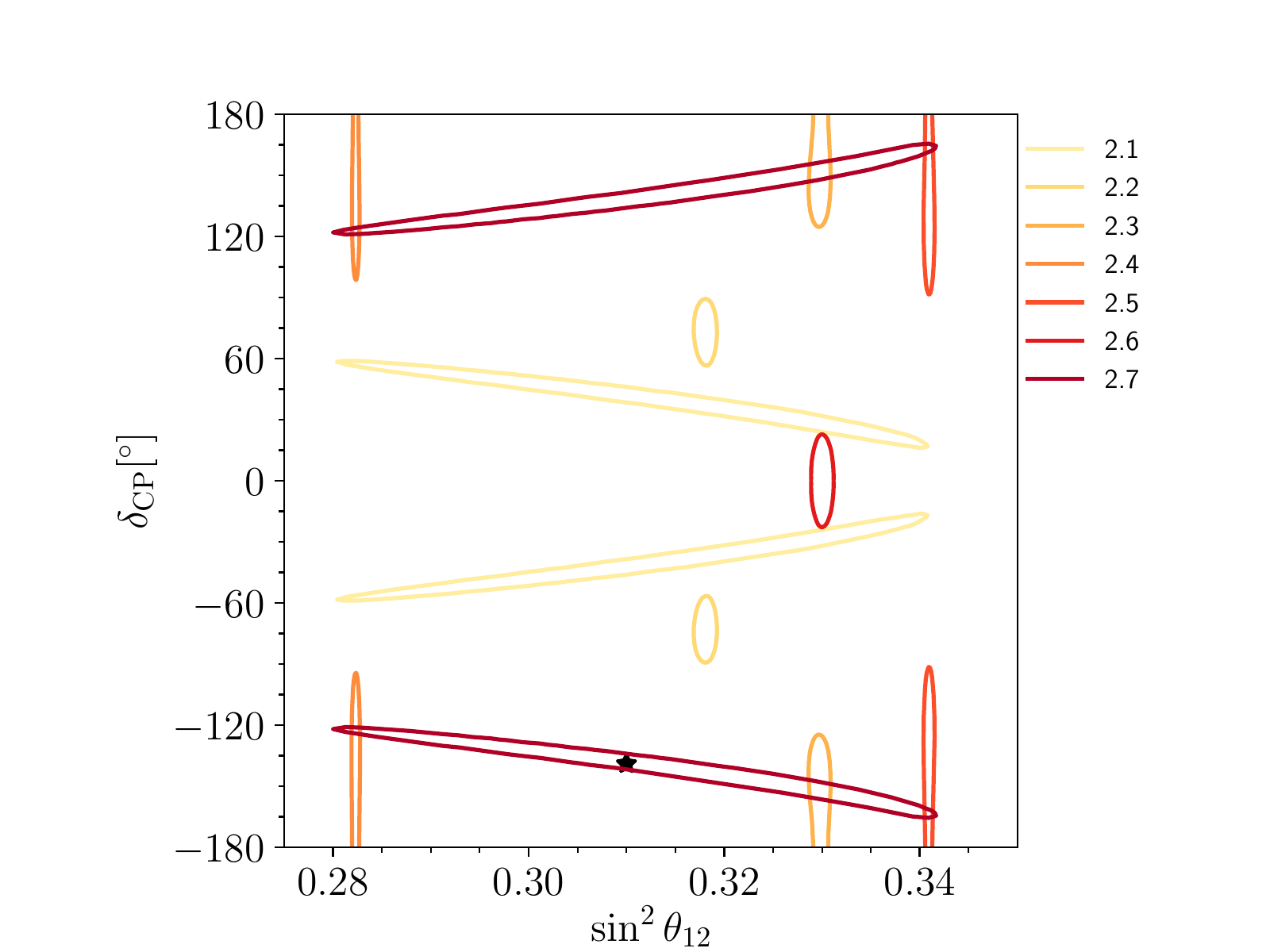} \\
\includegraphics[scale=0.5]{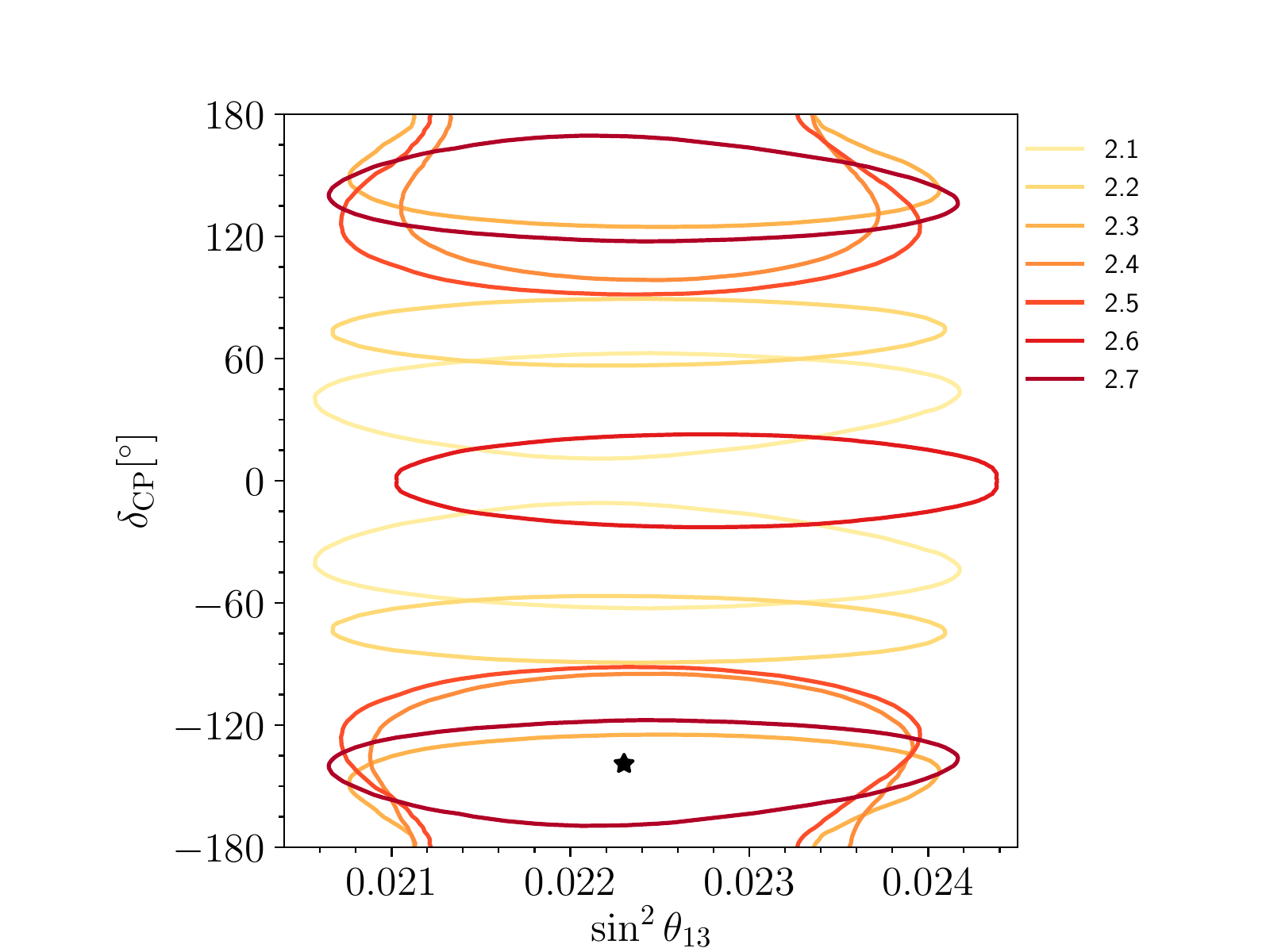}
\hspace{-30pt}
\includegraphics[scale=0.5]{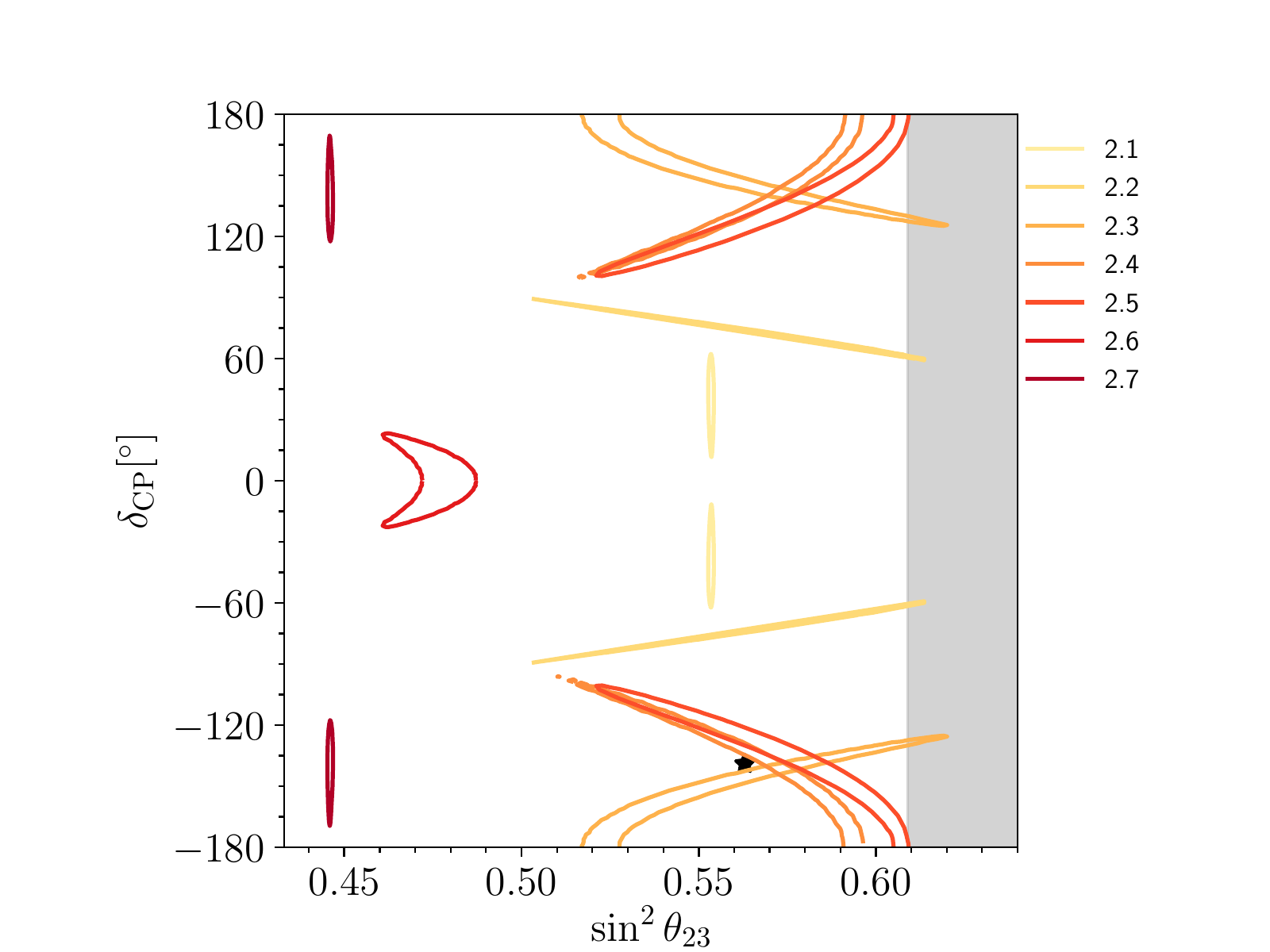}
\end{center}
\caption{Prediction of leptonic mixing parameters for the seven two-parameter models. The curves show the allowed contours of the leptonic mixing parameters for the $3\sigma$ intervals $\theta_{3\sigma}$ and $\phi_{3\sigma}$ of the respective model parameters $\theta$ and $\phi$ given in Table~\ref{tab:fits}. The white areas show the allowed $3\sigma$ regions of the leptonic mixing angles from fits to global neutrino oscillation data, whereas the gray-shaded areas show the corresponding excluded regions. A star (``$\star$'') indicates the present best-fit values of the leptonic mixing parameters from global neutrino oscillation data. {\it Top-left panel:} $\sin^2 \theta_{13}$ -- $\sin^2 \theta_{12}$. {\it Top-right panel:} $\sin^2 \theta_{12}$ -- $\sin^2 \theta_{23}$. {\it Middle-left panel:} $\sin^2 \theta_{13}$ -- $\sin^2 \theta_{23}$. {\it Middle-right panel:} $\sin^2 \theta_{12}$ -- $\delta_{\rm CP}$. {\it Bottom-left panel:} $\sin^2 \theta_{13}$ -- $\delta_{\rm CP}$. {\it Bottom-right panel:} $\sin^2 \theta_{23}$ -- $\delta_{\rm CP}$.}
\label{fig:pred_twoparamod}
\end{figure}

In Figs.~\ref{fig:pred_oneparamod} and \ref{fig:pred_twoparamod}, we show the predictions of the leptonic mixing parameters in different planes for the eleven one- and the seven two-parameter models, respectively. To produce Fig.~\ref{fig:pred_oneparamod}, we vary the model parameter $\theta$ in the 3$\sigma$ interval for the one-parameter models as given in Table~\ref{tab:fits} and obtain the values of the leptonic mixing angles for each value of $\theta$ using the relations given in Tables~\ref{tab:FHZ1}--\ref{tab:LD2}. We plot the obtained leptonic mixing parameters against each other. Since all relations between $\theta$ and the angles are rather simple, this method is sufficient to describe the allowed parameter space of the one-parameter models.
On the other hand, for the two-parameter models, some of the relations between the free parameters $\theta$ and $\phi$ and the leptonic mixing parameters are simpler and some of them are more complex. Therefore, to generate Fig.~\ref{fig:pred_twoparamod}, we generate values of $\theta$ and $\phi$ with a probability density proportional to $e^{-\chi^2/2}$ for each model, where $\chi^2$ is the function given in Eq.~\eqref{eq:chisq2p}. We then draw the smallest possible regions in the predicted parameter space containing 95~\% of the generated combinations. For the cases, where the models predict a direct relation between the two parameters plotted, e.g., if $\theta_{23}$ can be written as a function of $\theta_{13}$ only, we draw a curve on which 95~\% of the generated parameter combinations lie.

From Fig.~\ref{fig:pred_oneparamod} for the eleven one-parameter models, we note that the allowed values of (i)~$\sin^2 \theta_{12}$ (i.e., the values within the $3\sigma$ intervals of the model parameter) for Models~1.8, 1.9, and 1.11 (see top panel), (ii)~$\sin^2 \theta_{23}$ for Models~1.10 and 1.11 (see middle panel), and 
(iii)~$\sin^2 \theta_{12}$ and/or $\sin^2 \theta_{23}$ for Models~1.8--1.11 (see bottom panel) lie totally outside their individual $3\sigma$ regions from the global fit of NuFIT~4.1. From this figure, we also see that Models~1.1, 1.2, 1.4, 1.8, and 1.10 predict $\theta_{23} > 45^\circ$, Models~1.3, 1.5, 1.6, and 1.9 predict $\theta_{23} = 45^\circ$, and Models~1.7 and 1.11 predict $\theta_{23} < 45^\circ$. In conclusion, Models~1.8--1.11 predict values of the leptonic mixing angles which fall outside the current $3\sigma$ regions from the global fit of NuFIT~4.1. In addition, the values of $\chi^2_{\min}$ for Models~1.6--1.11 are all above 11.83 (see Table~\ref{tab:fits} or \ref{tab:fits2}), which corresponds to $3\sigma$ for 2~d.o.f.,\footnote{Note that the one-parameters models are fitted to three data points, so in this case, the number of degrees of freedom is $3 - 1 = 2$.} whereas for Models~1.1--1.5, the corresponding values are all below $3\sigma$. Therefore, we deem Models~1.6--1.11 as excluded at $3\sigma$ or more by the current data. Let us understand why, among the allowed one-parameter models (i.e., Models~1.1--1.5), Model~1.1 gives the best fit to the data and Model~1.5 the worst. From Fig.~\ref{fig:pred_oneparamod}, we note that these models do not give any restrictions on $\sin^2\theta_{13}$ (except for Model~1.4), and therefore, the fits mainly depend on the pulls of the predictions for $\sin^2\theta_{12}$ and $\sin^2\theta_{23}$. The distance between the prediction of a given model and the best-fit point in the model parameter space basically determines the value of $\chi^2_{\min{}}$ of that model in terms of the pulls of the three different leptonic mixing angles. Thus, this distance increases with increasing model number, and therefore, Model~1.1 leads to the best fit and Model~1.5 to the worst.

Next, from Fig.~\ref{fig:pred_twoparamod} for the seven two-parameter models, we observe that all seven models predict leptonic mixing angles within the $3\sigma$ regions of the global fit of NuFIT~4.1, since none of the contours displayed in the six panels lie totally outside these $3\sigma$ regions. Furthermore, it is interesting to note that Models~2.6 and 2.7 predict the lower octant (LO) of $\theta_{23}$ (see top-right, middle-left, and bottom-right panels), whereas Models~2.1--2.5 predict the higher octant (HO). In addition, the intervals of $\delta_{\rm CP}$ predicted by the different two-parameter models are: $\delta_{\rm CP} \in (\pm 11.6^\circ,\pm 62.2^\circ)$ (Model~2.1), $\delta_{\rm CP} \in (\pm 58.7^\circ,\pm 87.6^\circ)$ (Model~2.2), $\delta_{\rm CP} \in (\pm 125.2^\circ,\pm 180^\circ]$ (Model~2.3), $\delta_{\rm CP} \in (\pm 96.0^\circ,\pm 180^\circ]$ (Model~2.4),  $\delta_{\rm CP} \in (\pm 100.5^\circ,\pm 180^\circ]$ (Model~2.5), $\delta_{\rm CP} \in [0,\pm 24.9^\circ)$ (Model~2.6), and $\delta_{\rm CP} \in (\pm 117.6^\circ,\pm 169.5^\circ)$ (Model~2.7). This means that Models~2.3--2.6 are compatible with CP conservation, Model~2.2 with close to maximal CP violation, whereas Models~2.1 and 2.7 are compatible with neither CP conservation nor maximal CP violation. Now, let us understand why different two-parameter models give different fits to the data of NuFIT~4.1. Here, we also note that the two-parameter models do not put any restrictions on $\sin^2\theta_{13}$, and therefore, the fits depend on their predictions for $\sin^2\theta_{12}$ and $\sin^2\theta_{23}$. From the top-right panel of Fig.~\ref{fig:pred_twoparamod}, we see that Model 2.1~predicts $\sin^2\theta_{12}$ and $\sin^2\theta_{23}$ closest to the current best-fit point, whereas Model~2.7 predicts $\sin^2\theta_{12}$ and $\sin^2\theta_{23}$ furthest from the current best-fit point. This is the reason why Model~2.1 gives the best fit to the data and Model~2.7 the worst. From this panel, it is also easy to understand why the fits of Models 2.2--2.6 lie in between the fits of Models~2.1 and 2.7. Note that among the seven two-parameter models, Models 2.6 and 2.7 have values of $\chi^2_{\min{}}$ larger than 9, i.e., $3\sigma$ for 1~d.o.f.

Thus, confronting the 18 one- and two-parameter lepton flavor models with global neutrino oscillation data by NuFIT~4.1, we conclude that Models~1.1--1.5 and 2.1--2.5 are allowed at $3\sigma$, whereas Models~1.6--1.11, 2.6, and 2.7 are excluded at $3\sigma$ or more by the current data. For the analysis of the models with ESSnuSB, we consider only those models that are allowed by the current data at $3\sigma$. Therefore, in the next three sections, we will address these ten models, i.e., Models~1.1--1.5 and 2.1--2.5, with ESSnuSB.

%===============
\section{Experimental Setup of ESSnuSB}
\label{sec:essnusb}
%===============

The ESSnuSB experiment \cite{Baussan:2013zcy,Wildner:2015yaa} is a proposed long-baseline neutrino oscillation experiment in Sweden and mainly designed to measure potential leptonic CP violation with excellent precision. In our work, we use exactly the same configuration of ESSnuSB as was used to generate the results of Ref.~\cite{Blennow:2019bvl}. In the original experimental setup of ESSnuSB, the second peak of the neutrino oscillation probability was identified to be optimal to maximize the CP violation discovery potential. The source of neutrinos is a beam of power 5~MW capable of delivering protons of energy 2.5~GeV corresponding to $2.7 \times 10^{23}$~protons on target per year, situated at the European Spallation Source (ESS) in Lund, Sweden. The detector is 1~Mt MEMPHYS-like water-Cherenkov detector \cite{Agostino:2012fd} located at a distance of 540~km away from the source in the mine in Garpenberg, Sweden. We assume a total running time of 10~years with 5 years in neutrino mode and 5 years in antineutrino mode. In our analysis, we also consider an identical near detector having mass of 0.1~kt and located at a distance of 500~m away from the source. For the near detector, the fluxes are simulated at a distance of 1~km from the target station, whereas for the far detector, they are calculated at 100~km. We use correlated systematics between the near and the far detectors. The specification for the systematic errors is adopted from Ref.~\cite{Coloma:2012ji} and listed in Table~\ref{tab:systematics} for convenience. Furthermore, we implement the energy-dependent efficiencies as given in Fig.~8 of Ref.~\cite{Agostino:2012fd}.
\begin{table}[!t]
\centering
\renewcommand*{\arraystretch}{1.5}
\begin{tabular} {|c|c|}
\hline
Systematics &  Error \\
\hline
\hline
Fiducial volume of near detector & 0.5~\% \\
Fiducial volume of far detector & 2.5~\% \\
Flux error for $\nu$ & 7.5~\% \\
Flux error for $\bar{\nu}$ & 15~\% \\
Neutral current background & 7.5~\% \\
Cross section $\times$ efficiency QE & 15~\% \\
Ratio $\nu_e/\nu_{\mu}$ QE & 11~\% \\
\hline
\end{tabular}
\caption{Considered values of the systematic errors. The abbreviation QE stands for ``quasi-elastic''. 
}
\label{tab:systematics}
\end{table}

%===============
\section{Details of Simulation and Statistical Analysis}
\label{sec:stat}
%===============

All of our numerical simulations are performed using the GLoBES software \cite{Huber:2004ka,Huber:2007ji} with the experimental description implemented as described in Section~\ref{sec:essnusb}.
In order to judge whether or not two models can be distinguished using the data of ESSnuSB, we follow the statistical procedure laid out in this section. In order to accomplish this, we introduce the chi-square function
\begin{equation}
\chi^2(\theta,D) = \chi^2_{\rm ESSnuSB}(\theta,D) + \chi^2_0(\theta)\,,
\end{equation}
where $\chi^2_{\rm ESSnuSB}$ is the chi-square function for the ESSnuSB data alone, $\chi^2_0$ is the prior chi-square coming from already performed experiments, $\theta$ is the set of parameter values in a particular model, and $D$ is the data obtained in ESSnuSB. For $\chi^2_{\rm ESSnuSB}$, we use the default GLoBES chi-square function
\begin{equation}
\chi^2_{\rm ESSnuSB}(\theta,D) = \sum_i \left[\bar D_i(\theta) - D_i + D_i \ln\left(\frac{D_i}{\bar D_i(\theta)}\right)\right]\,,
\end{equation}
where $D_i$ is the number of events in bin $i$ and $\bar D_i(\theta)$ is the prediction for bin $i$ of the model being tested given the parameter set $\theta$. We assume $\chi^2_0$ to be the same prior as used in Section~\ref{sec:confronting}.

In order to answer the question of whether Model~A can be excluded if Model~B is true, we assume that the data gathered in ESSnuSB will be the Asimov data~\cite{Cowan:2010js} corresponding to Model~B being true with particular parameter values $\theta^{\rm B}_{\rm true}$. We then minimize $\chi^2(\theta^{\rm A},D)$ for the predictions of Model~A in order to find the best-fit parameters of Model~A and take this $\chi^2$ as a measure of whether or not Model~A could be ruled out if Model~B were true. In the cases, where the true Model~B is taken to be a one-parameter model, we present our results as a function of the true parameter $\theta^{\rm B}_{\rm true}$, and when it is taken to be a model with two parameters, our results are presented as the profiled $\chi^2$, i.e., as a function of one of the true parameters minimized over the other.

Note that, since we include the current priors in the $\chi^2$ function, it may occur that even if Model~B is true, Model~A provides a better fit to the ESSnuSB data together with the prior. This will be particularly true when the assumed true Model~B is disfavored relative to Model~A by current data. The interpretation of such a result should therefore be that the data of ESSnuSB will not be able to provide sufficient discriminatory power to overcome the fact that Model~B is disfavored by current data. On the other hand, if $\chi^2$ of the test Model~A is significantly larger than that of the assumed true Model~B, then Model~A could be ruled out by the ESSnuSB result if Model~B is the model implemented in Nature.

In addition to testing models against each other, we simulate the case where the oscillation parameters take particular values and consider the level at which the different models could be rejected in such a case. In order to do this, we follow the very same procedure as outlined above with the difference that all models are compared to the global $\chi^2_{\rm min}$, i.e., we minimize $\chi^2$ with respect to all of the leptonic mixing parameters ($\theta_{12}$, $\theta_{13}$, $\theta_{23}$, and $\delta_{\rm CP}$) and compare this to $\chi^2_{\rm min}$ in each model. Note that the number of d.o.f.~is taken to be 3 for the one-parameter models and 2 for the two-parameter models as the models are being compared to the case where all four mixing parameters (three angles and $\delta_{\rm CP}$) are left free. In all cases, we fix the mass-squared differences to their current best-fit values from Table~\ref{tab:NuFit}.

%===============
\section{Results: Addressing Flavor Models with ESSnuSB}
\label{sec:results}
%===============

\begin{figure}
\vspace{-3cm}
\hspace{5pt}
\begin{tabular}{ll}
\includegraphics[scale=0.48]{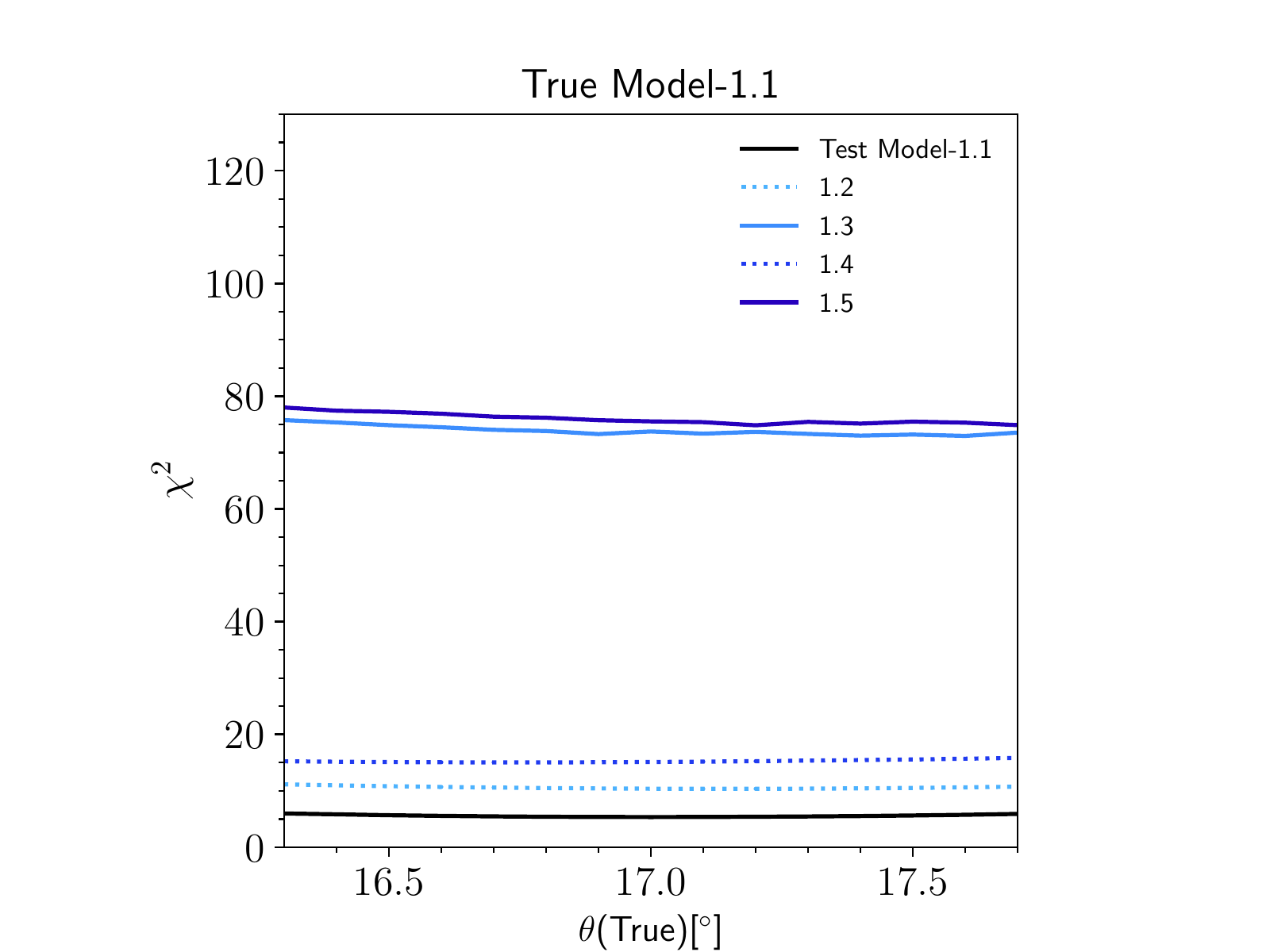}
\hspace{-30pt}
\includegraphics[scale=0.48]{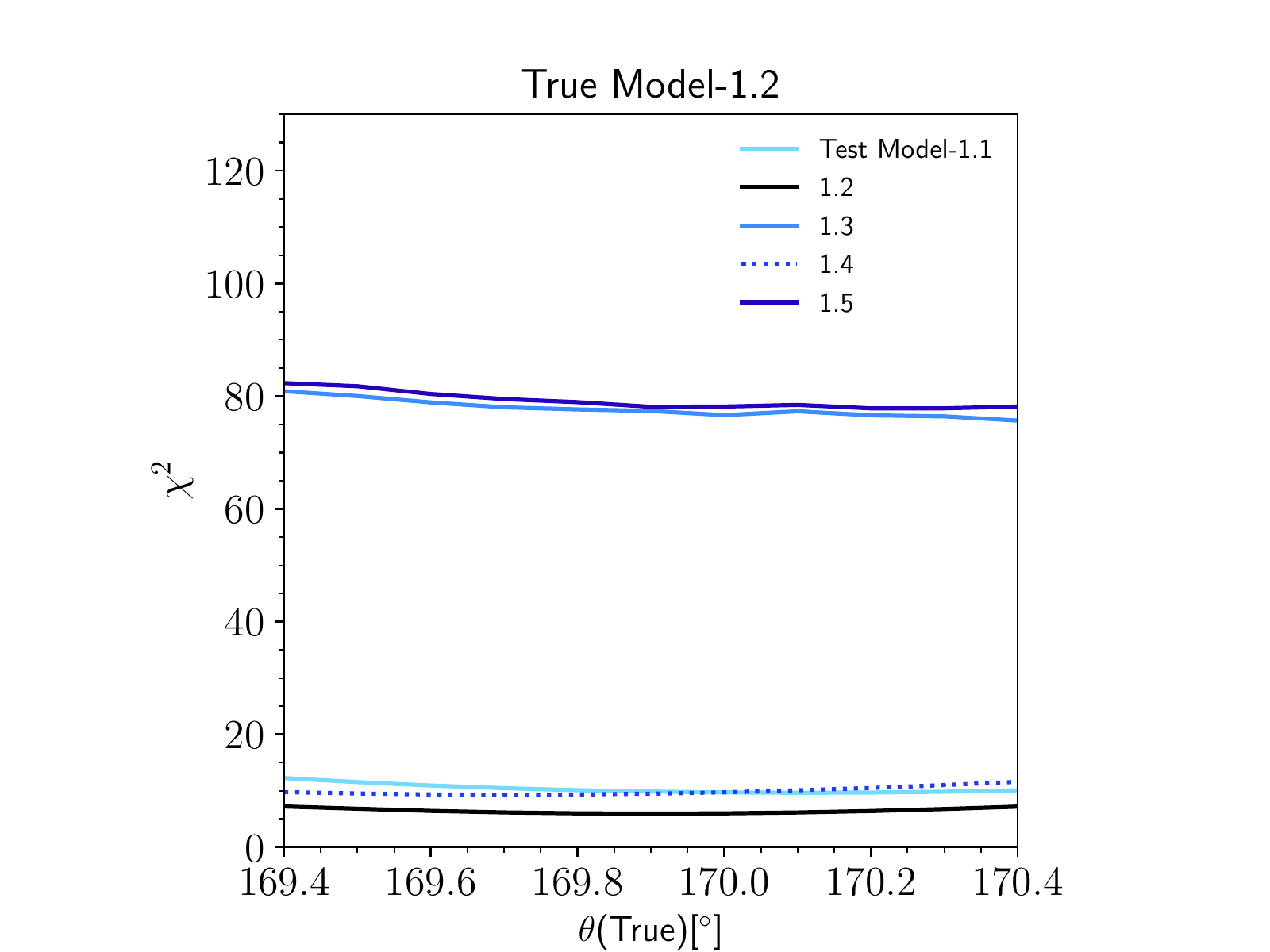} \\[1mm]
\includegraphics[scale=0.48]{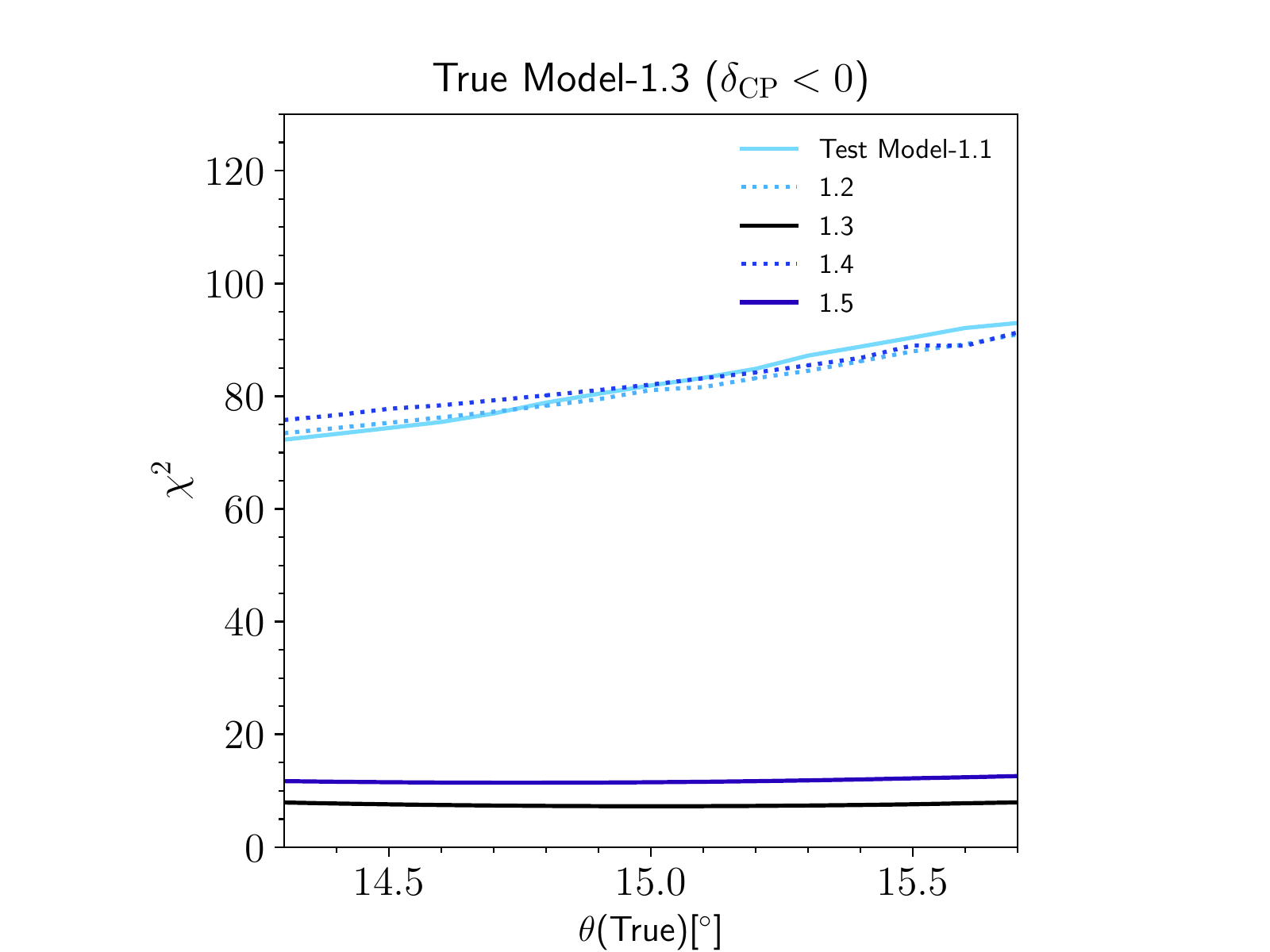}
\hspace{-30pt}
\includegraphics[scale=0.48]{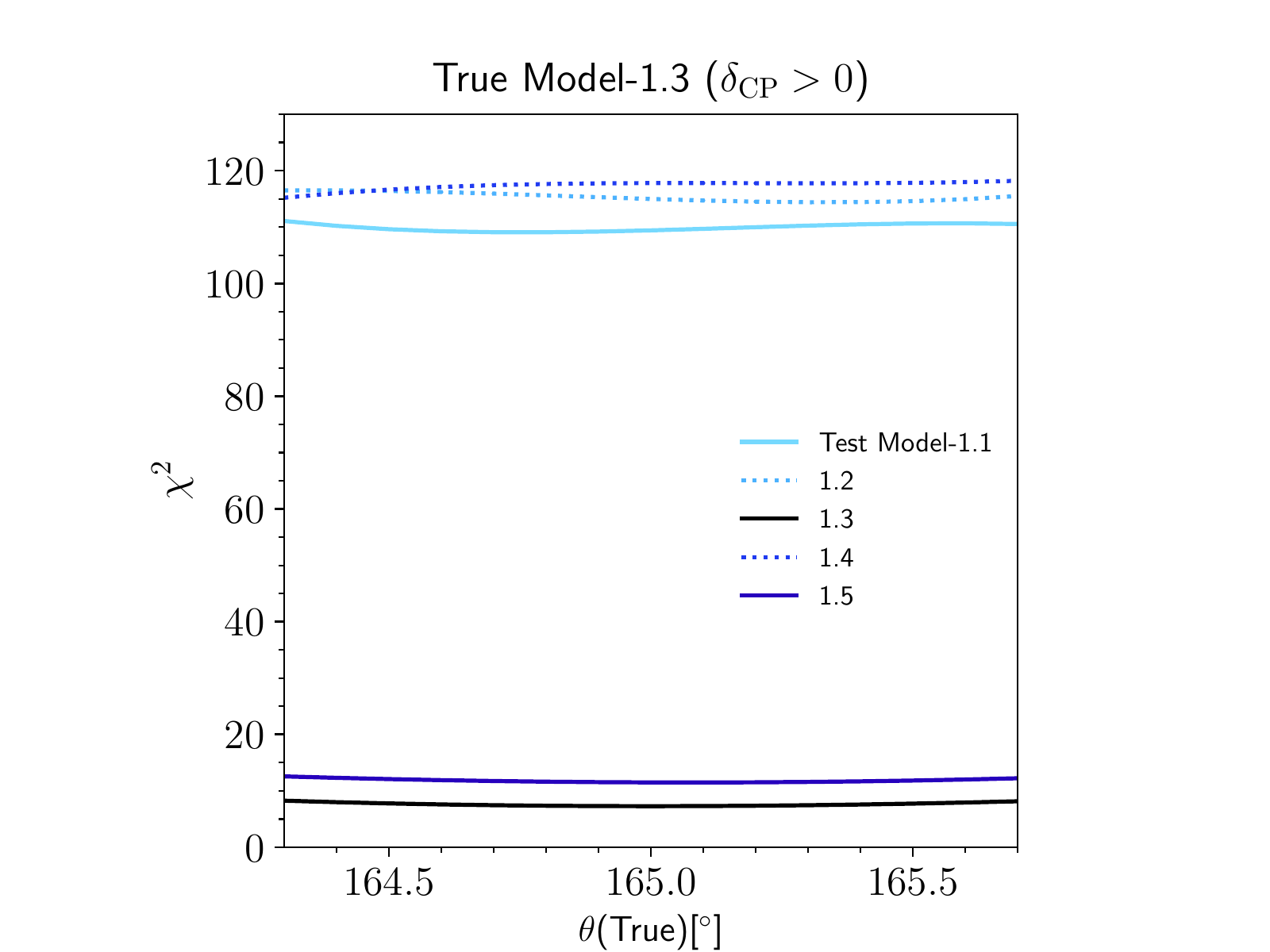} \\[1mm]
\multicolumn{2}{c}{\includegraphics[scale=0.48]{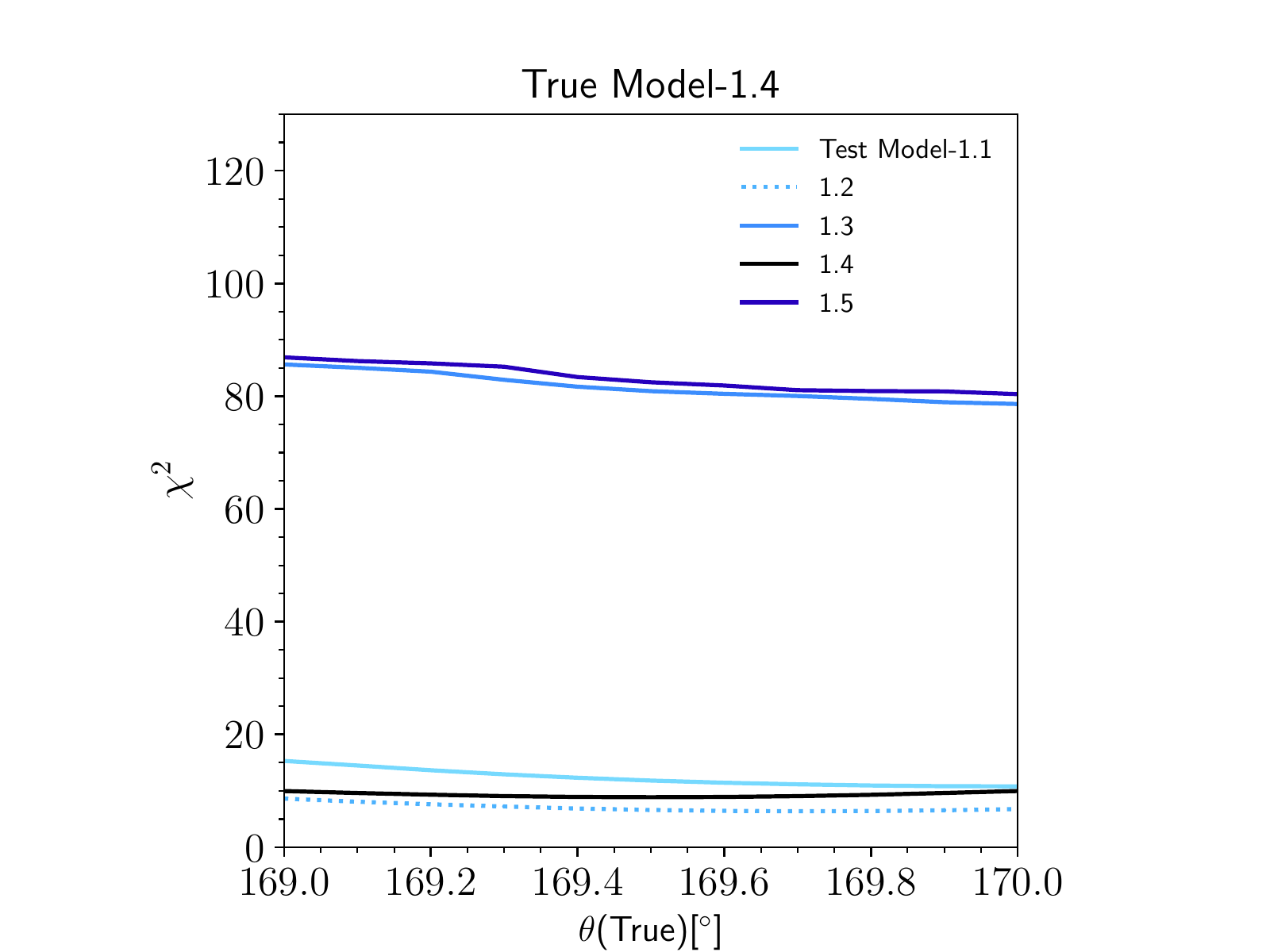}} \\[1mm]
\includegraphics[scale=0.48]{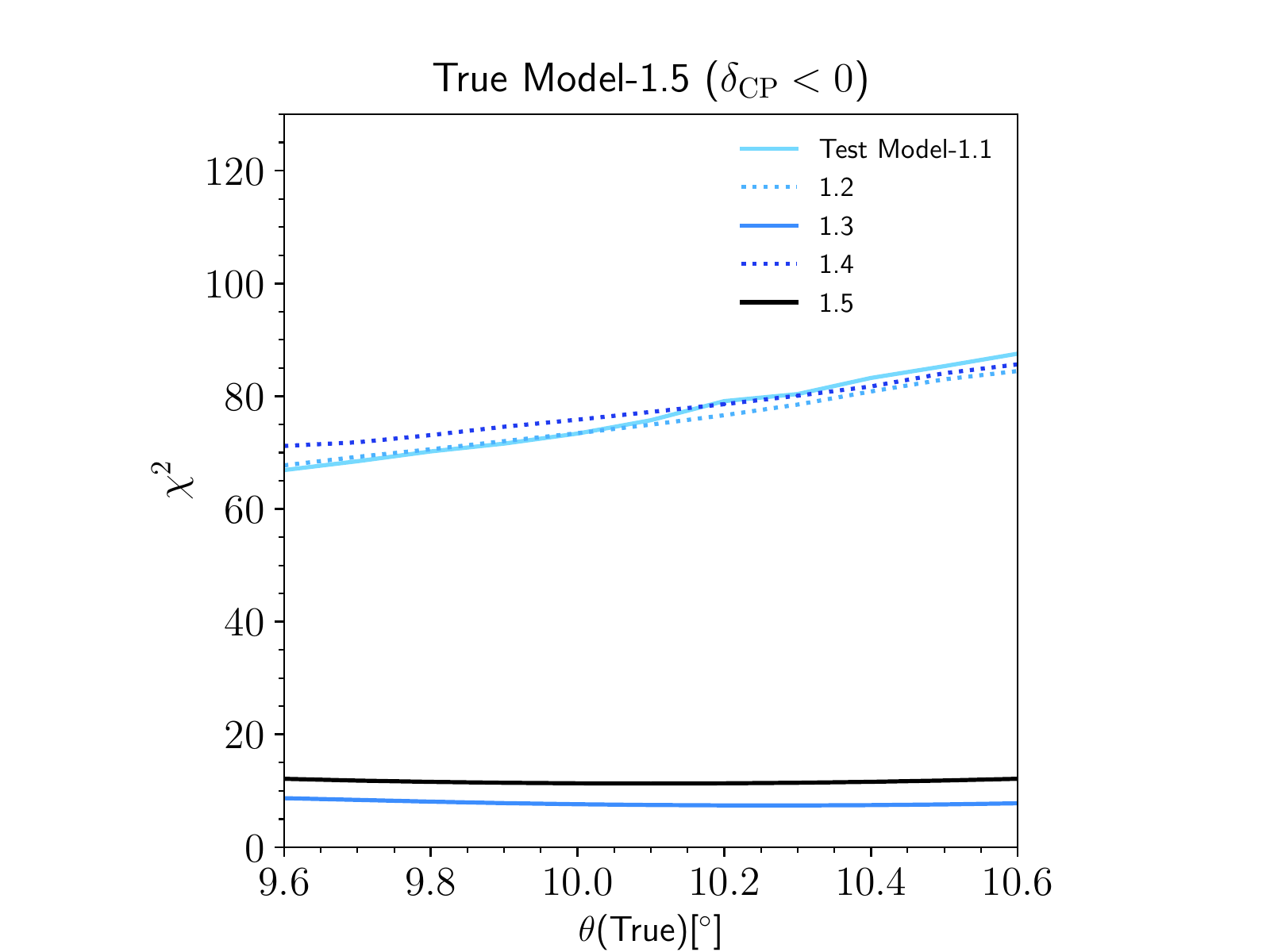}
\hspace{-30pt}
\includegraphics[scale=0.48]{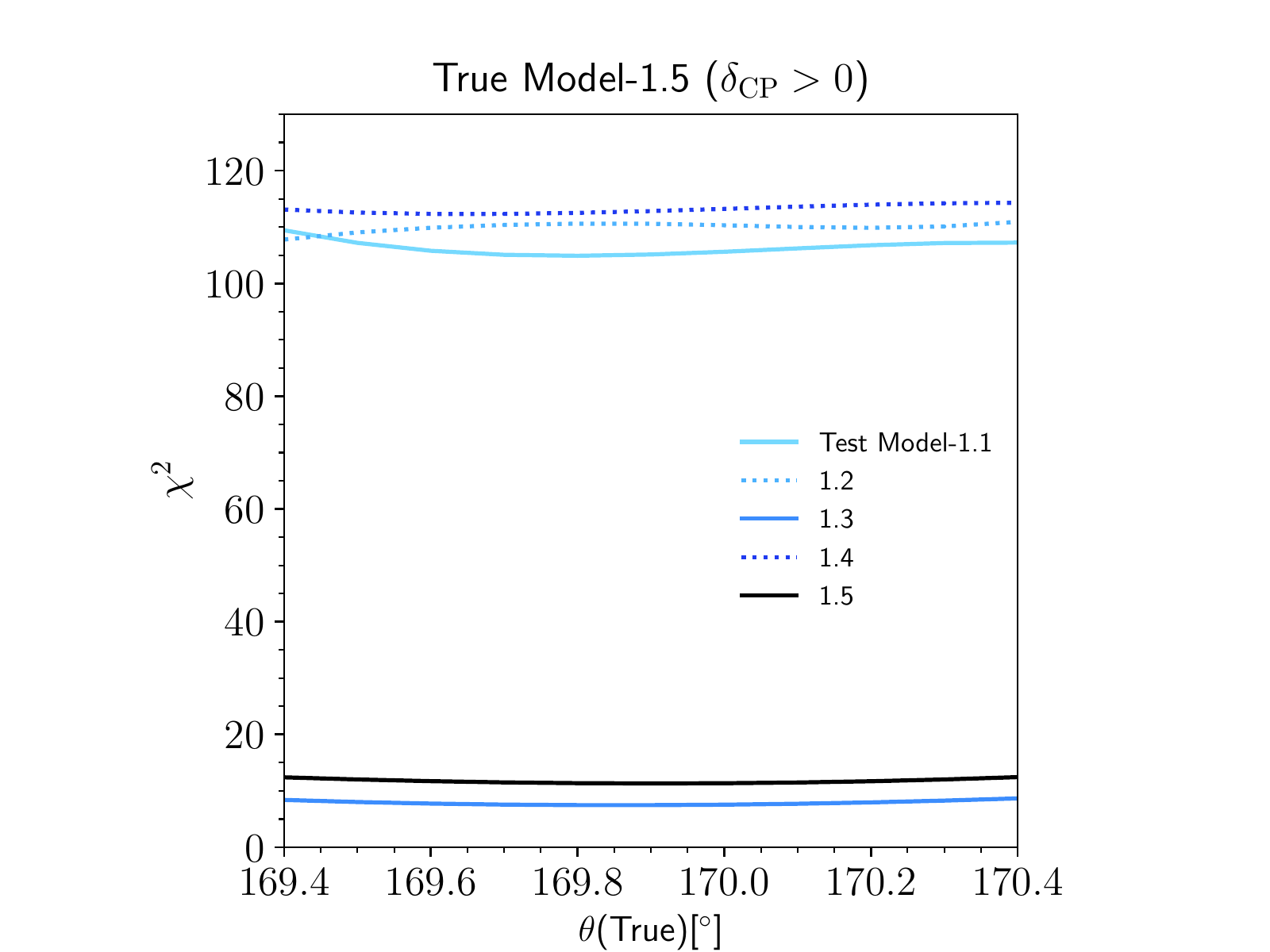} 
\end{tabular}
\caption{Comparison among the five allowed one-parameter models as test models for ESSnuSB using priors. Each panel shows the quantity $\chi^2_\mathrm{min}$, given a true model, as a function of the model parameter $\theta$ in its $3\sigma$ interval $\theta_{3\sigma}$ presented in Table~\ref{tab:fits}. In each panel, $\chi^2_\mathrm{min}$ for the given true model is displayed with a black solid curve (when acting as a test model), whereas it is marked for the five test models by curves corresponding to their respective model colors.}
\label{fig:comp_oneparamod}
\end{figure}

In \Fig~\ref{fig:comp_oneparamod}, we show the comparison of the one-parameter models~1.1--1.5, where each panel corresponds to different assumed true models and the horizontal axes represent the assumed true parameter value in the true model. The different curves show the resulting $\chi^2_{\rm min}$ in the different models, including both the prior and the Asimov data from ESSnuSB. In each panel, the curve of the assumed true model is drawn in black in order to highlight it. Note that, due to the addition of the prior, $\chi^2$ of the true model is not equal to zero despite using the Asimov data for ESSnuSB. For each model, $\chi^2_{\rm min}$ is therefore equal to the minimum of the prior for that model. This also means that it is possible for an assumed true model to give a worse fit than another model, even in the case where the ESSnuSB data are generated from that model. The interpretation of this should be that the ESSnuSB data will not be sufficient to overcome the preference for the other model that is present in the current data.

For the one-parameter models, we see from \Fig~\ref{fig:comp_oneparamod} that the models essentially split into two different groups, one group including Models~1.1, 1.2, and 1.4 and another including Models~1.3 and 1.5. The reason is that Models~1.1, 1.2, and 1.4 predict $\sin\delta_{\rm CP} = 0$, whereas Models~1.3 and 1.5 predict $|\sin\delta_{\rm CP}|= 1$. If the ESSnuSB data is generated assuming a model from one group, then $\chi^2$ of all models in that group remains more or less the same, whereas the models from the other group are strongly disfavored due to the ESSnuSB precision to $\delta_{\rm CP}$. For example, looking at the top-left panel, where Model~1.1 is assumed to be true, $\chi^2$ of Models~1.1, 1.2, and 1.4 remain essentially the same regardless of the assumed true value of the parameter $\theta$, while the other models would be excluded at a level of around $7\sigma$, where we use $\sqrt{\Delta\chi^2}$, where $\Delta \chi^2$ is the difference in $\chi^2_{\rm min}$ between the models, as the sensitivity estimator~\cite{Blennow:2013oma}. Correspondingly, Model~1.1 would be strongly excluded when data is generated from any of Models~1.3 and 1.5. 

\begin{figure}[t!]
\vspace{-2cm}
\hspace{-10pt}
\begin{tabular}{ll}
\includegraphics[scale=0.55]{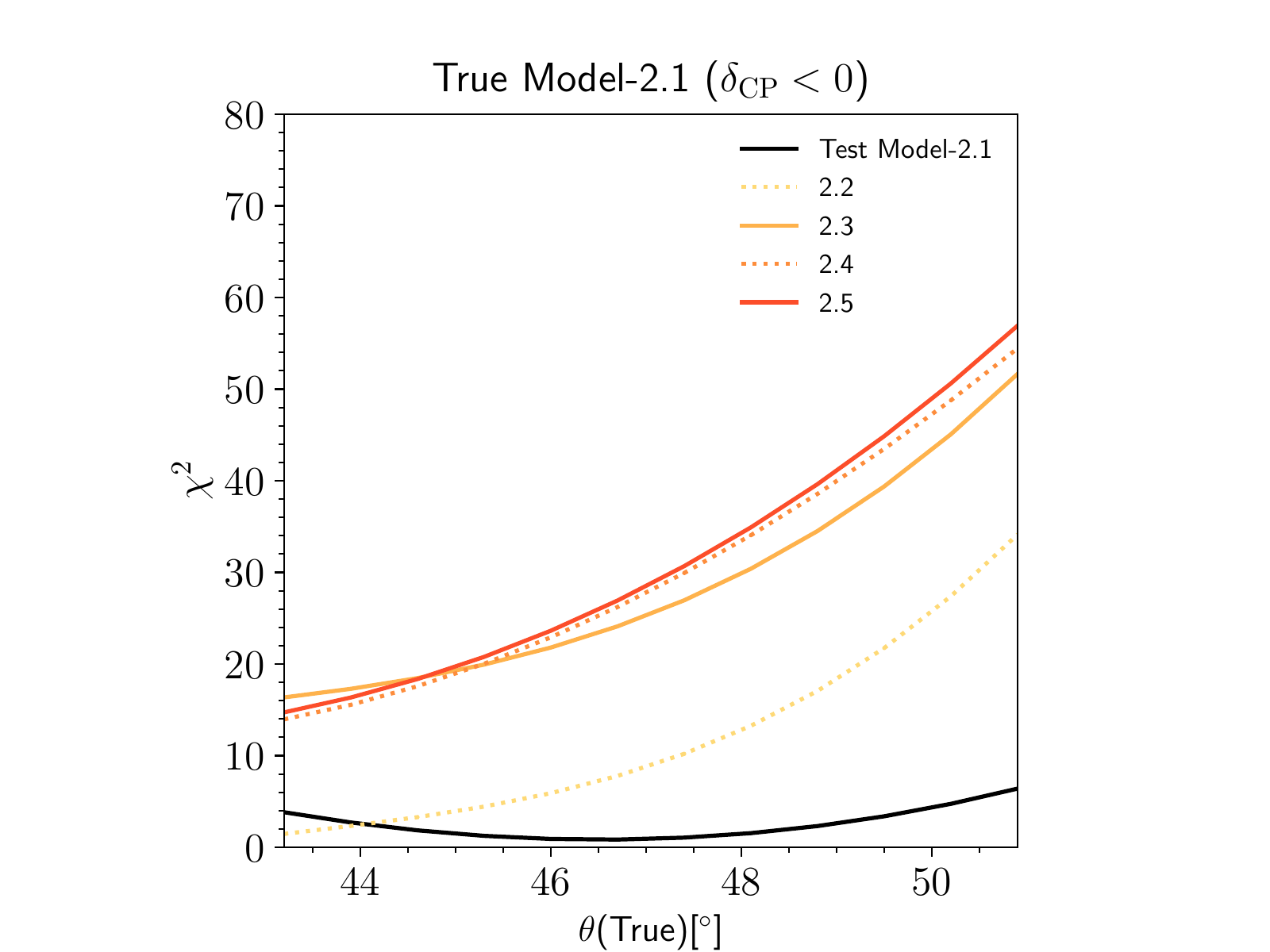}
\hspace{-50pt}
\includegraphics[scale=0.55]{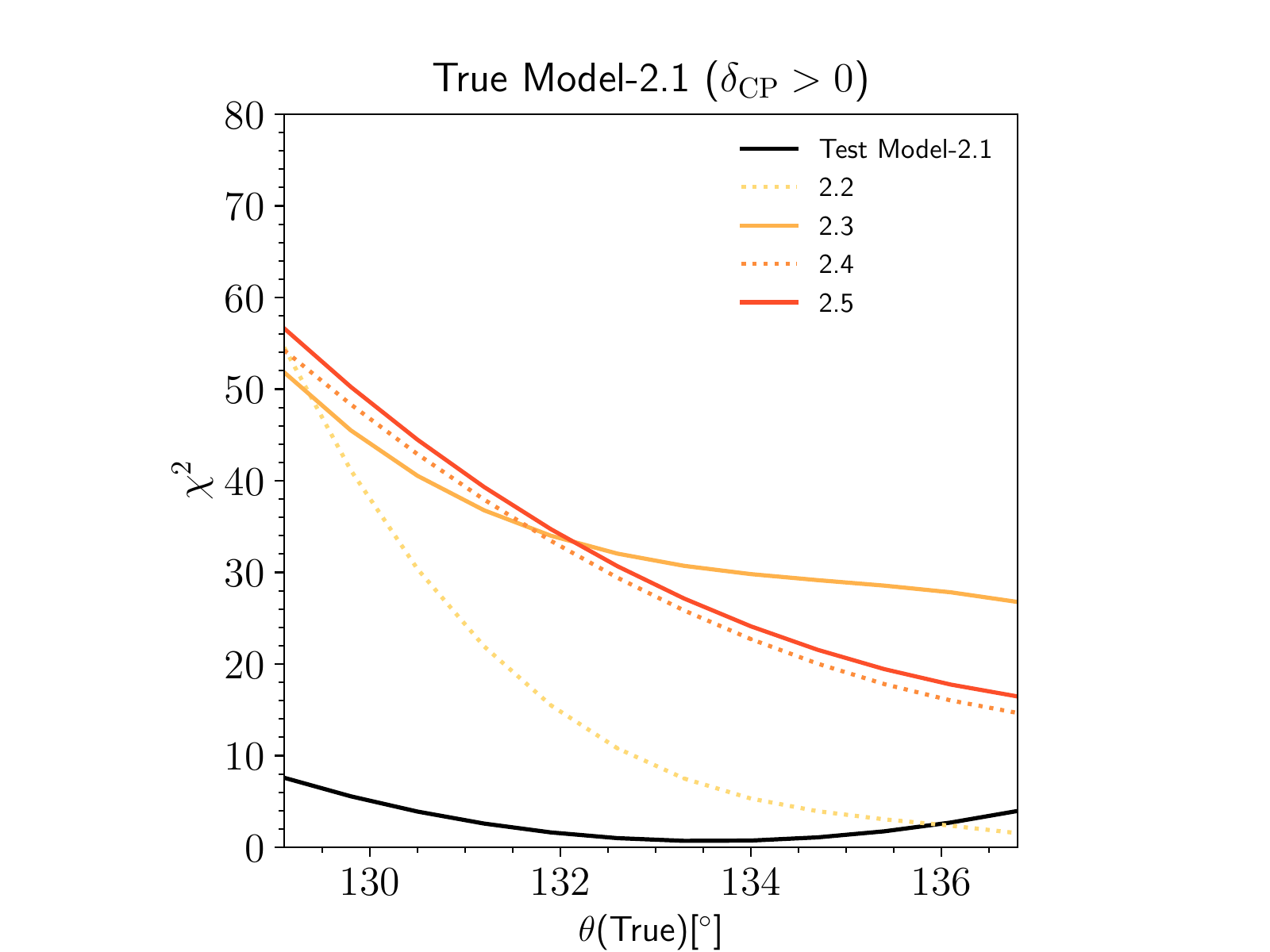} \\[1mm]
\includegraphics[scale=0.55]{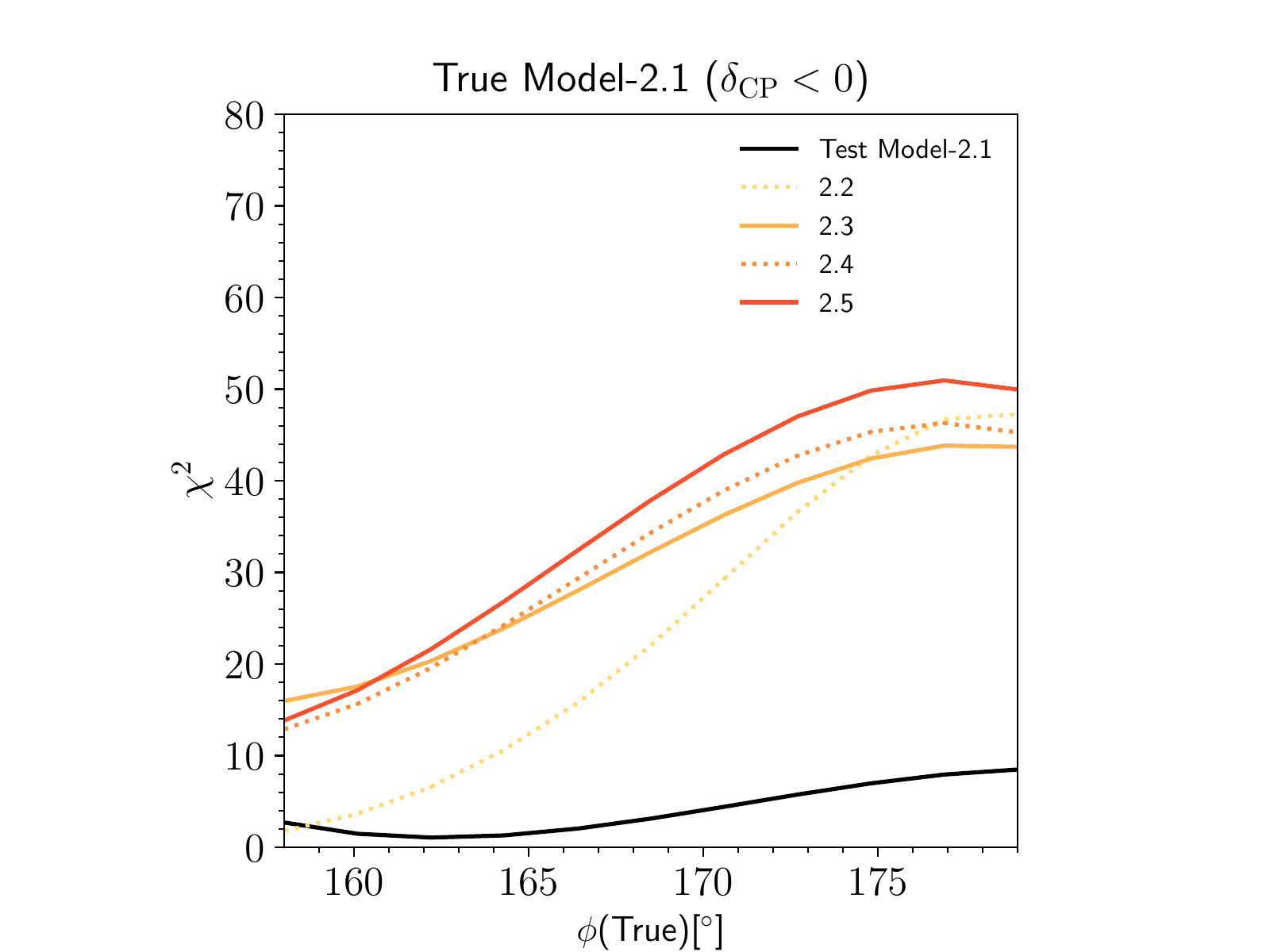}
\hspace{-50pt}
\includegraphics[scale=0.55]{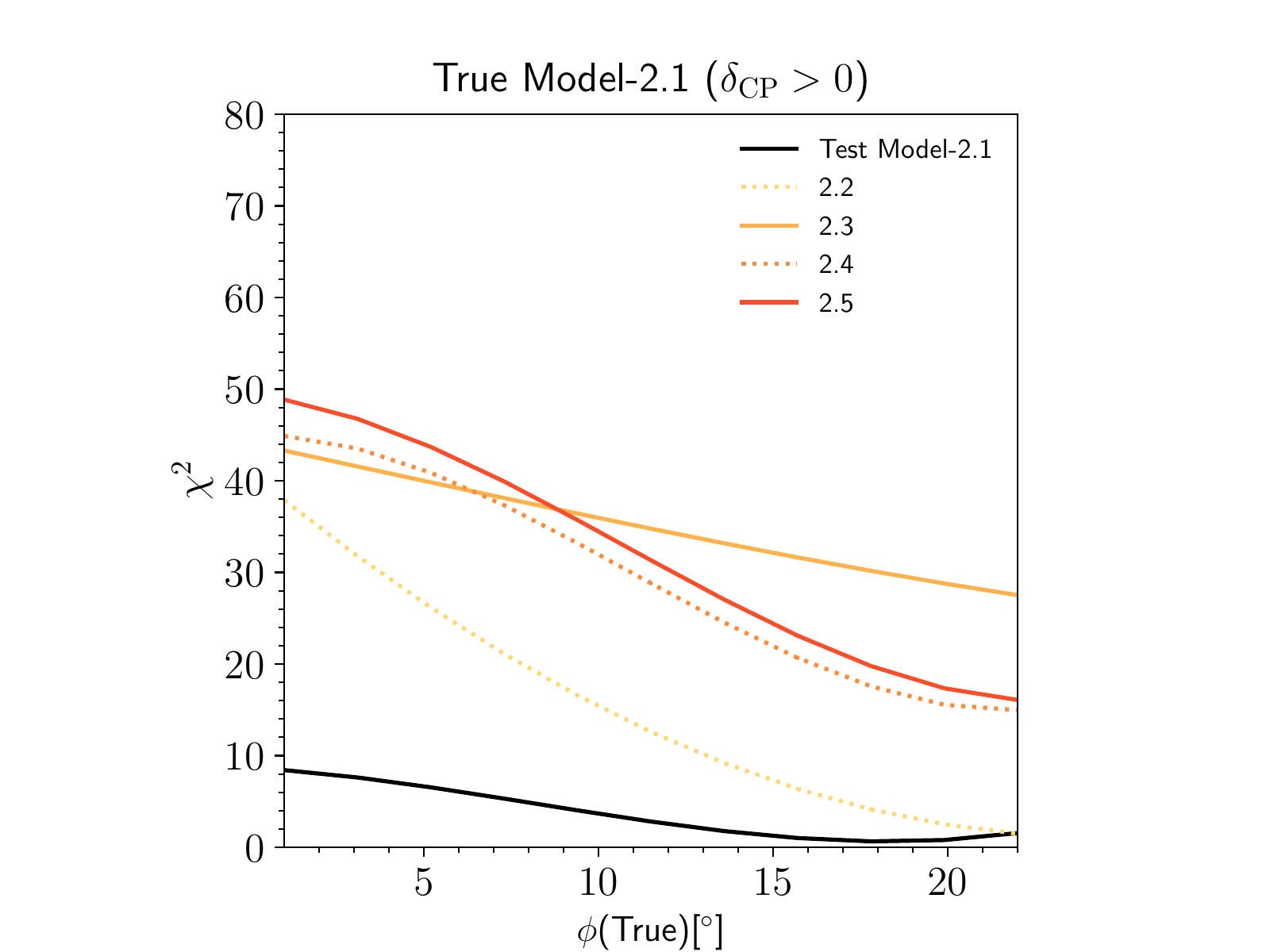}
\end{tabular}
\caption{Comparison among the five allowed two-parameter models as test models for ESSnuSB using priors in the cases the model parameters $\phi$ (upper row) and $\theta$ (lower row) are minimized and corresponding to the first (left column) and the second (right column) best-fit points in Table~\ref{tab:fits} for Model~2.1 as the true model. For description of each panel, see~Fig.~\ref{fig:comp_oneparamod}.}
\label{fig:comp_twoparamod_21}
\end{figure}

\begin{figure}[t!]
\vspace{-2cm}
\hspace{-10pt}
\begin{tabular}{ll}
\includegraphics[scale=0.55]{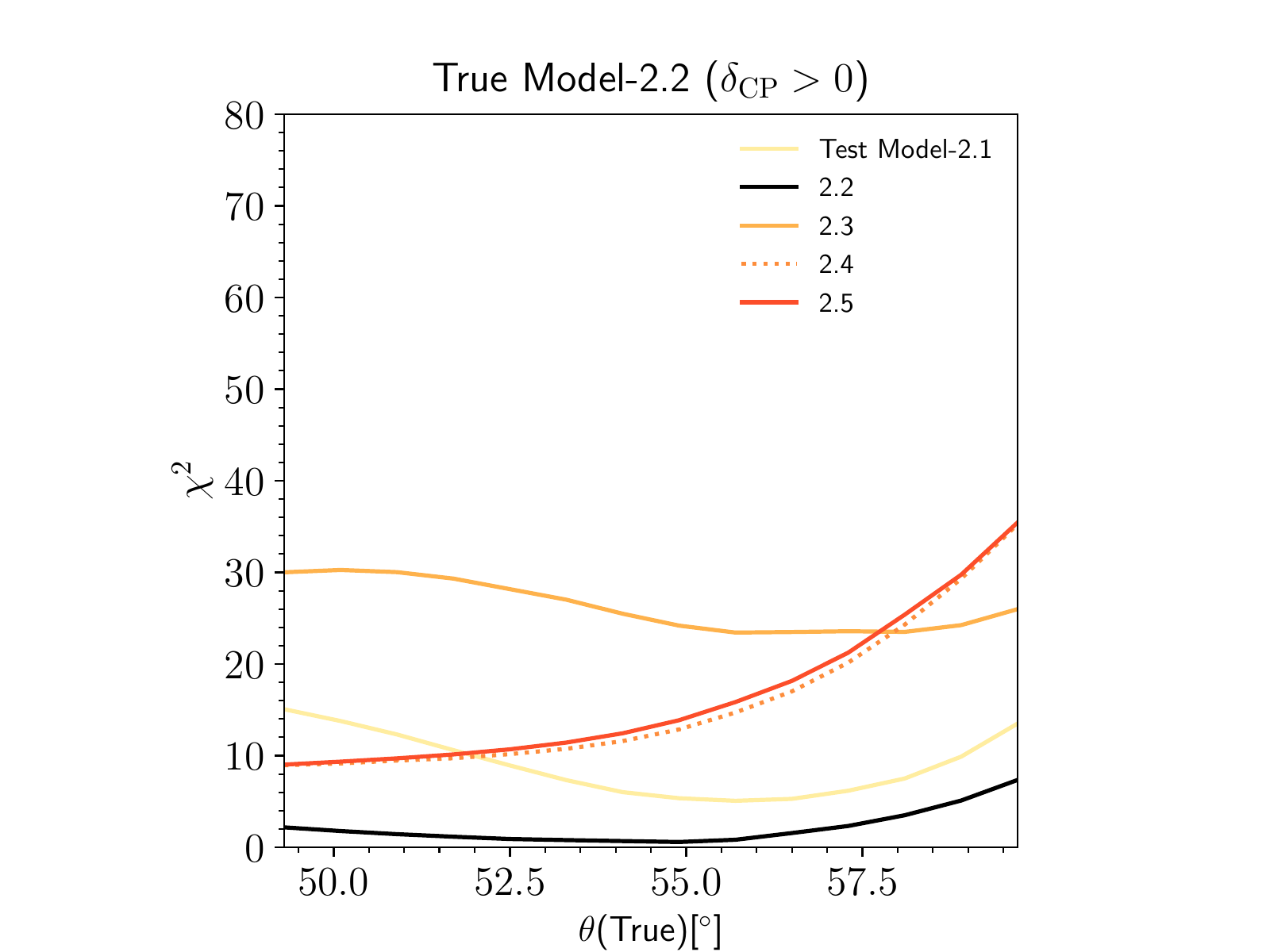}
\hspace{-50pt}
\includegraphics[scale=0.55]{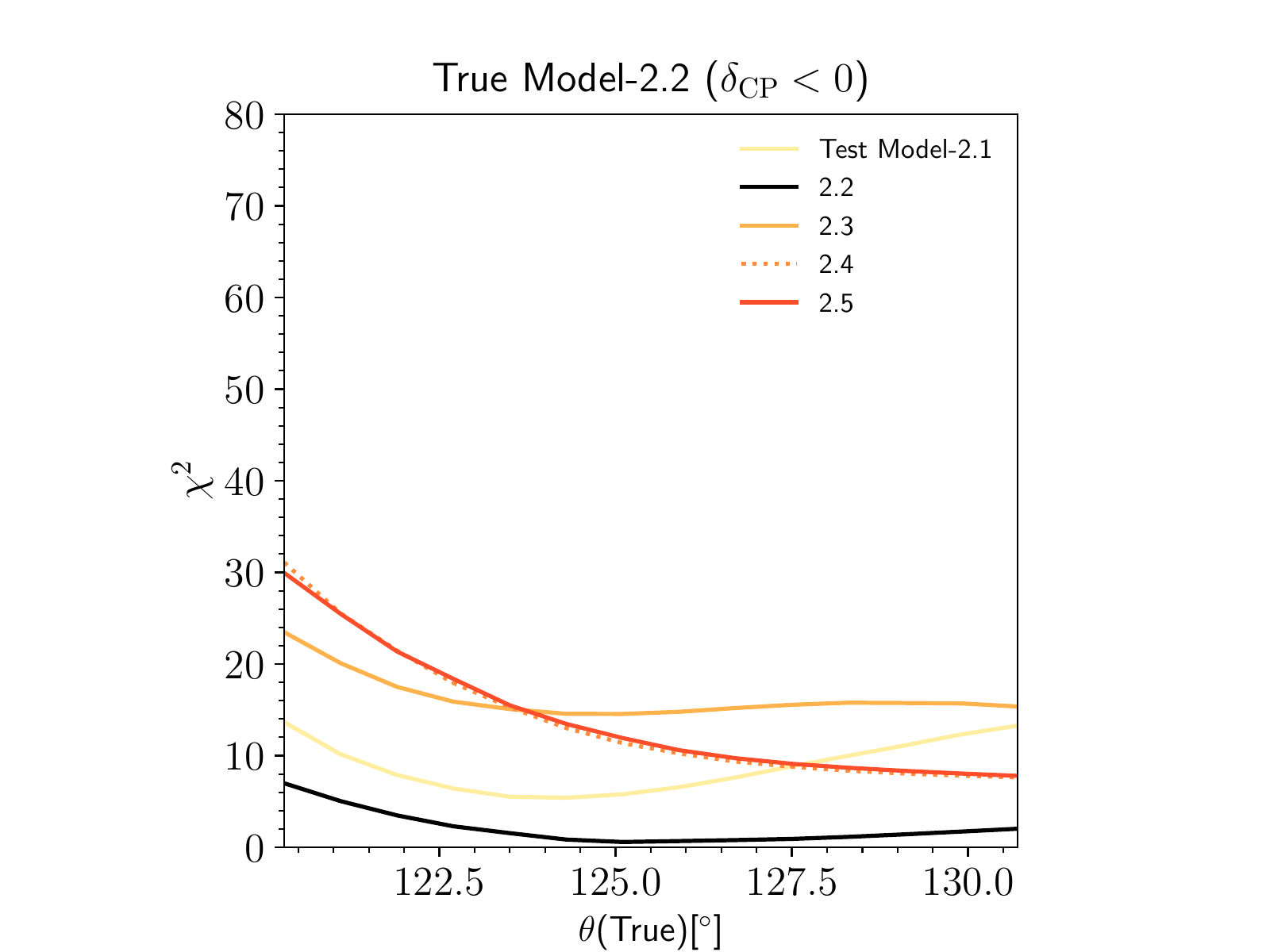} \\[1mm]
\includegraphics[scale=0.55]{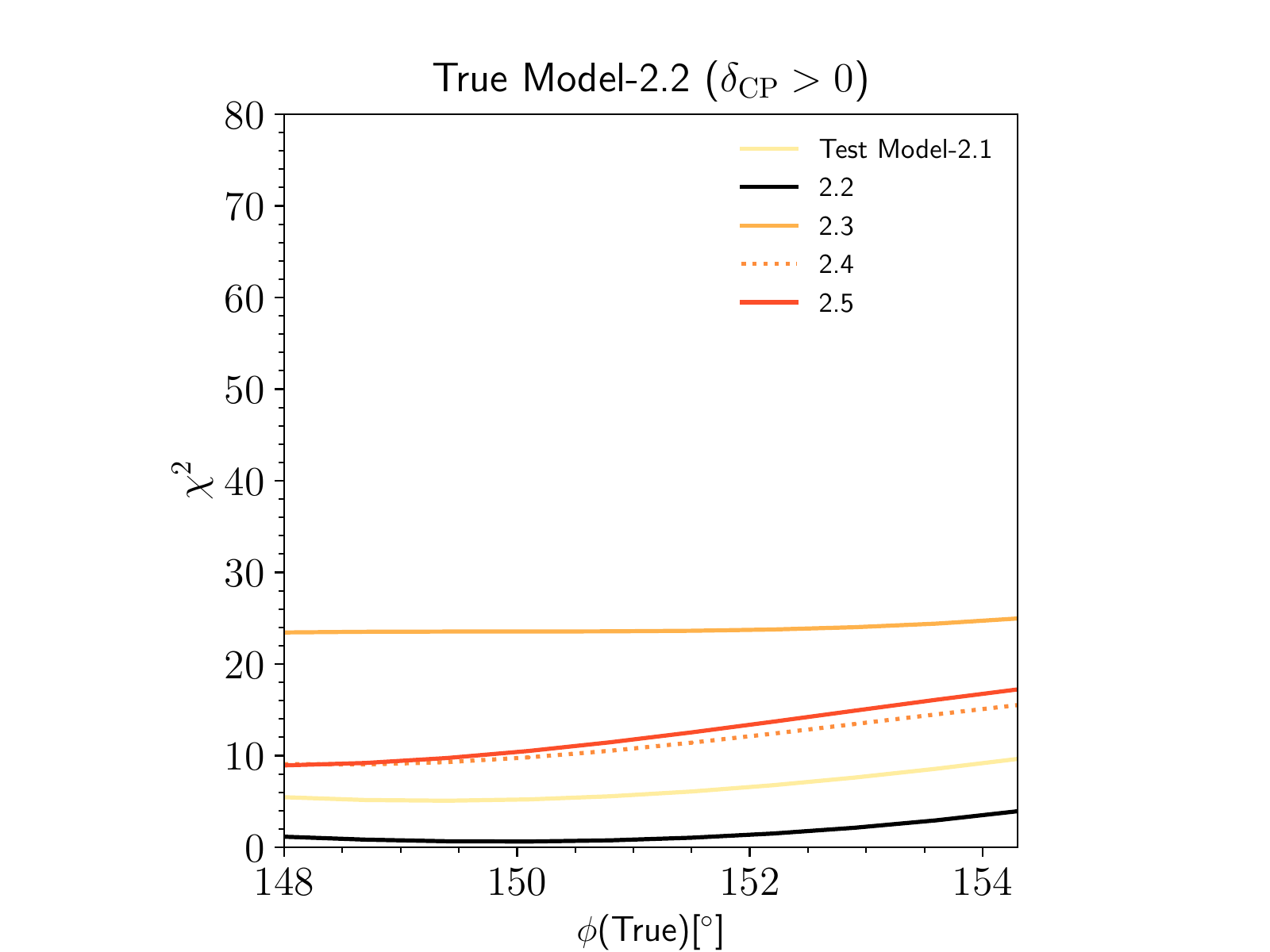}
\hspace{-50pt}
\includegraphics[scale=0.55]{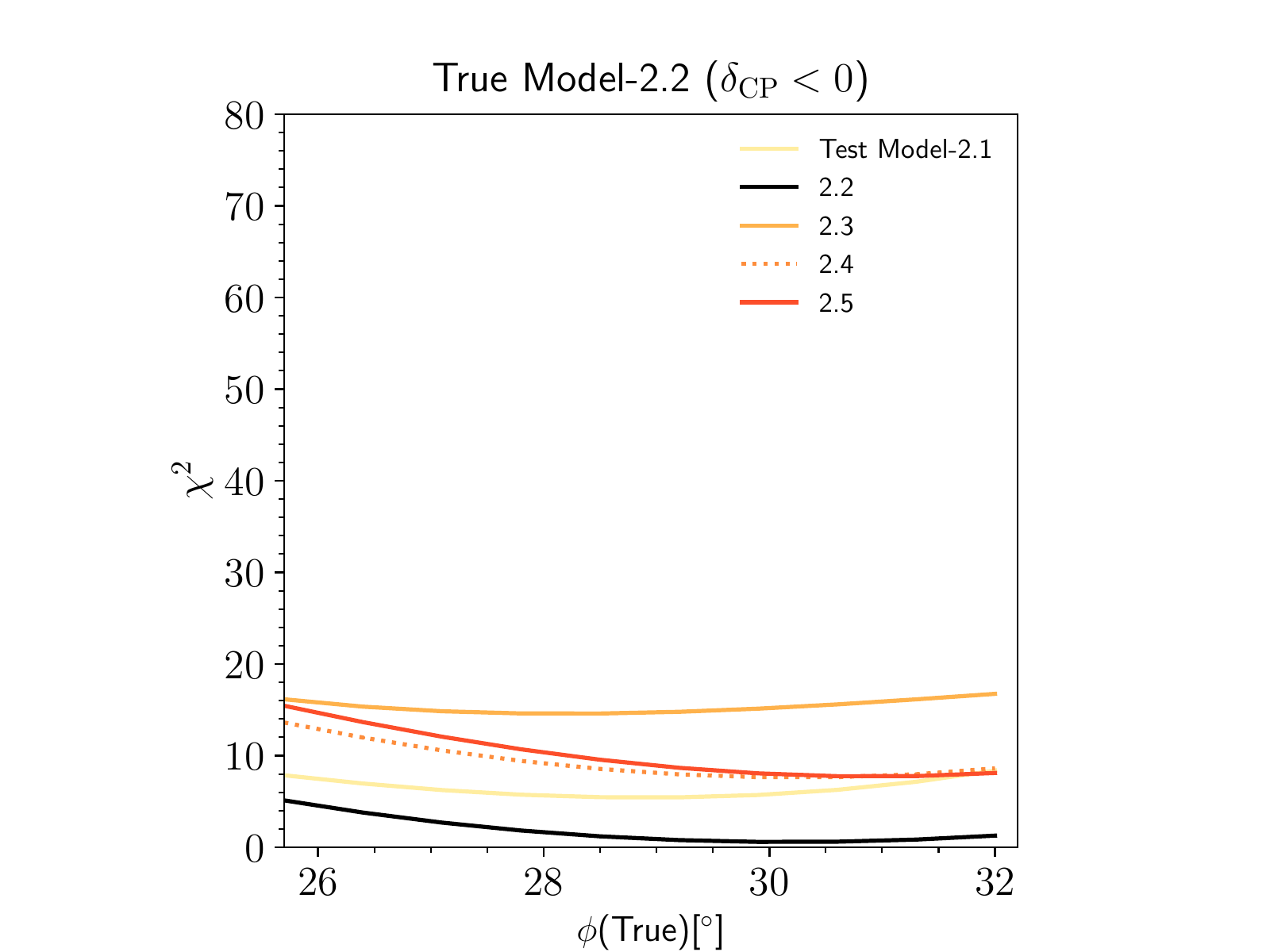}
\end{tabular}
\caption{Comparison among the five allowed two-parameter models as test models for ESSnuSB using priors in the cases the model parameters $\phi$ (upper row) and $\theta$ (lower row) are minimized and corresponding to the first (left column) and the second (right column) best-fit points in Table~\ref{tab:fits} for Model~2.2 as the true model. For description of each panel, see~Fig.~\ref{fig:comp_oneparamod}.}
\label{fig:comp_twoparamod_22}
\end{figure}

\begin{figure}[t!]
\vspace{-2cm}
\hspace{-10pt}
\begin{tabular}{ll}
\includegraphics[scale=0.55]{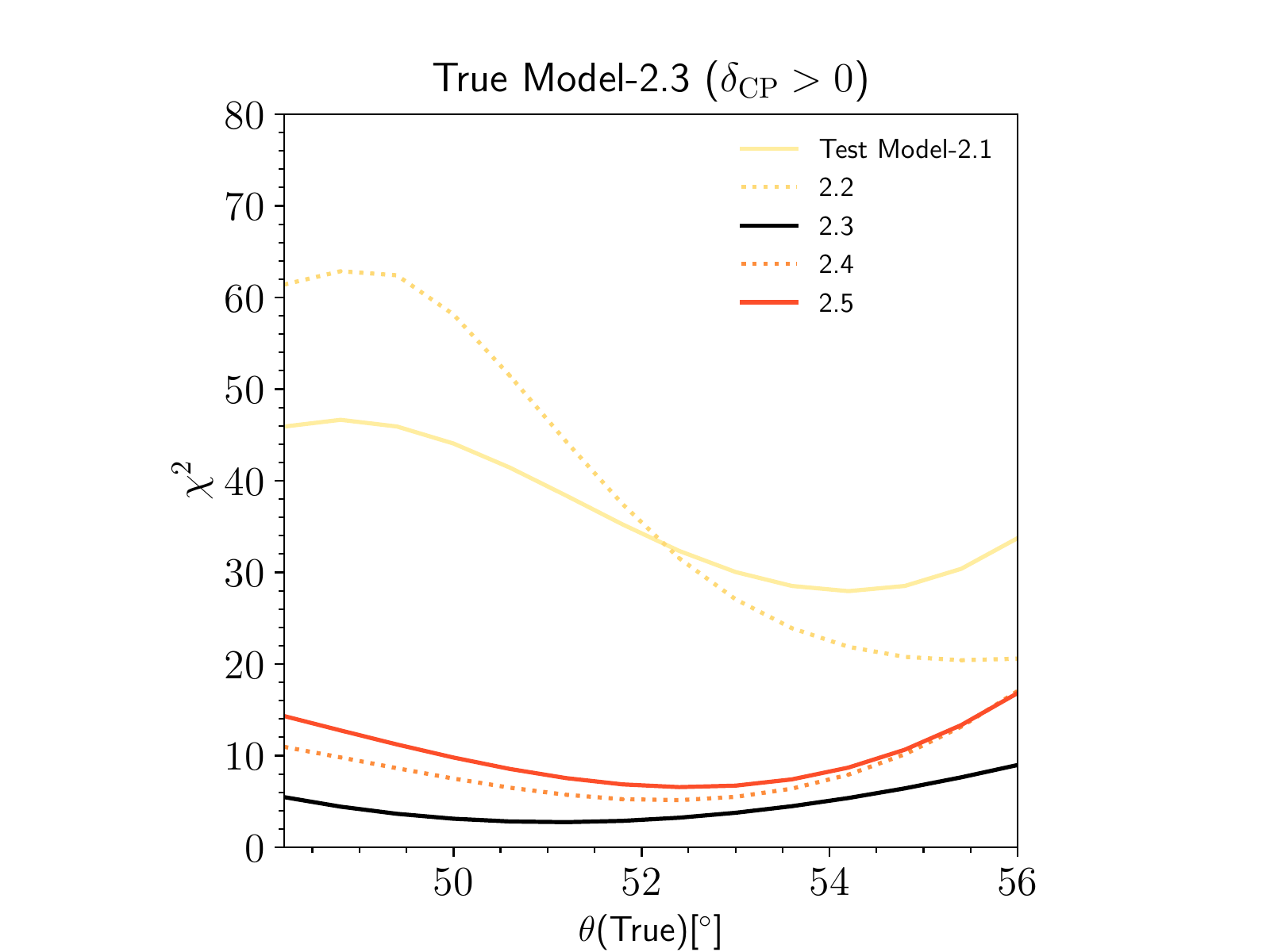}
\hspace{-50pt}
\includegraphics[scale=0.55]{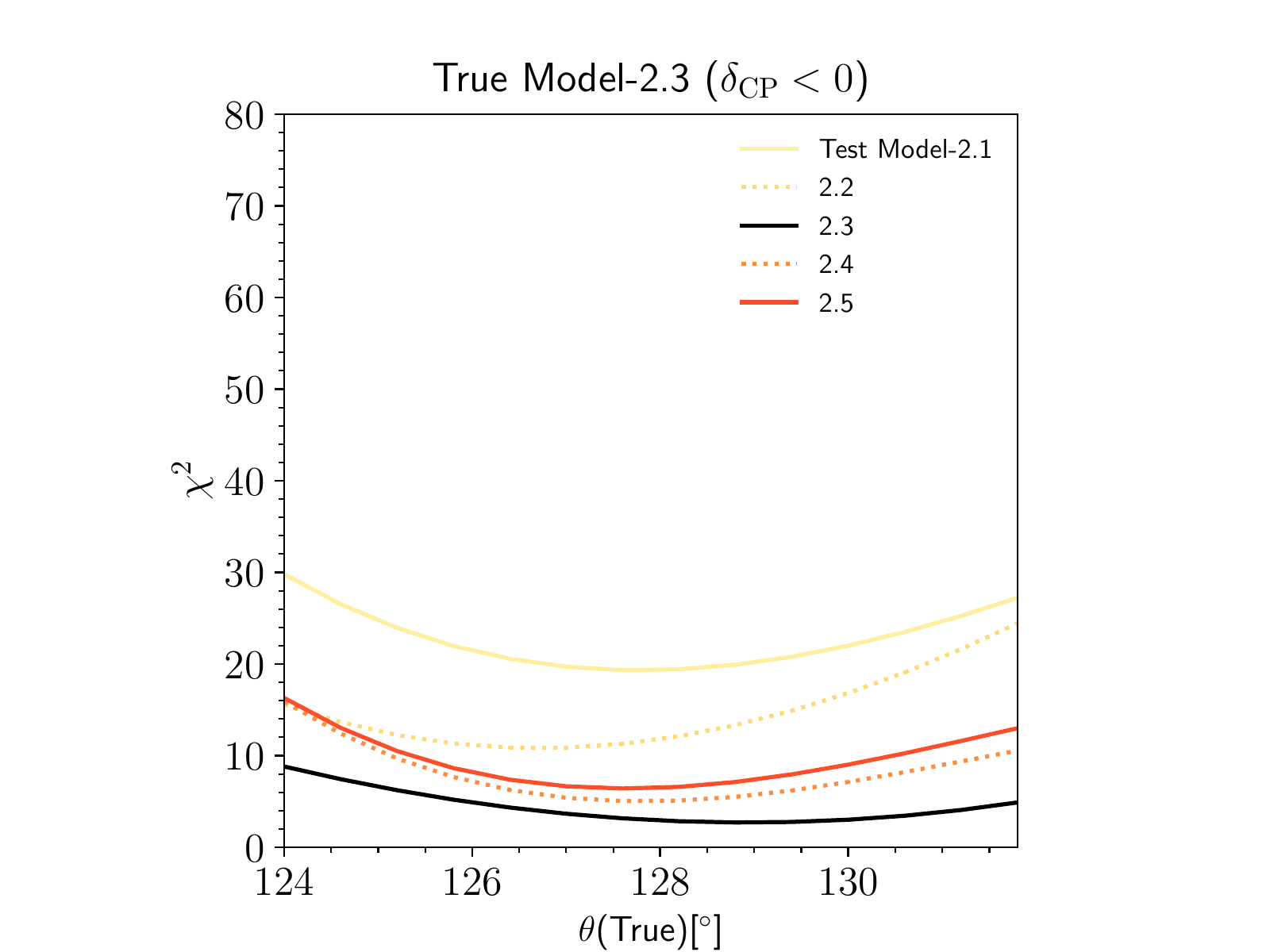} \\[1mm]
\includegraphics[scale=0.55]{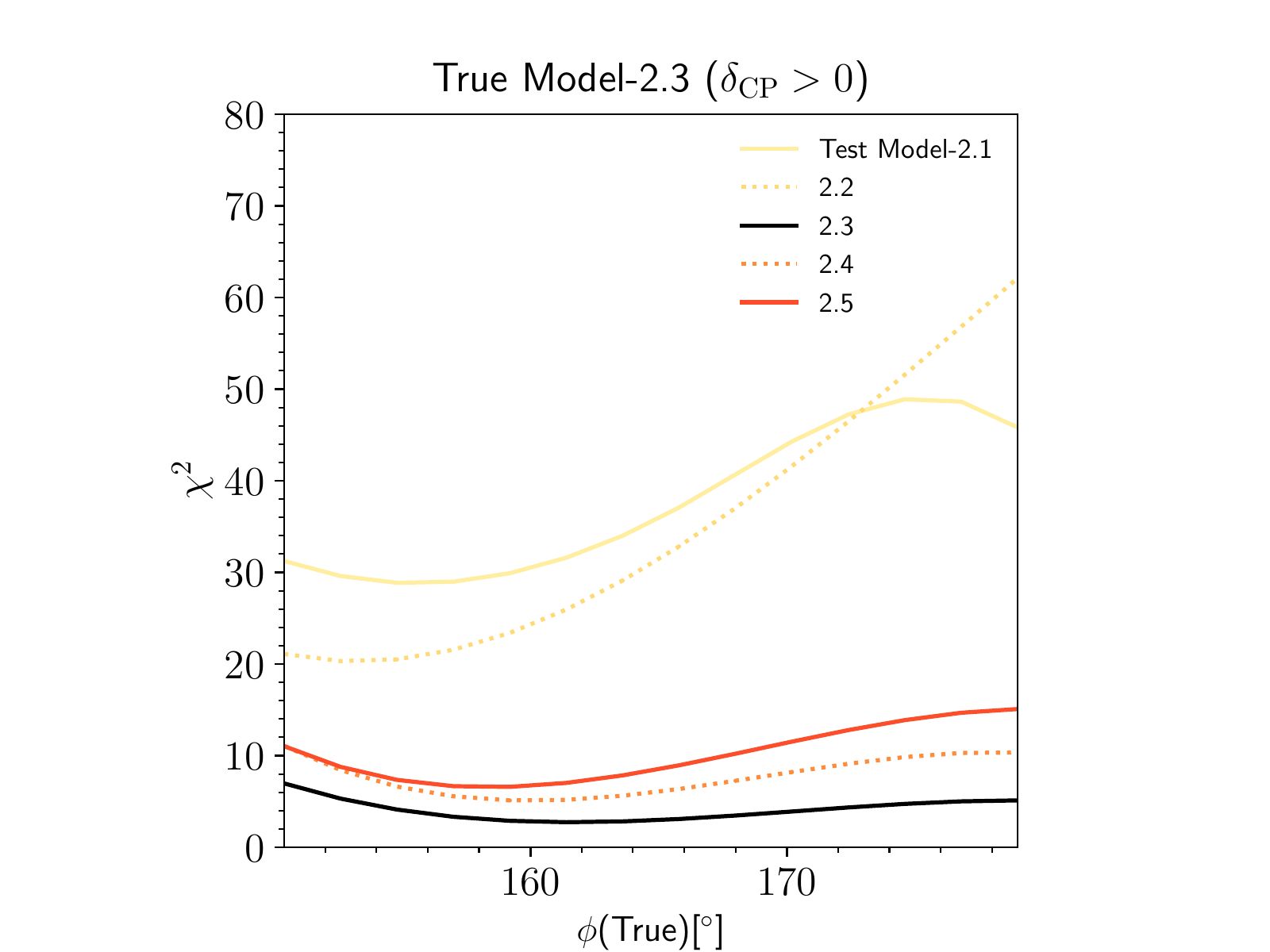}
\hspace{-50pt}
\includegraphics[scale=0.55]{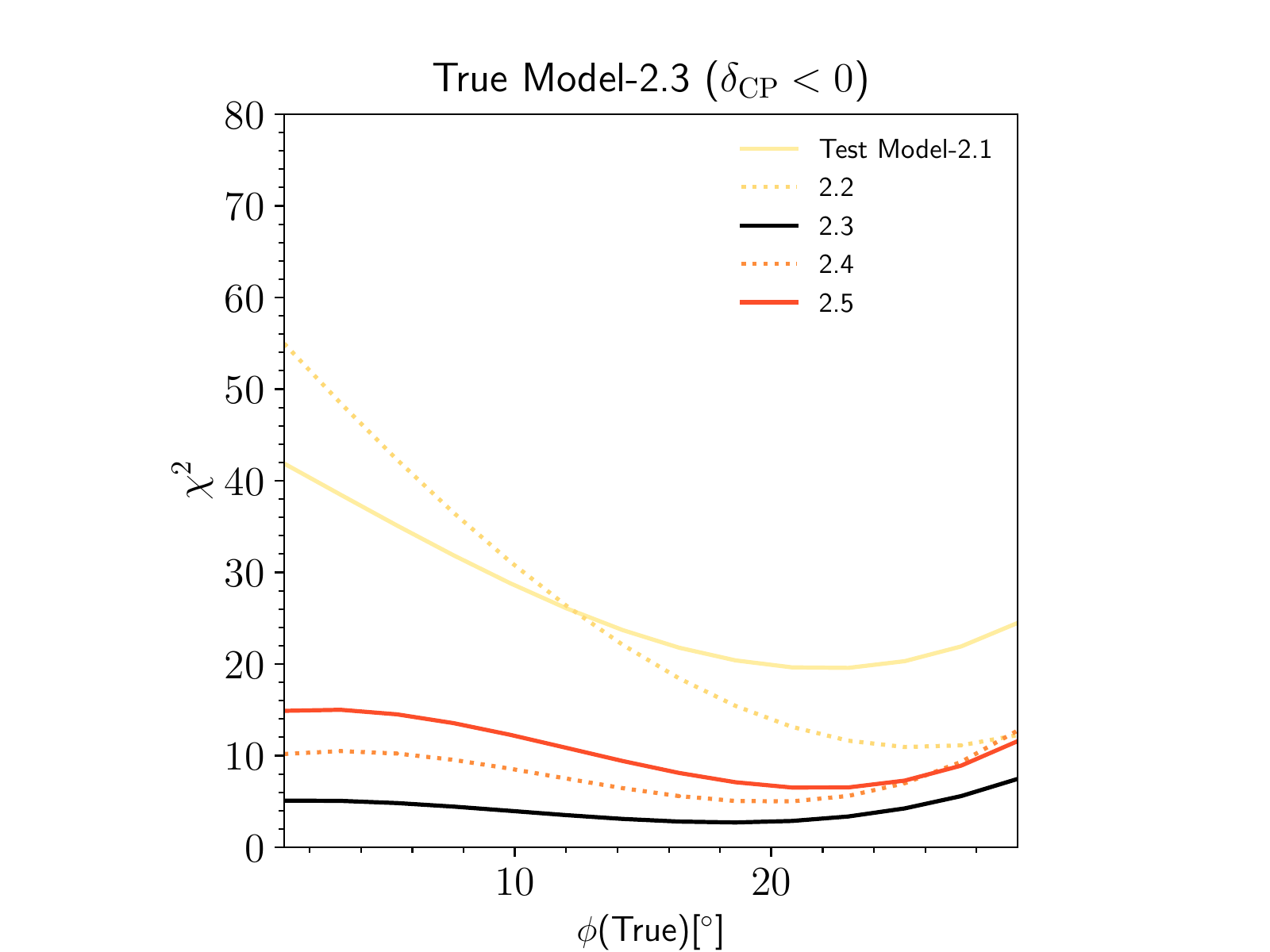}
\end{tabular}
\caption{Comparison among the five allowed two-parameter models as test models for ESSnuSB using priors in the cases the model parameters $\phi$ (upper row) and $\theta$ (lower row) are minimized and corresponding to the first (left column) and the second (right column) best-fit points in Table~\ref{tab:fits} for Model~2.3 as the true model. For description of each panel, see~Fig.~\ref{fig:comp_oneparamod}.}
\label{fig:comp_twoparamod_23}
\end{figure}

\begin{figure}[t!]
\vspace{-2cm}
\hspace{-10pt}
\begin{tabular}{ll}
\includegraphics[scale=0.55]{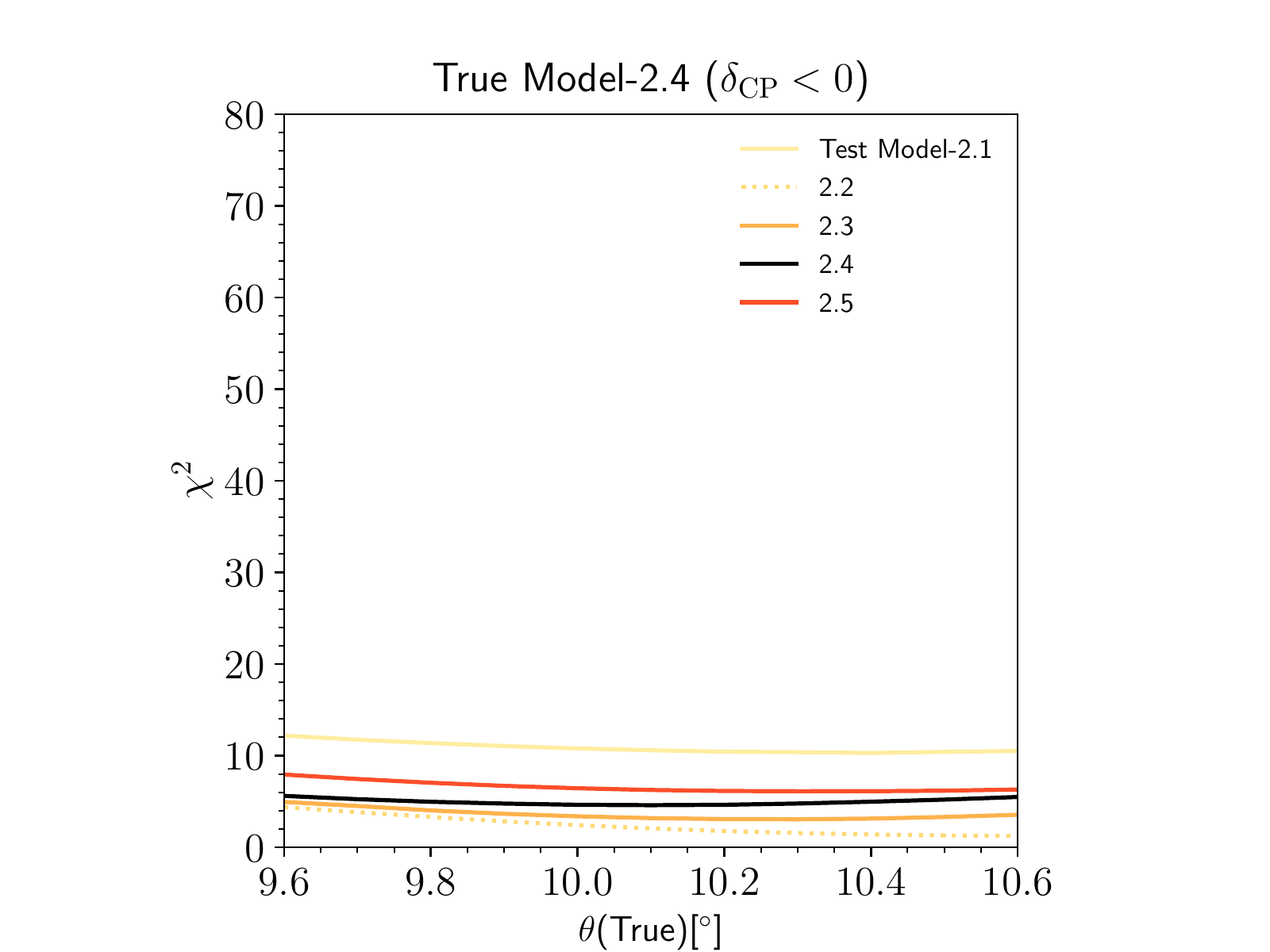}
\hspace{-50pt}
\includegraphics[scale=0.55]{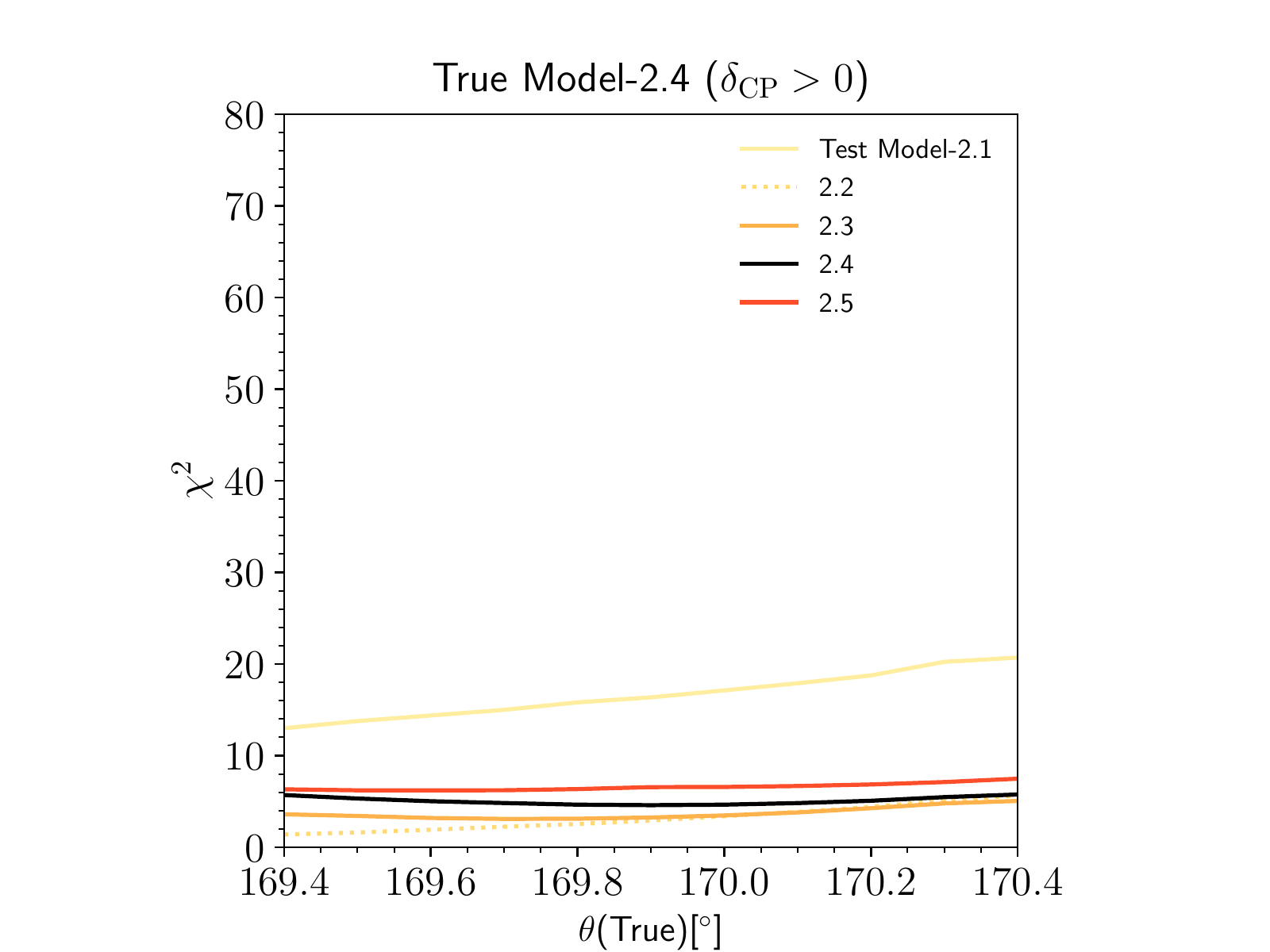} \\[1mm]
\includegraphics[scale=0.55]{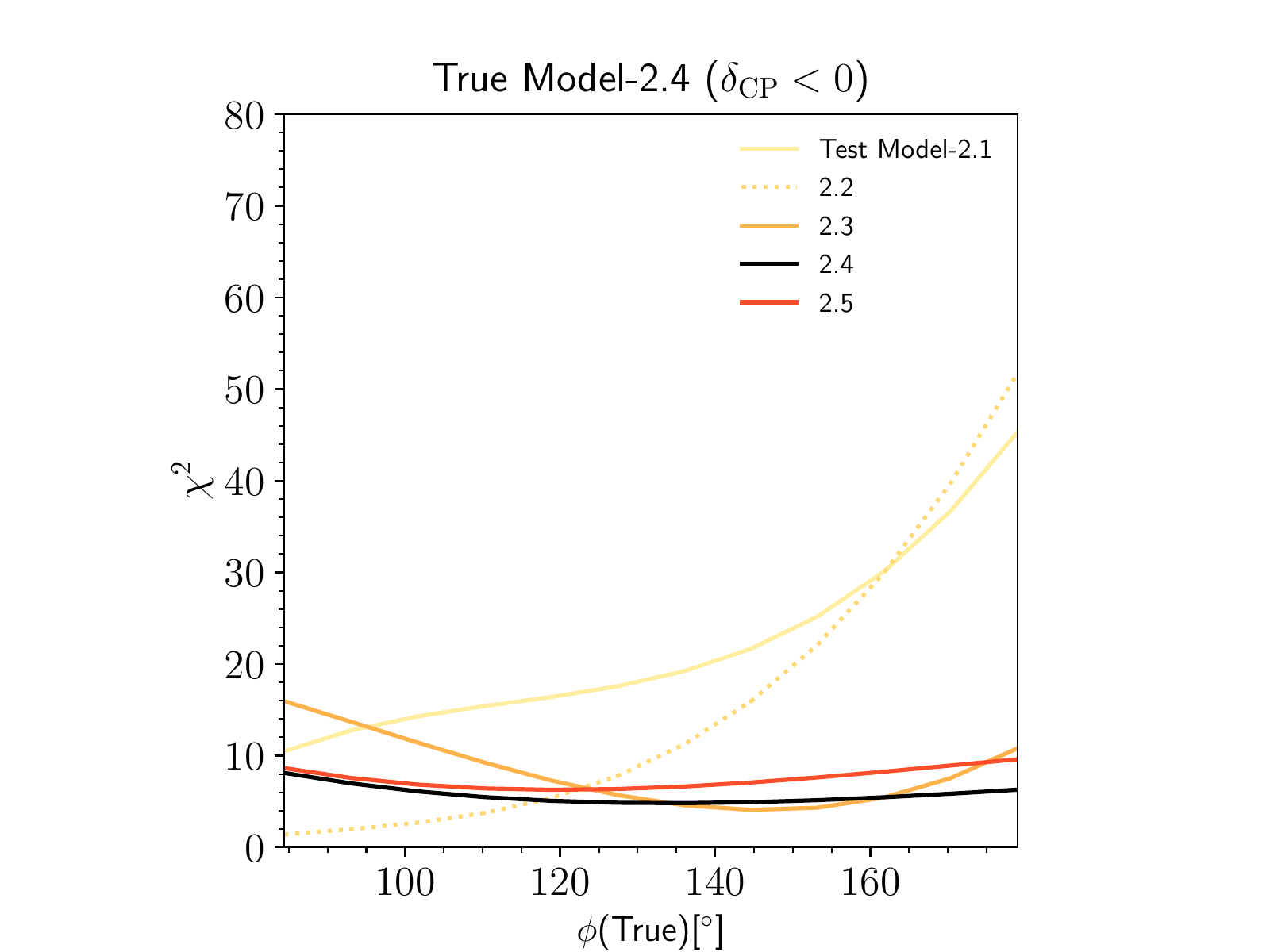}
\hspace{-50pt}
\includegraphics[scale=0.55]{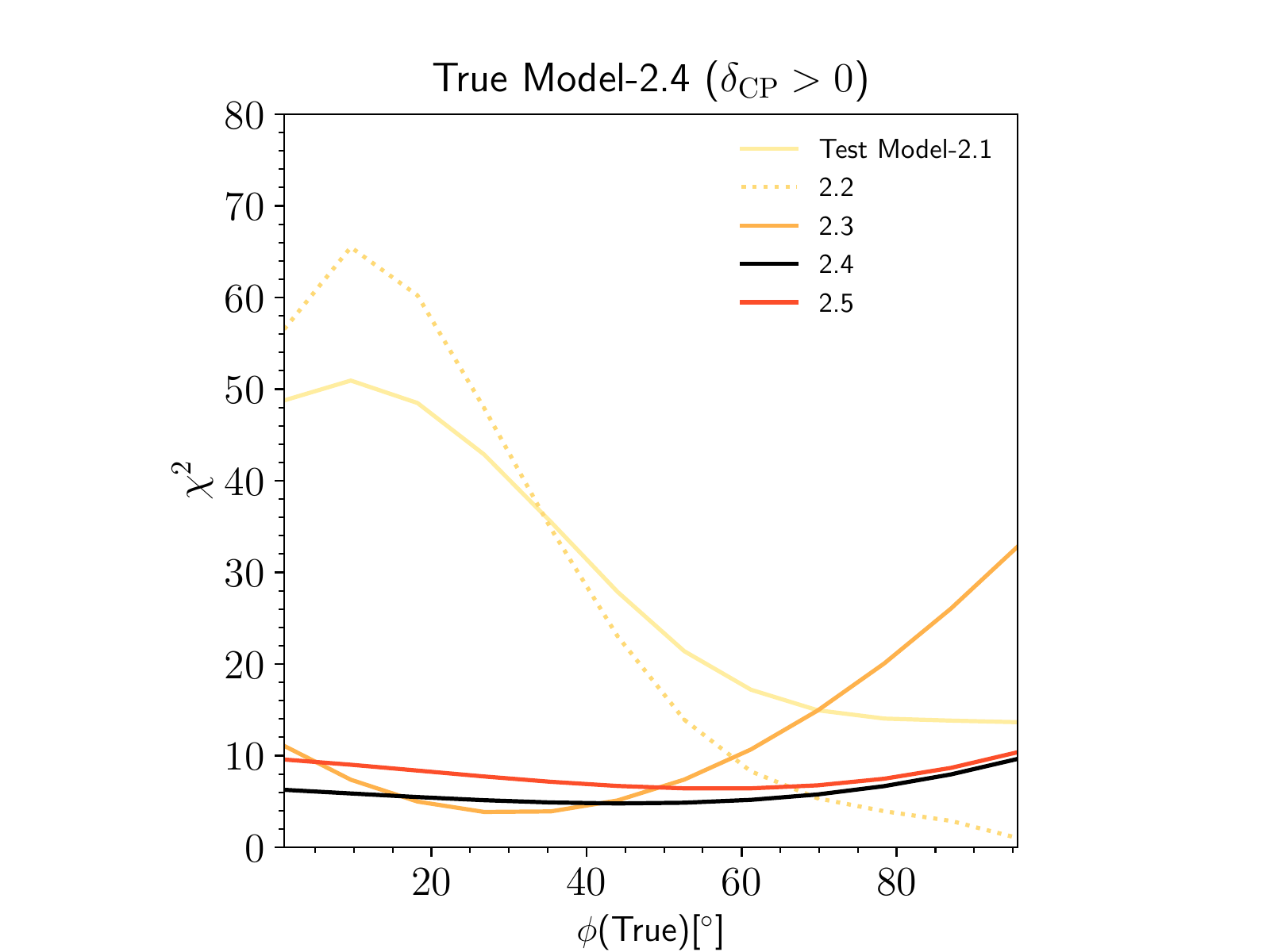}
\end{tabular}
\caption{Comparison among the five allowed two-parameter models as test models for ESSnuSB using priors in the cases the model parameters $\phi$ (upper row) and $\theta$ (lower row) are minimized and corresponding to the first (left column) and the second (right column) best-fit points in Table~\ref{tab:fits} for Model~2.4 as the true model. For description of each panel, see~Fig.~\ref{fig:comp_oneparamod}.}
\label{fig:comp_twoparamod_24}
\end{figure}

\begin{figure}[t!]
\vspace{-2cm}
\hspace{-10pt}
\begin{tabular}{ll}
\includegraphics[scale=0.55]{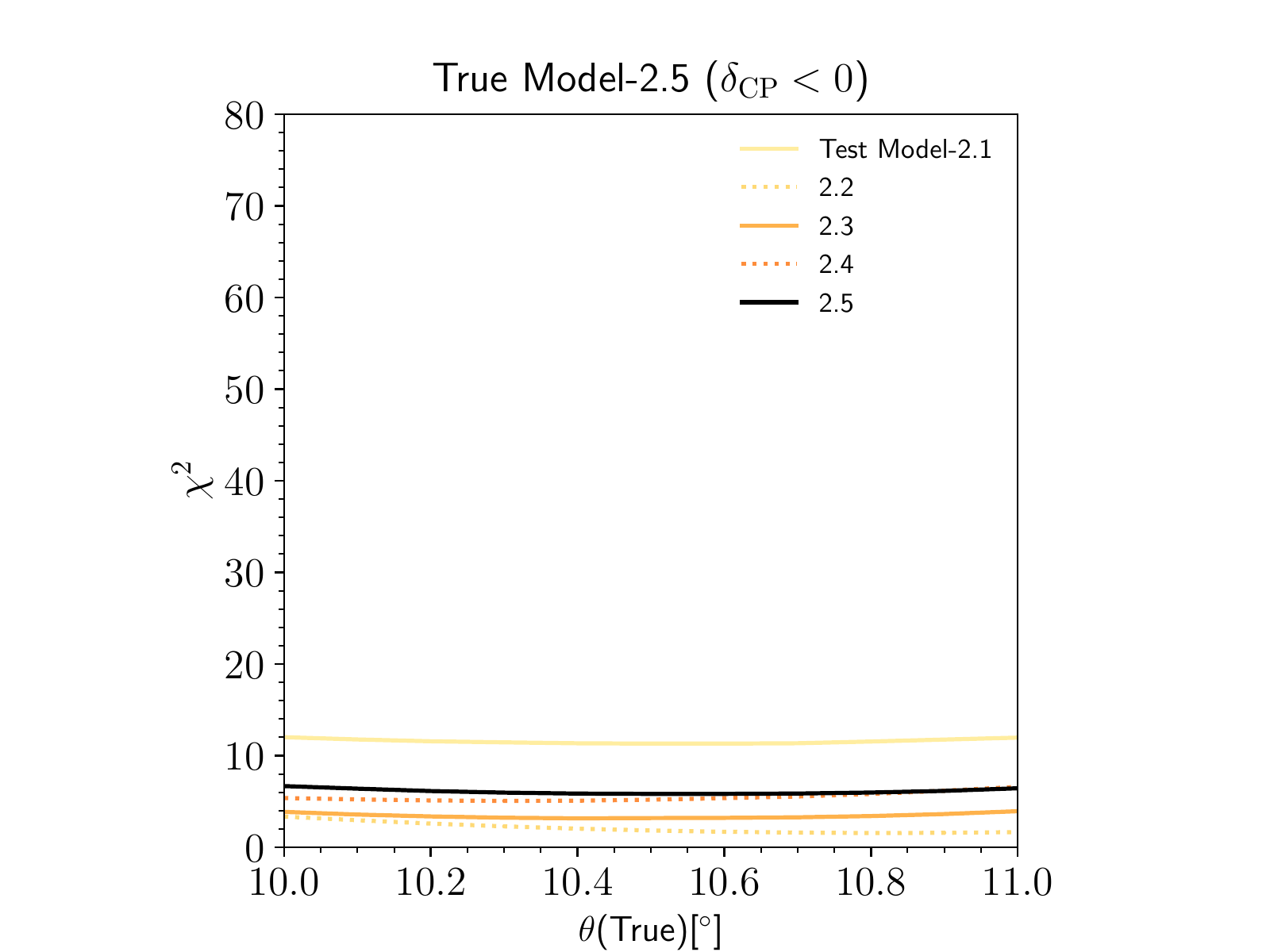}
\hspace{-50pt}
\includegraphics[scale=0.55]{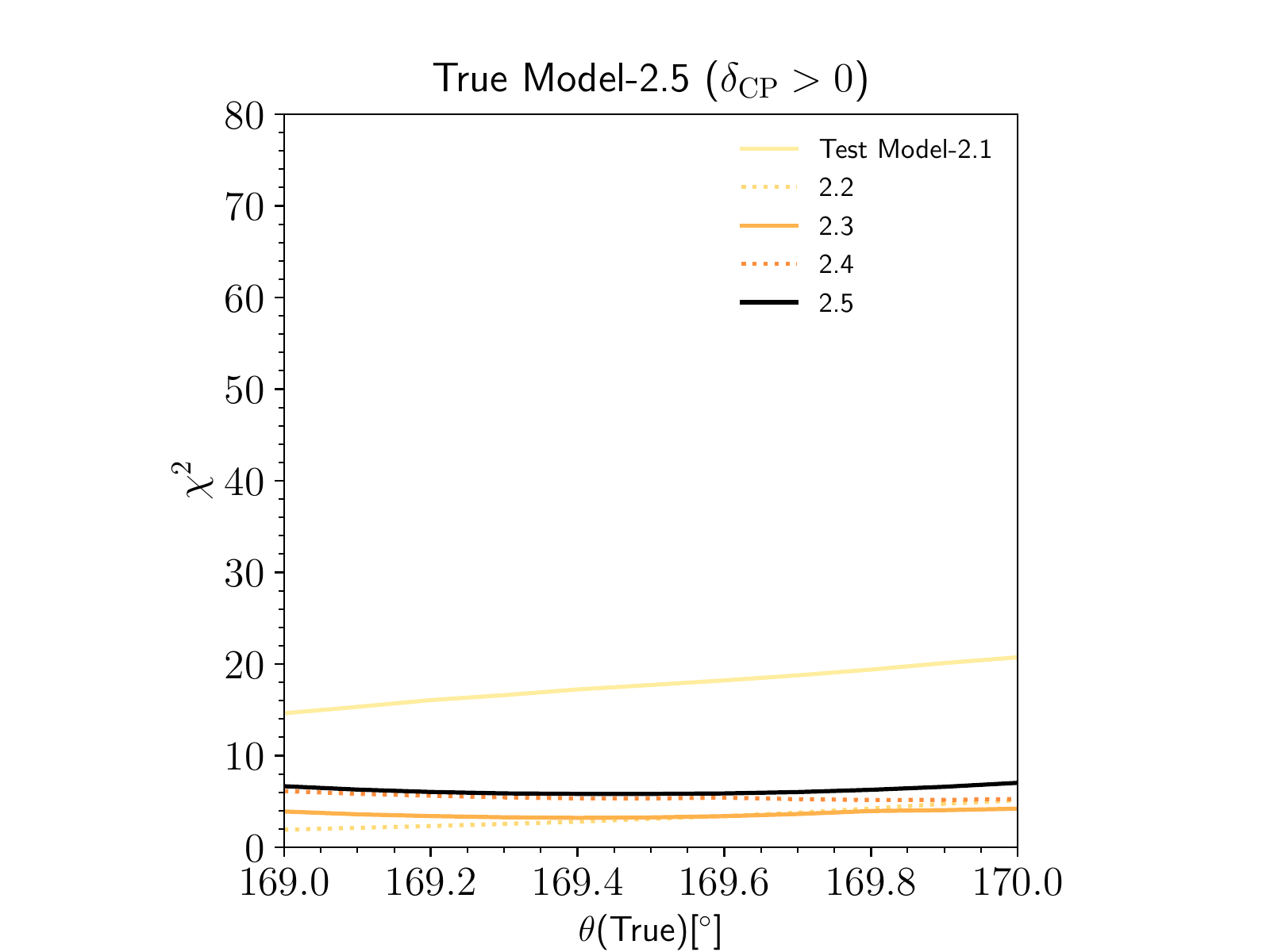} \\[1mm]
\includegraphics[scale=0.55]{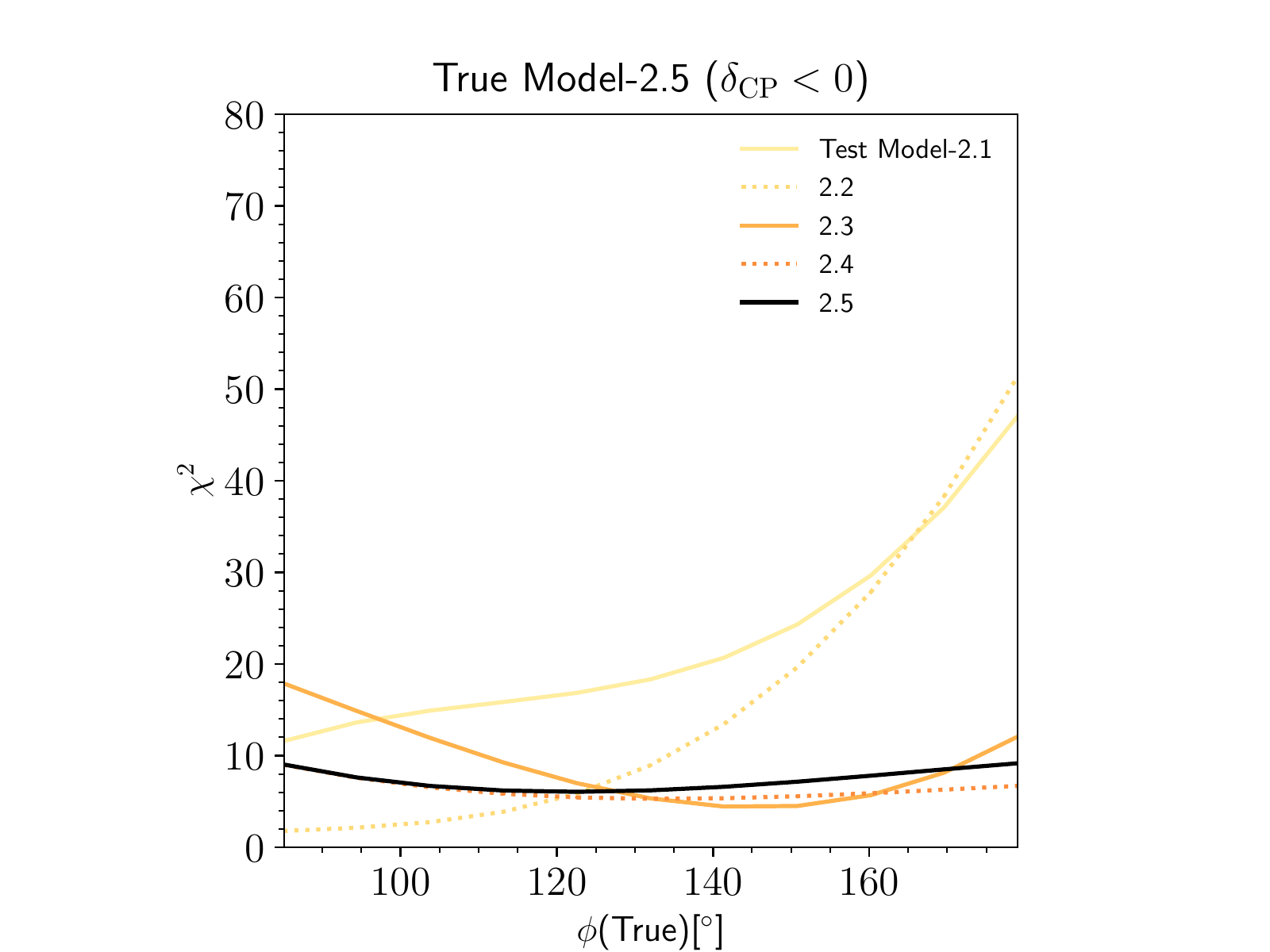}
\hspace{-50pt}
\includegraphics[scale=0.55]{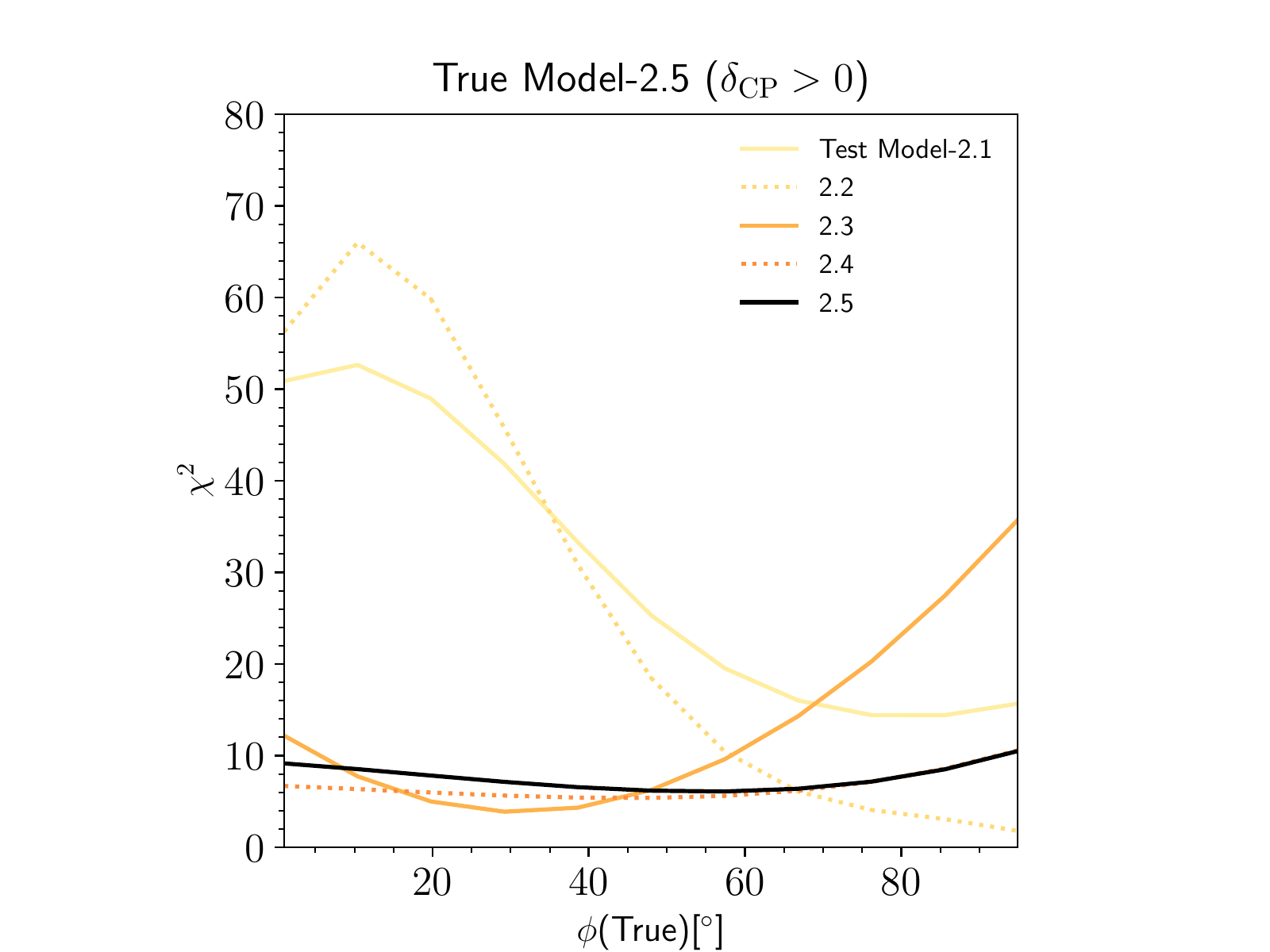}
\end{tabular}
\caption{Comparison among the five allowed two-parameter models as test models for ESSnuSB using priors in the cases the model parameters $\phi$ (upper row) and $\theta$ (lower row) are minimized and corresponding to the first (left column) and the second (right column) best-fit points in Table~\ref{tab:fits} for Model~2.5 as the true model. For description of each panel, see~Fig.~\ref{fig:comp_oneparamod}.}
\label{fig:comp_twoparamod_25}
\end{figure}

In \Figs~\ref{fig:comp_twoparamod_21}--\ref{fig:comp_twoparamod_25}, we present the results for the two-parameter models. As the two-parameter models have several parameters, each panel shows the results for particular choices of the true parameter shown on the horizontal axes profiled over the other parameter, e.g., in \Fig~\ref{fig:comp_twoparamod_21} (upper-left panel), the curves show how $\chi^2_{\rm min}$ of the different models behave as a function of the true $\theta$ in Model~2.1 when minimized over the true value of $\phi$. Unlike in the case of the one-parameter models, the two-parameter models do not split into different groups. Still, $\chi^2$ of the different models are highly dependent on the assumed true model, but also on the true parameter values assumed for those models. A remarkable feature is that, for the three models that provide the best fit to the current neutrino oscillation data, the ESSnuSB data make those models the preferred one if they are assumed as the true model for most of the parameter space. As such, the ESSnuSB data can also help in the rejection of other models, in particular those that are relatively close in $\chi^2$ at the present time, even for the two-parameter models where the regions of the leptonic mixing parameters are more extended than for the one-parameter models, cf.~\Figs~\ref{fig:pred_oneparamod} and~\ref{fig:pred_twoparamod}. The addition of the ESSnuSB data is sufficient to disfavor most other models at at least $3\sigma$ for Models~2.1--2.3. As discussed earlier, the two-parameter models give degenerate fits of $\theta$ and $\phi$ to current data, corresponding to $\delta_{\rm CP} < 0$ or $\delta_{\rm CP} > 0$, since the values of the other leptonic mixing parameters are the same. In other words, the two-parameter models are symmetric for the degenerate values of $\theta$ and $\phi$ apart from the prediction for $\delta_{\rm CP}$ and this feature is reflected in \Figs~\ref{fig:comp_twoparamod_21}--\ref{fig:comp_twoparamod_25}, where the exact symmetry is broken by the different phenomenology for different values of $\delta_{\rm CP}$ at ESSnuSB. For example, the symmetry is more prominent in Fig.~\ref{fig:comp_twoparamod_21} and less prominent in Fig.~\ref{fig:comp_twoparamod_23}. This can be understood, since for Model~2.1, the neutrino oscillation probabilities are not very different when $\delta_{\rm CP} \rightarrow -\delta_{\rm CP}$, while they are different for Model~2.3. Another interesting feature to note is that the sensitivities for Models~2.4 and 2.5 are very similar. The reason is that these two models predict similar values of all leptonic mixing parameters except for $\theta_{12}$.

\begin{figure}[t!]
\begin{center}
\includegraphics[scale=0.55]{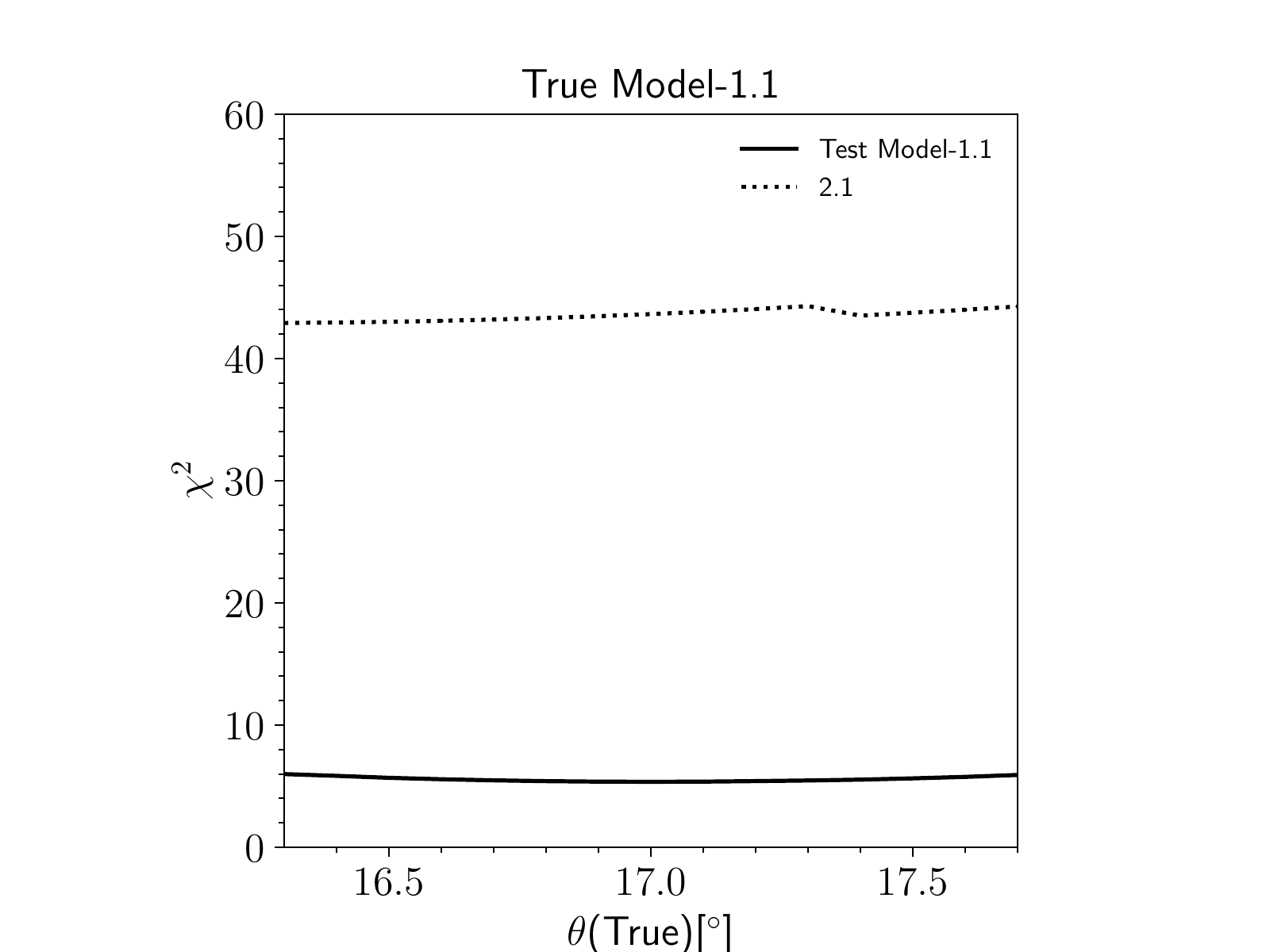}
\vspace{1mm}
\caption{Comparison between the best one-parameter model (Model 1.1) and the best two-parameter model (Model 2.1) for ESSnuSB using priors. The panel shows the quantity $\chi^2$, given Model 1.1 as the true model, as a function of the model parameter $\theta$ in its $3\sigma$ interval $\theta_{3\sigma}$ presented in Table~\ref{tab:fits}. The value of $\chi^2_\mathrm{min}$ for Model 1.1 is displayed with a black solid curve (when acting as a test model), whereas it is marked for Model 2.1 (i.e., the test model) by a black dotted curve.}
\label{fig:comp_1_2}
\end{center}
\end{figure}

\begin{figure}[t!]
\hspace{-10pt}
\begin{tabular}{ll}
\includegraphics[scale=0.55]{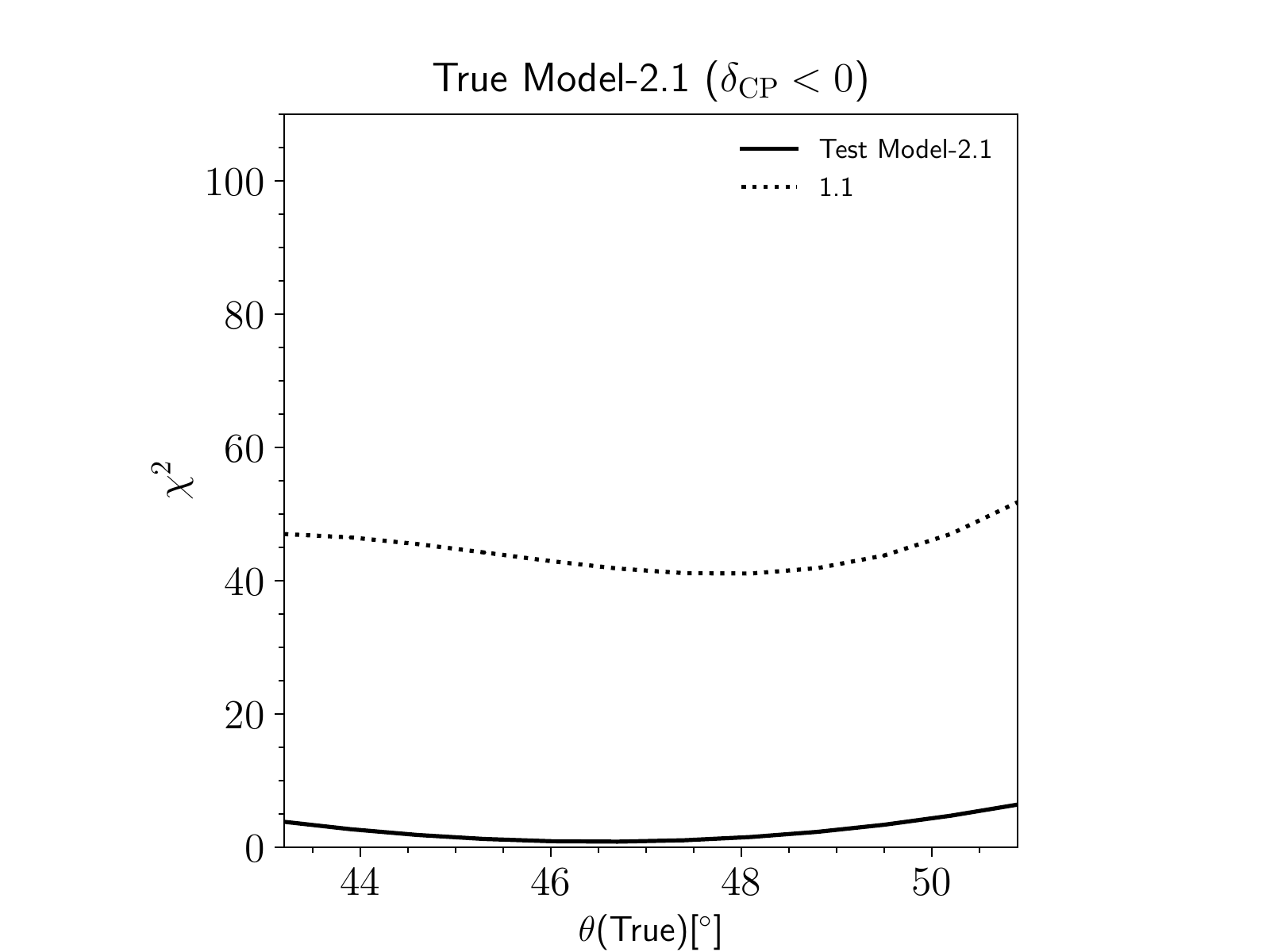}
\hspace{-50pt}
\includegraphics[scale=0.55]{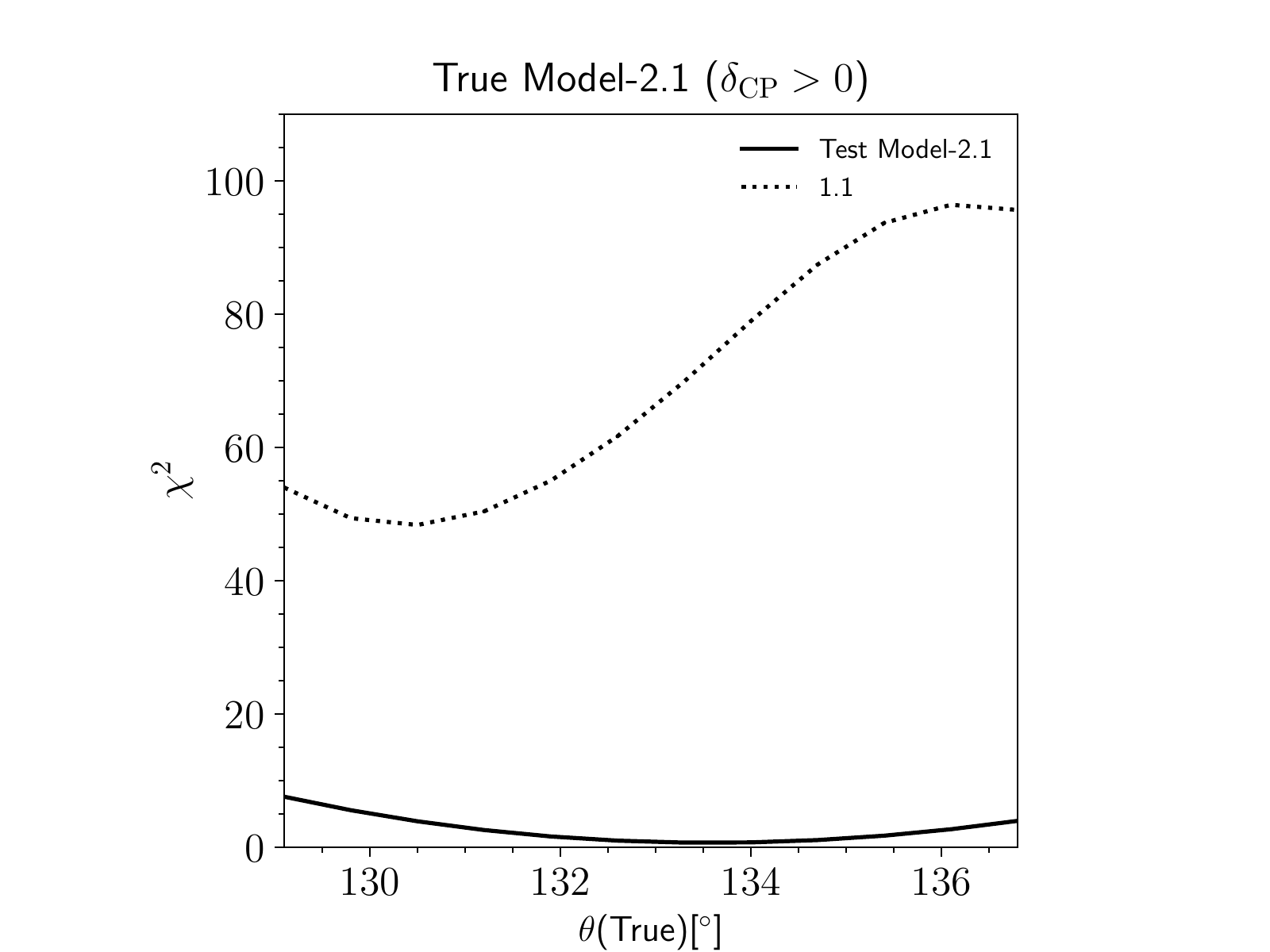}\\[1mm]
\includegraphics[scale=0.55]{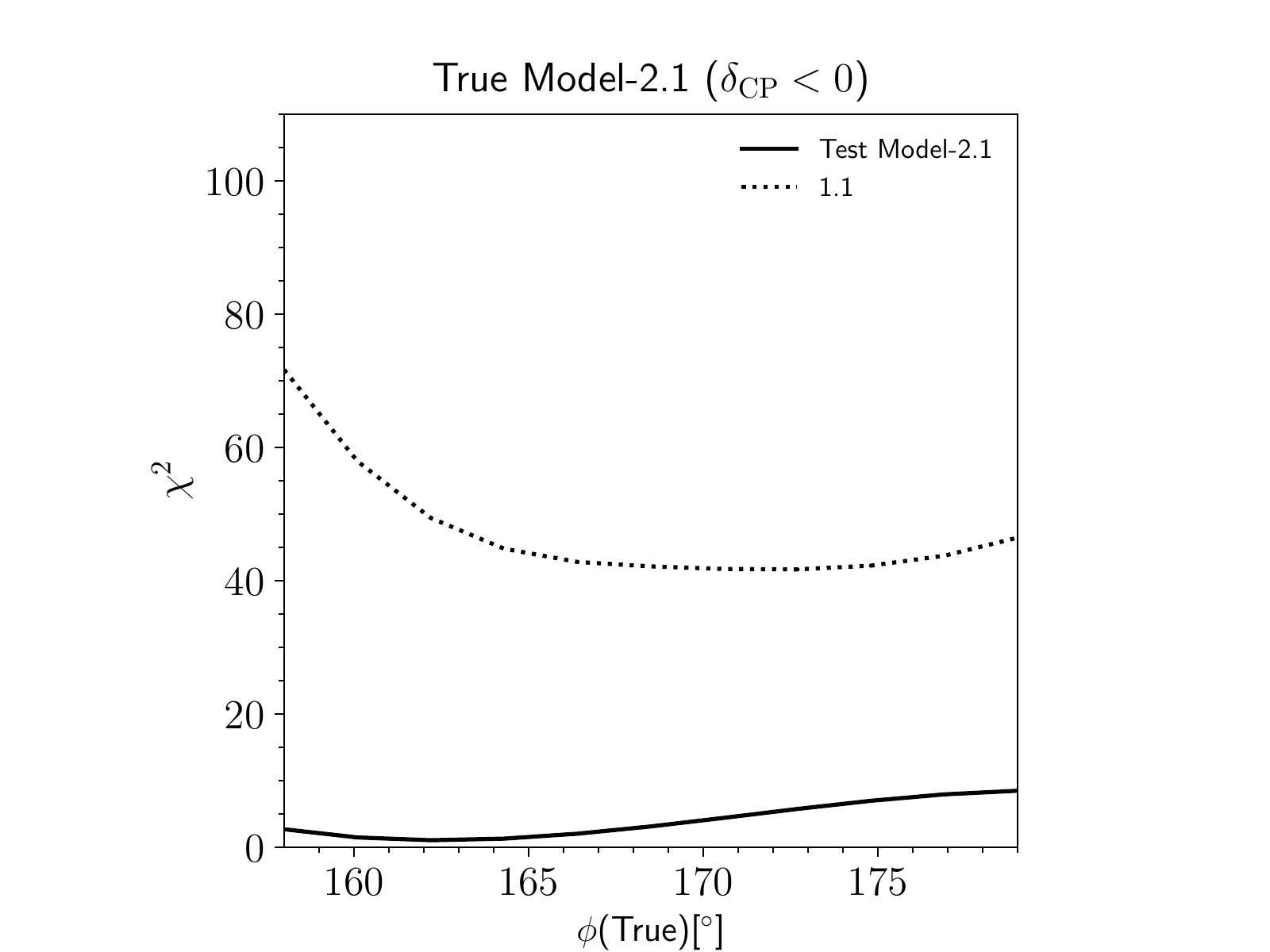}
\hspace{-50pt}
\includegraphics[scale=0.55]{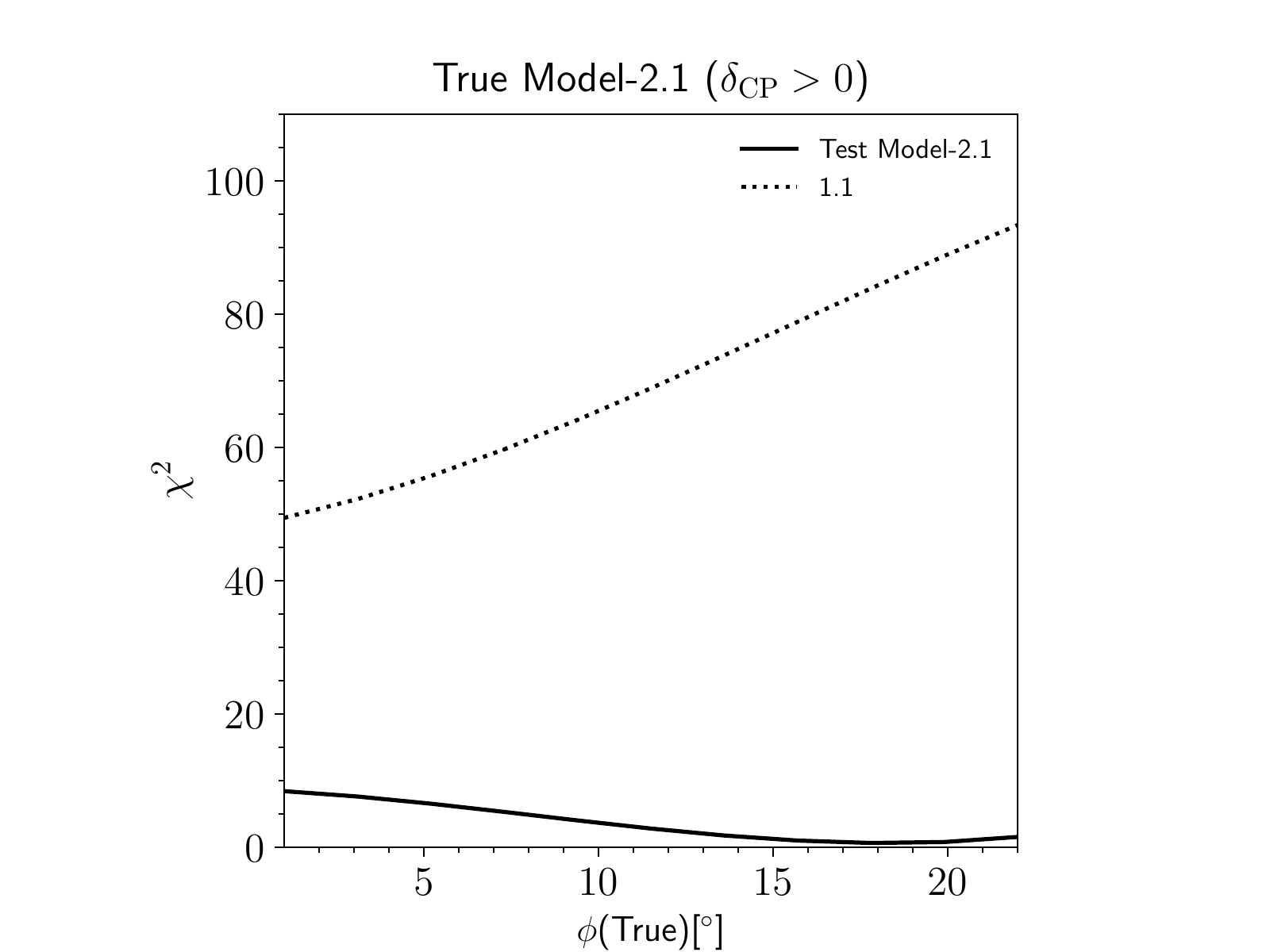}
\end{tabular}
\caption{Comparison between the best two-parameter model (Model 2.1) and the best one-parameter model (Model 1.1) for ESSnuSB using priors. The upper (lower) panels show the quantity $\chi^2_\mathrm{min}$, given Model~2.1  as the true model, as function of the model parameter $\theta$ ($\phi$) in its $3\sigma$ interval $\theta_{3\sigma}$ ($\phi_{3\sigma}$) presented in Table~\ref{tab:fits}. In the panels, $\chi^2_\mathrm{min}$ for Model 2.1 is displayed with black solid curves (when acting as a test model), whereas it is marked for Model 1.1 (i.e., the test model) by black dotted curves.}
\label{fig:comp_2_1}
\end{figure}

It is also possible to compare models with different number of parameters. As performing the full analysis using all ten models from the one- and two-parameter cases would result in figures that would be extremely cluttered, we present the comparison of the one- and two-parameter models that are the current best fits to the available data, i.e., Models~1.1 and~2.1. The resulting comparisons are shown in \Figs~\ref{fig:comp_1_2} and~\ref{fig:comp_2_1} assuming Model~1.1 and~2.1 to be the true model, respectively. From these figures, we can see that the two models could be separated by around $6\sigma$ or more, depending on the true model parameters. The reason for this is that Model~1.1 predicts a value of $\delta_{\rm CP}$ of $180^\circ$, whereas the values predicted by Model~2.1 are spread, but closer to 0 than $180^\circ$. Thus, the measurement of $\delta_{\rm CP}$ by ESSnuSB will be able to provide a good discriminator between the models.

\begin{figure}[t!]
\vspace{-1cm}
\hspace{-10pt}
\begin{tabular}{ll}
\includegraphics[scale=0.55]{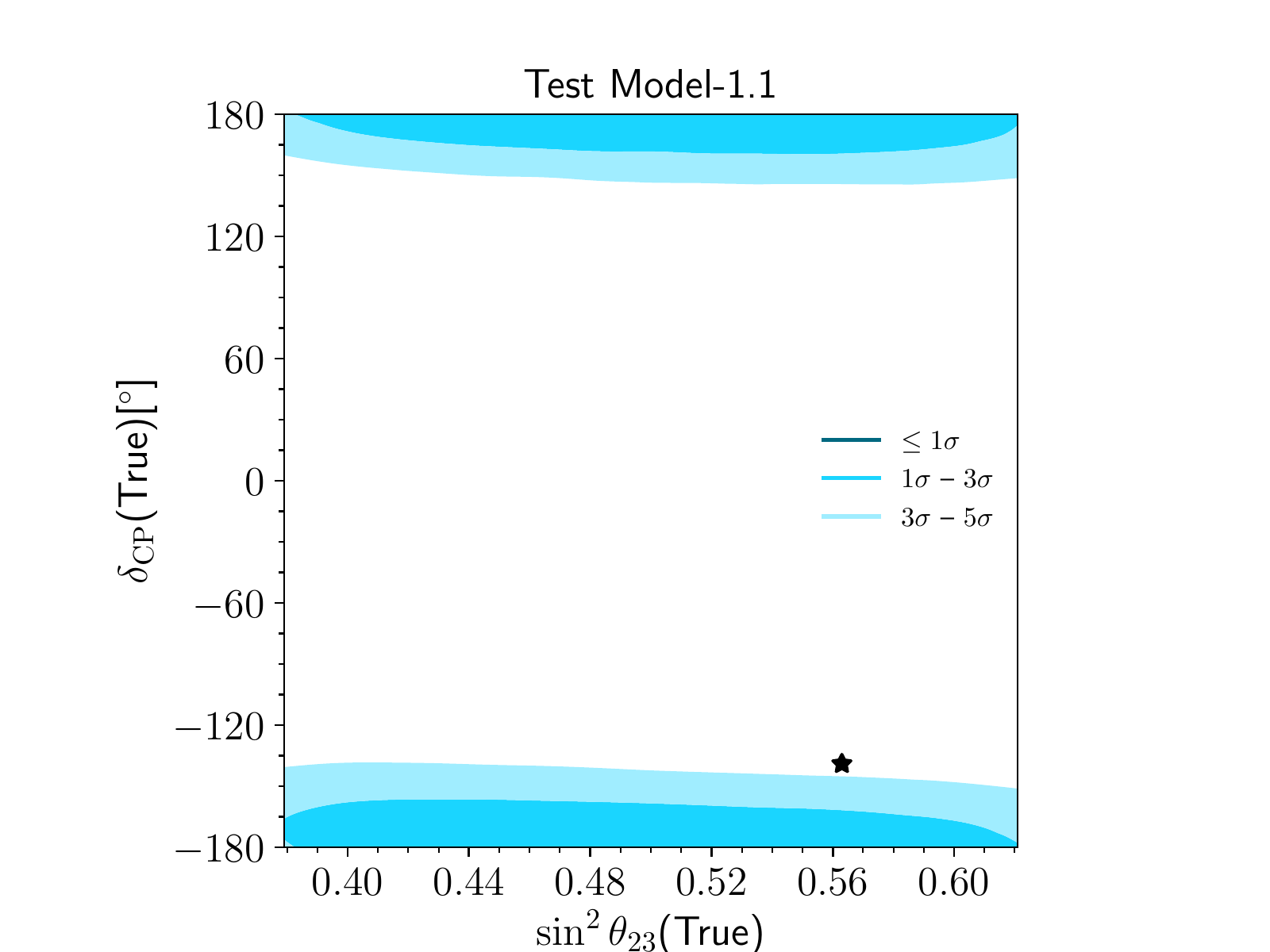}
\hspace{-50pt}
\includegraphics[scale=0.55]{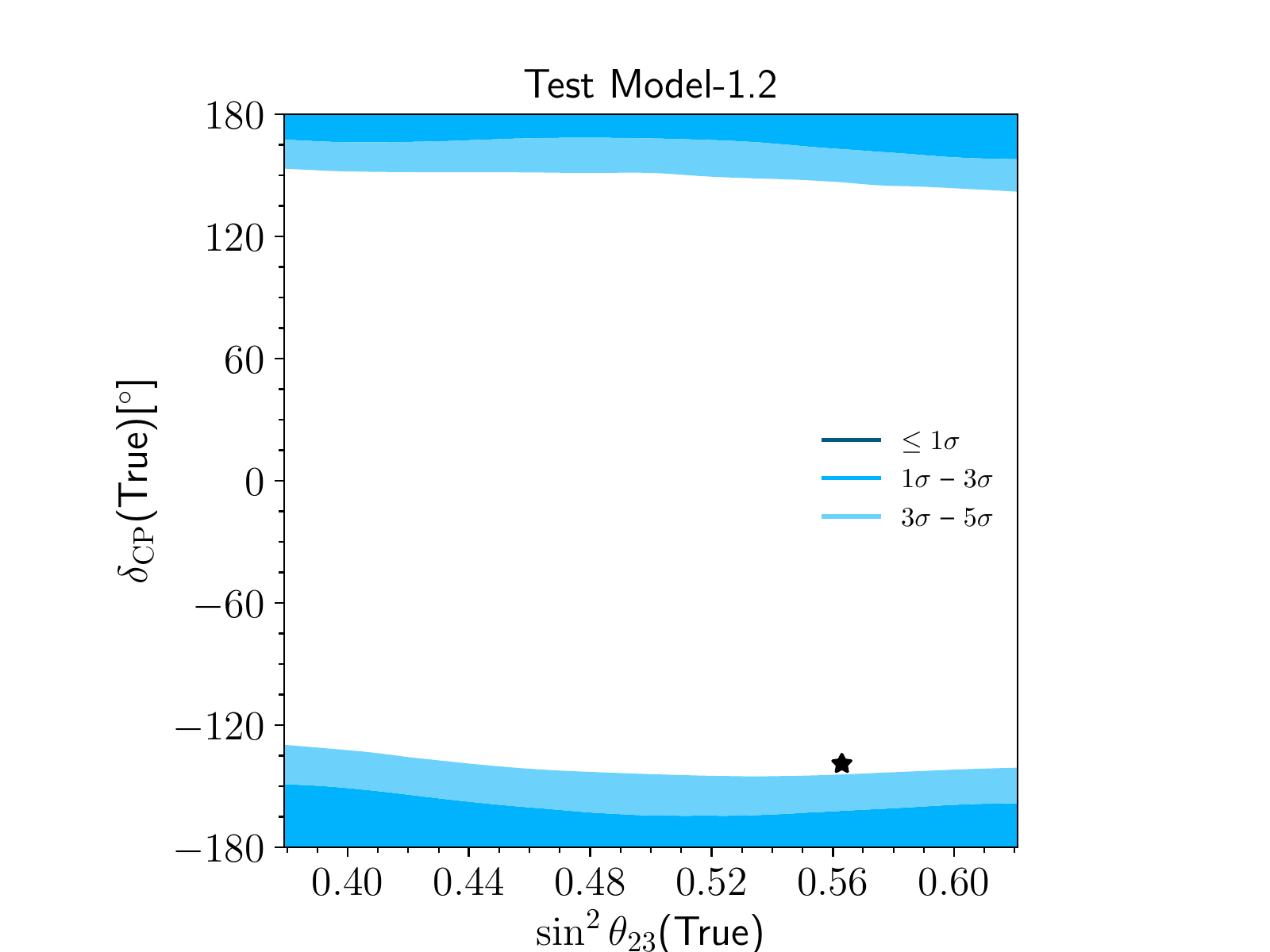} \\[1mm]
\includegraphics[scale=0.55]{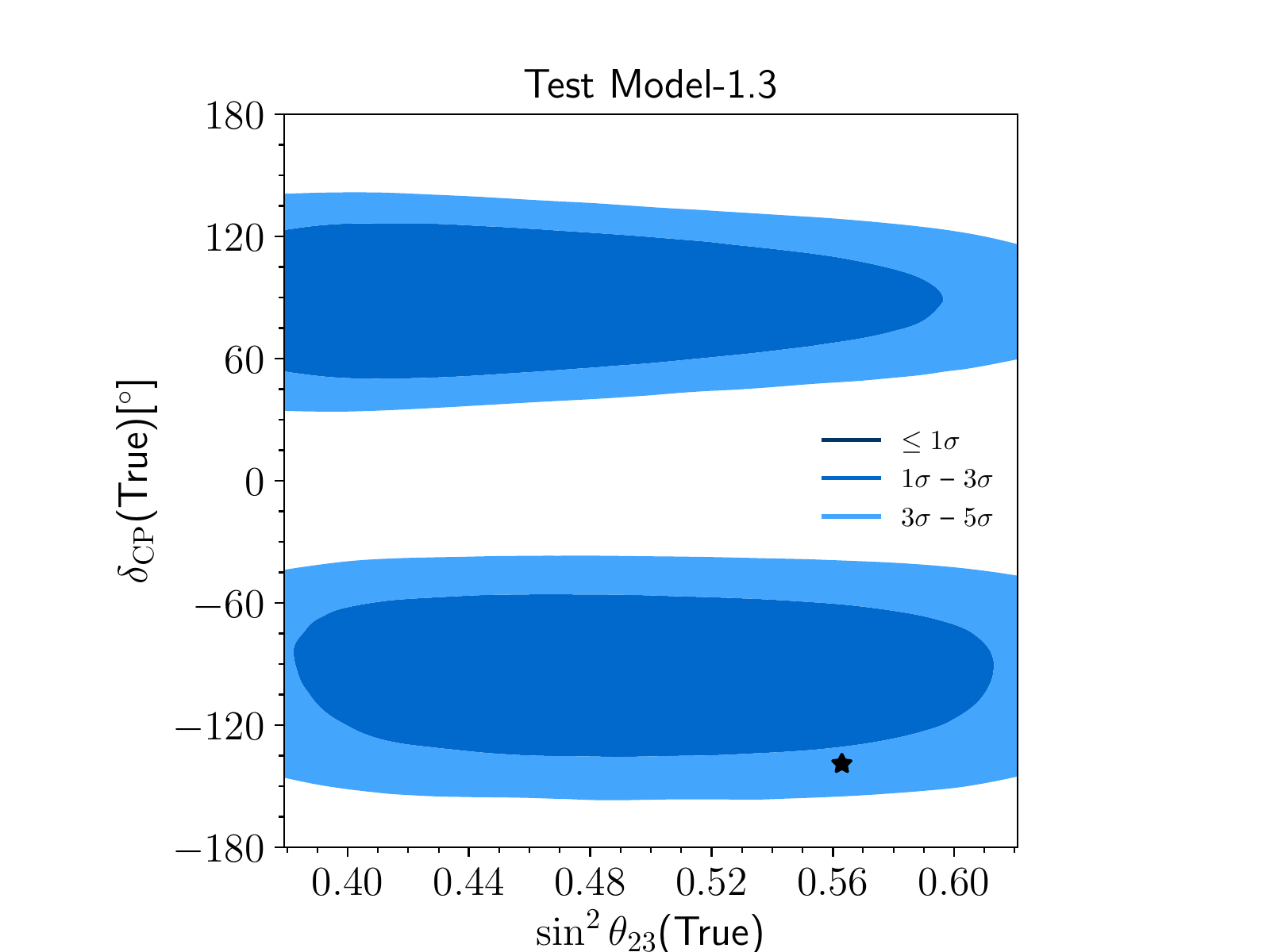}
\hspace{-50pt}
\includegraphics[scale=0.55]{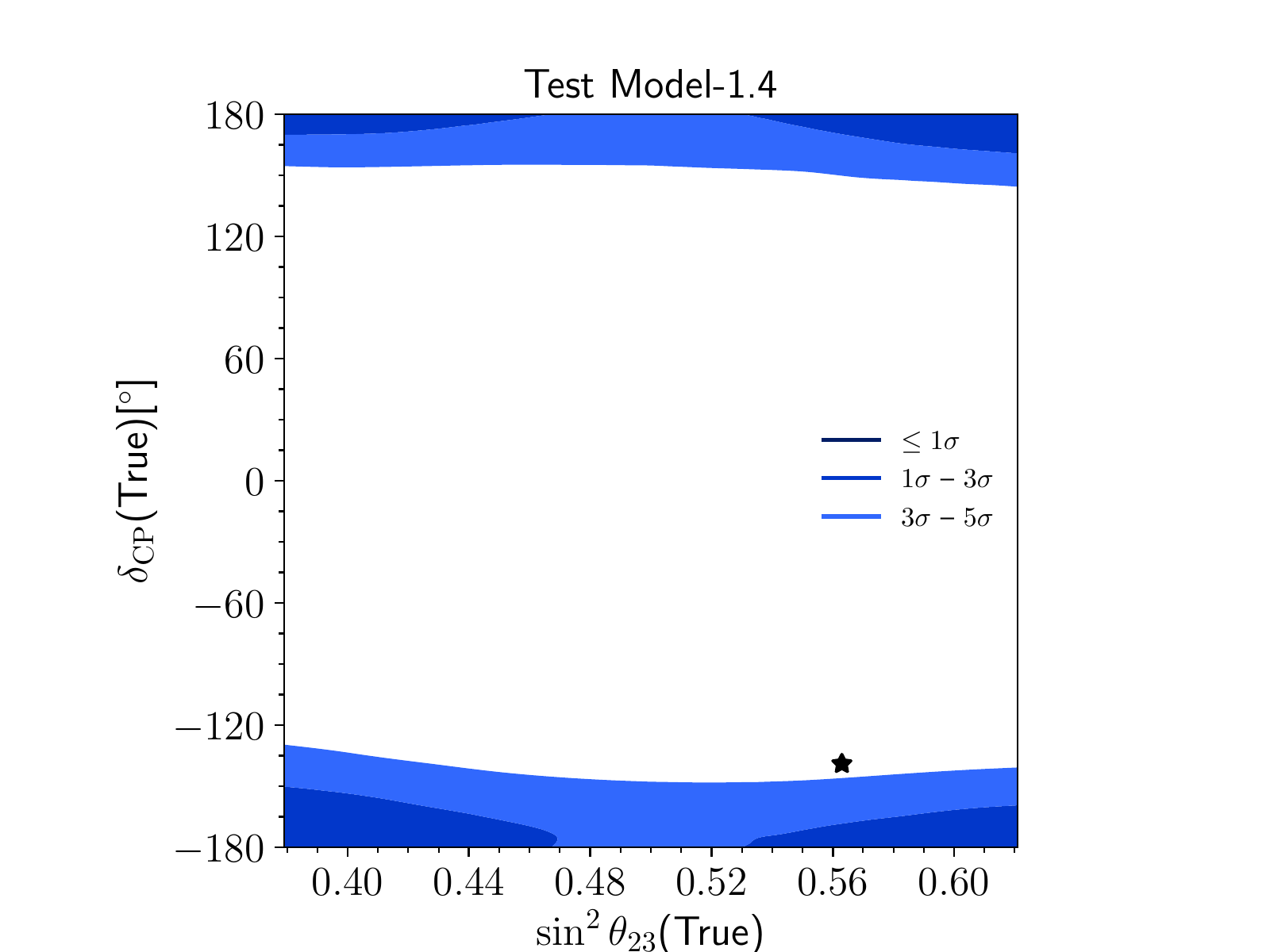} \\[1mm]
\multicolumn{2}{c}{\includegraphics[scale=0.55]{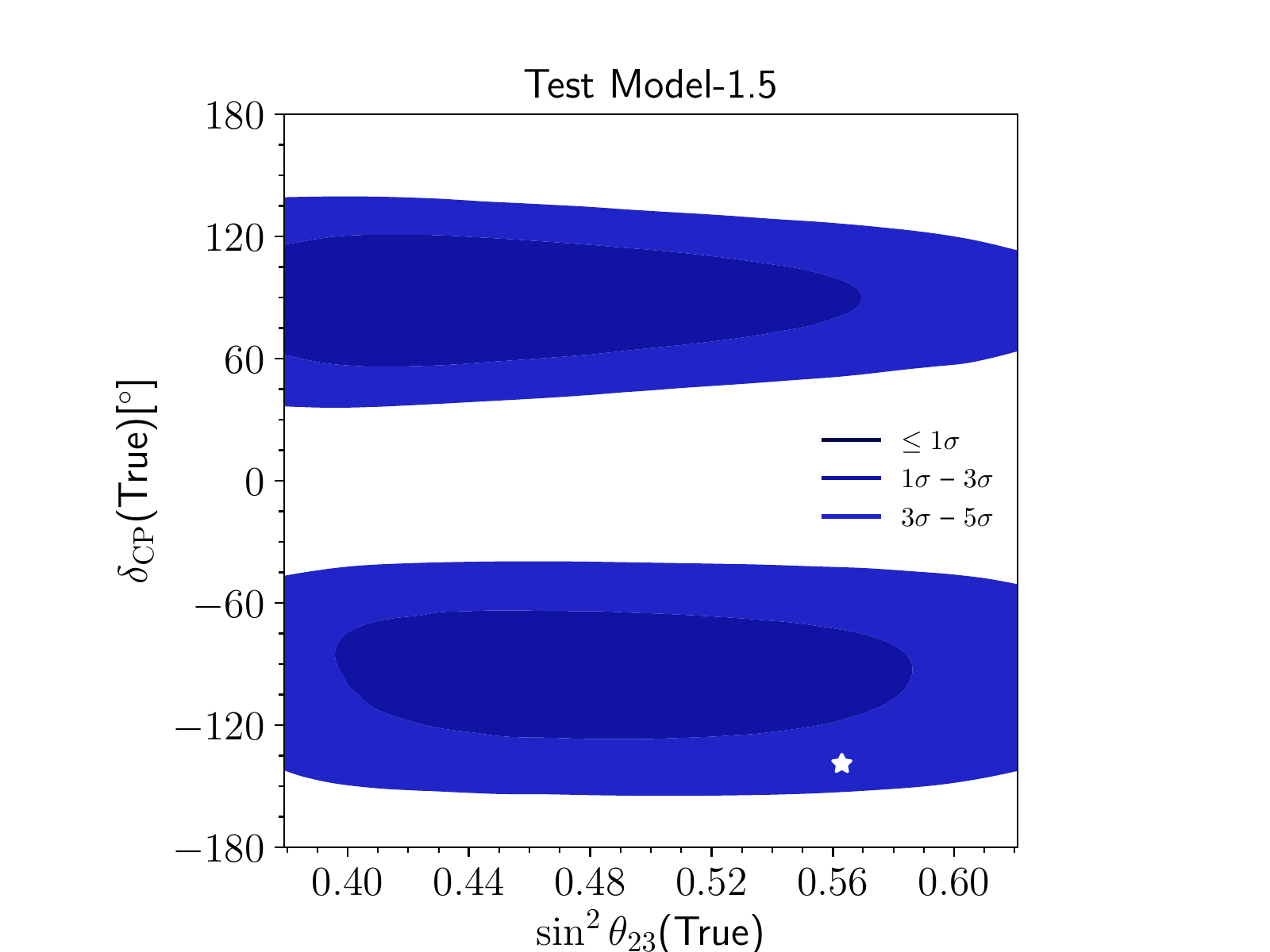}}
\end{tabular}
\caption{Compatibility of the five allowed one-parameter models with any potentially true values of $\sin^2 \theta_{23}$ and $\delta_{\rm CP}$ for ESSnuSB. For the other neutrino oscillation parameters, we assume the following true values $\sin^2 \theta_{12} = 0.310$, $\sin^2 \theta_{13} = 0.02237$, $\Delta m_{21}^2 = 7.39 \cdot 10^{-5} \, {\rm eV}^2$, and $\Delta m_{31}^2 = 2.528 \cdot 10^{-3} \, {\rm eV}^2$. A star (``$\star$'') indicates the present best-fit values of $\sin^2 \theta_{23}$ and $\delta_{\rm CP}$ from global neutrino oscillation data. The contours correspond to the indicated number of sigmas for 3 d.o.f.}
\label{fig:t23delta_oneparamod}
\end{figure}

In Fig.~\ref{fig:t23delta_oneparamod}, we present the capability of ESSnuSB to exclude the one-parameter models in the $\sin^2\theta_{23}$(true)--$\delta_{\rm CP}$(true) plane. For the other neutrino oscillation parameters, we assume the following true values $\sin^2 \theta_{12} = 0.310$, $\sin^2 \theta_{13} = 0.02237$, $\Delta m_{21}^2 = 7.39 \cdot 10^{-5} \, {\rm eV}^2$, and $\Delta m_{31}^2 = 2.528 \cdot 10^{-3} \, {\rm eV}^2$. If the true values of the leptonic mixing parameters chosen by Nature fall in the shaded region, then the model under test will be compatible with the Asimov data at the shown confidence level. We observe that, for all five models, no model would survive at $1 \sigma$ and the region of true parameters for which the models would survive at $3 \sigma$ shrink in size as we progress from Model~1.1 to Model~1.5. The reason is the following: as we go from Model~1.1 to Model~1.5, the fit of the models to the current data becomes worse, and therefore, they can be excluded at higher confidence level by ESSnuSB, in particular as they are relatively bad fits at the present time with the current best-fit values for the leptonic mixing parameters. Since Models~1.1, 1.2, and 1.4 predict $\delta_{\rm CP} = 180^\circ$, the regions where these models cannot be ruled out are around $\delta_{\rm CP}{\rm (true)} = 180^\circ$, and since Models~1.3 and 1.5 predict  $\delta_{\rm CP} = \pm 90^\circ$, the regions remain close to $\delta_{\rm CP}{\rm (true)} = \pm 90^\circ$, which is due to the excellent $\delta_{\rm CP}$ measurement capability of ESSnuSB. Therefore, if the true value of $\delta_{\rm CP}$ is close to the present best-fit point (indicated by stars), then Models~1.1, 1.2, and 1.4 can be excluded at more than $5\sigma$, whereas Models~1.3 and 1.5 will still be consistent with the current data at $5 \sigma$. From the different panels, we also note that for Models~1.3--1.5, for some of the true values of $\sin^2\theta_{23}{\rm (true)}$ the models could also be excluded by ESSnuSB at $3 \sigma$, but this is not the case at $5 \sigma$. This is due to the relatively poor capability of ESSnuSB to measure the leptonic mixing angle $\theta_{23}$. Furthermore, we have checked that the global $\chi^2_{\rm min}$ is larger at $\delta_{\rm CP}{\rm (true)} = 90^\circ$ and smaller at $\delta_{\rm CP}{\rm (true)} = -90^\circ$ in the lower octant of $\theta_{23}{\rm (true)}$. For Models~1.3 and 1.5, this is the reason why the exclusion around $\delta_{\rm CP}{\rm (true)} = -90^\circ$ is stronger than around $\delta_{\rm CP}{\rm (true)} = 90^\circ$.

\begin{figure}[t!]
\vspace{-1cm}
\hspace{-10pt}
\begin{tabular}{ll}
\includegraphics[scale=0.55]{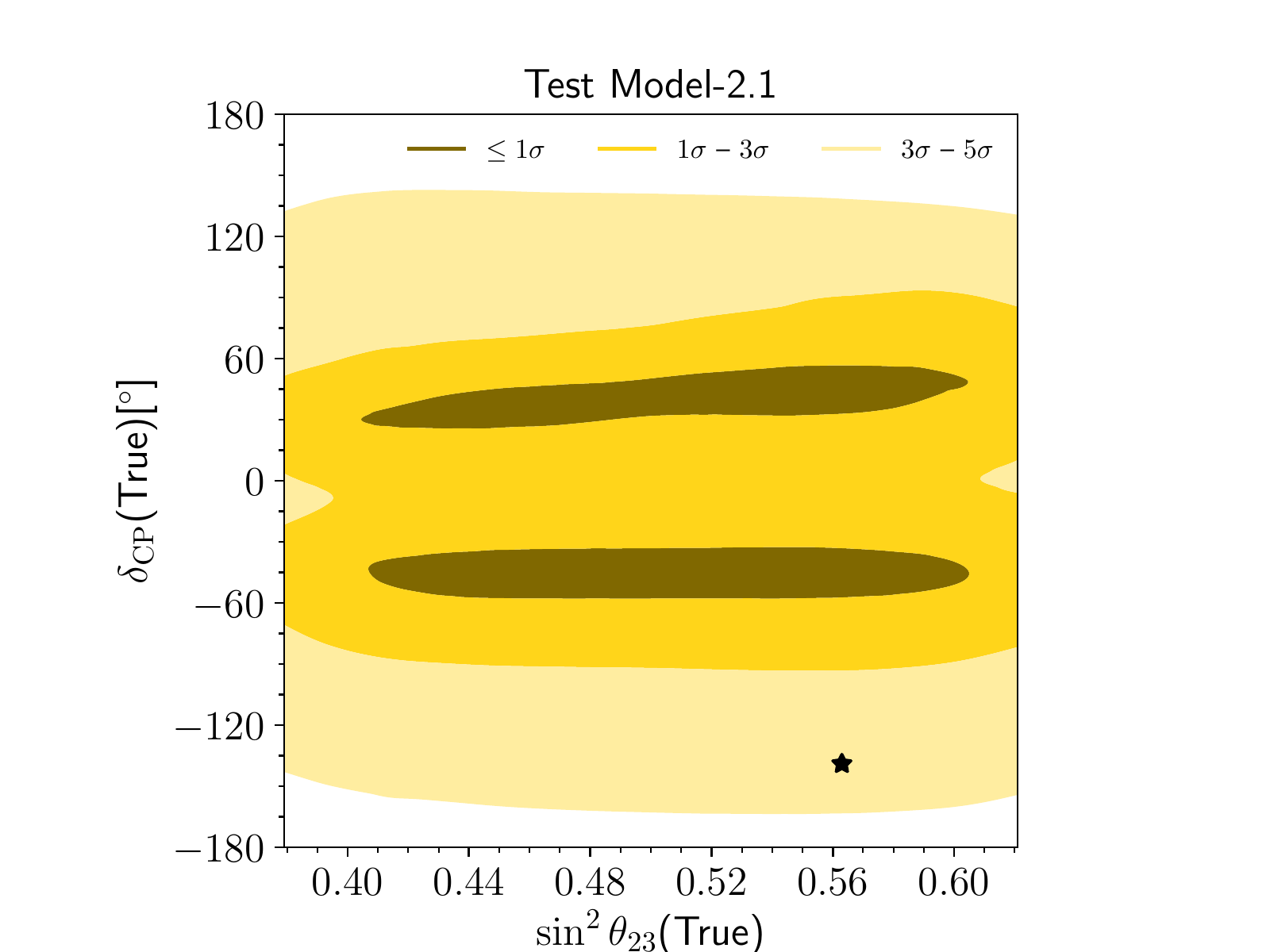}
\hspace{-50pt}
\includegraphics[scale=0.55]{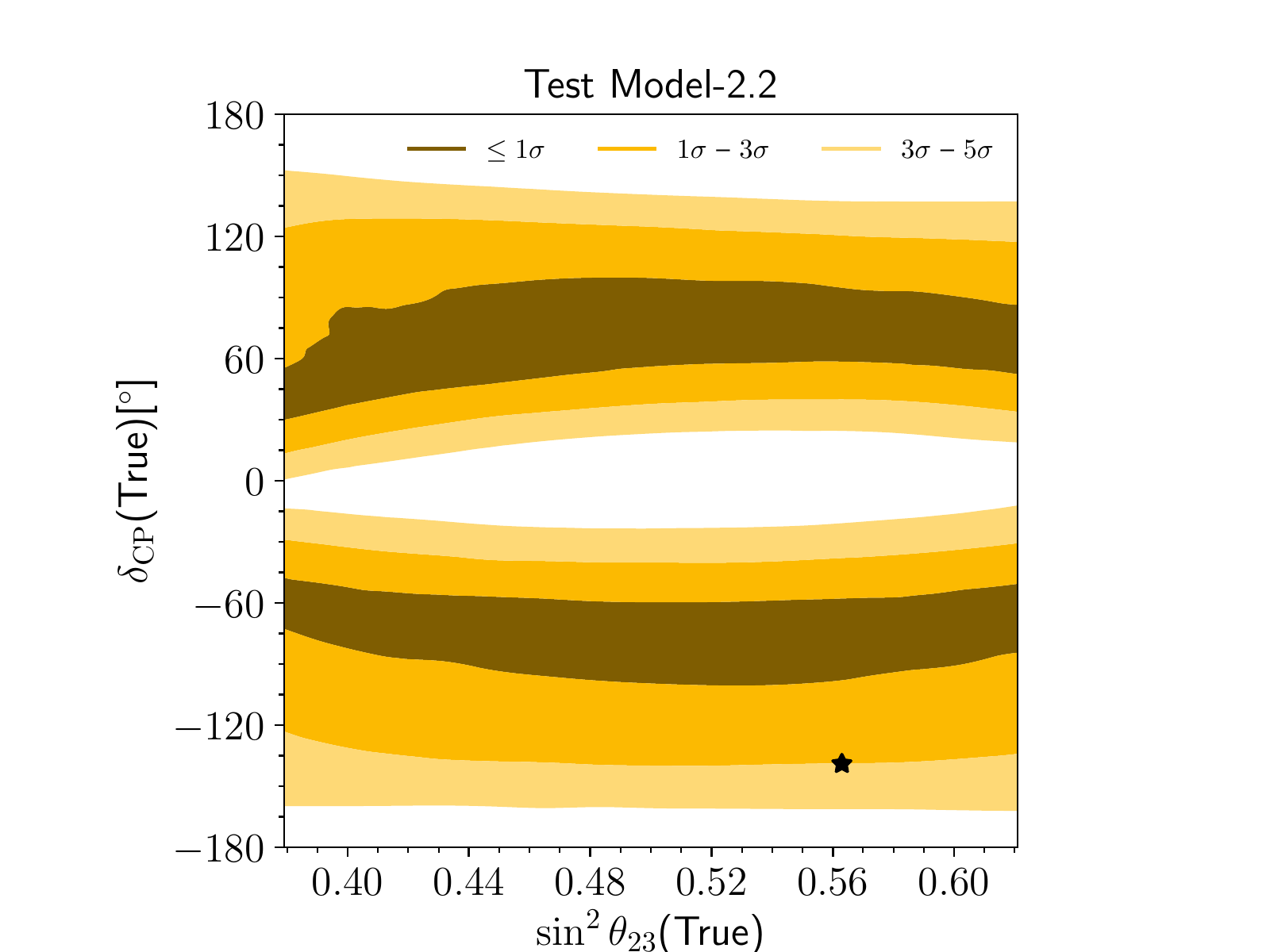} \\[1mm]
\includegraphics[scale=0.55]{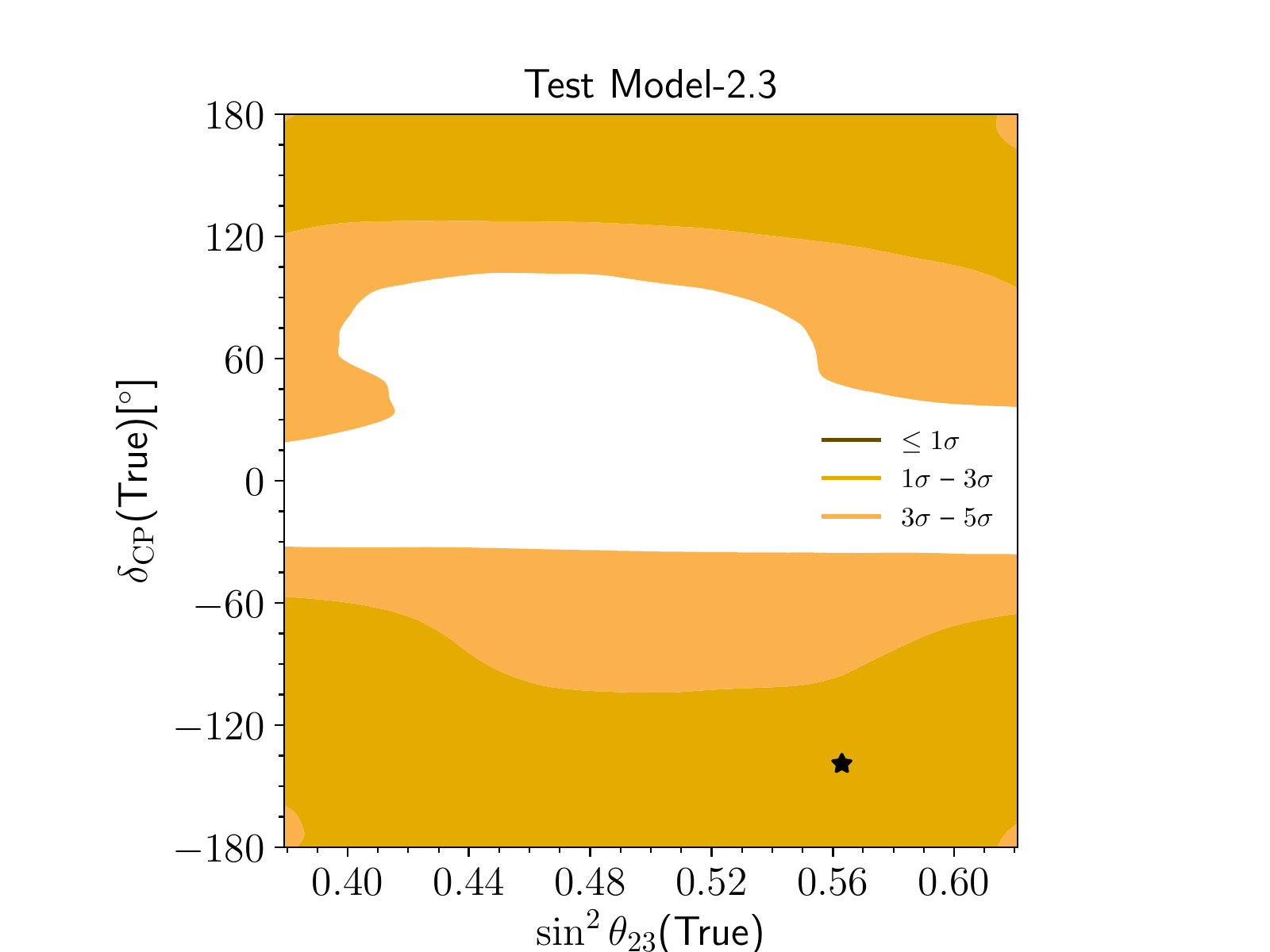}
\hspace{-50pt}
\includegraphics[scale=0.55]{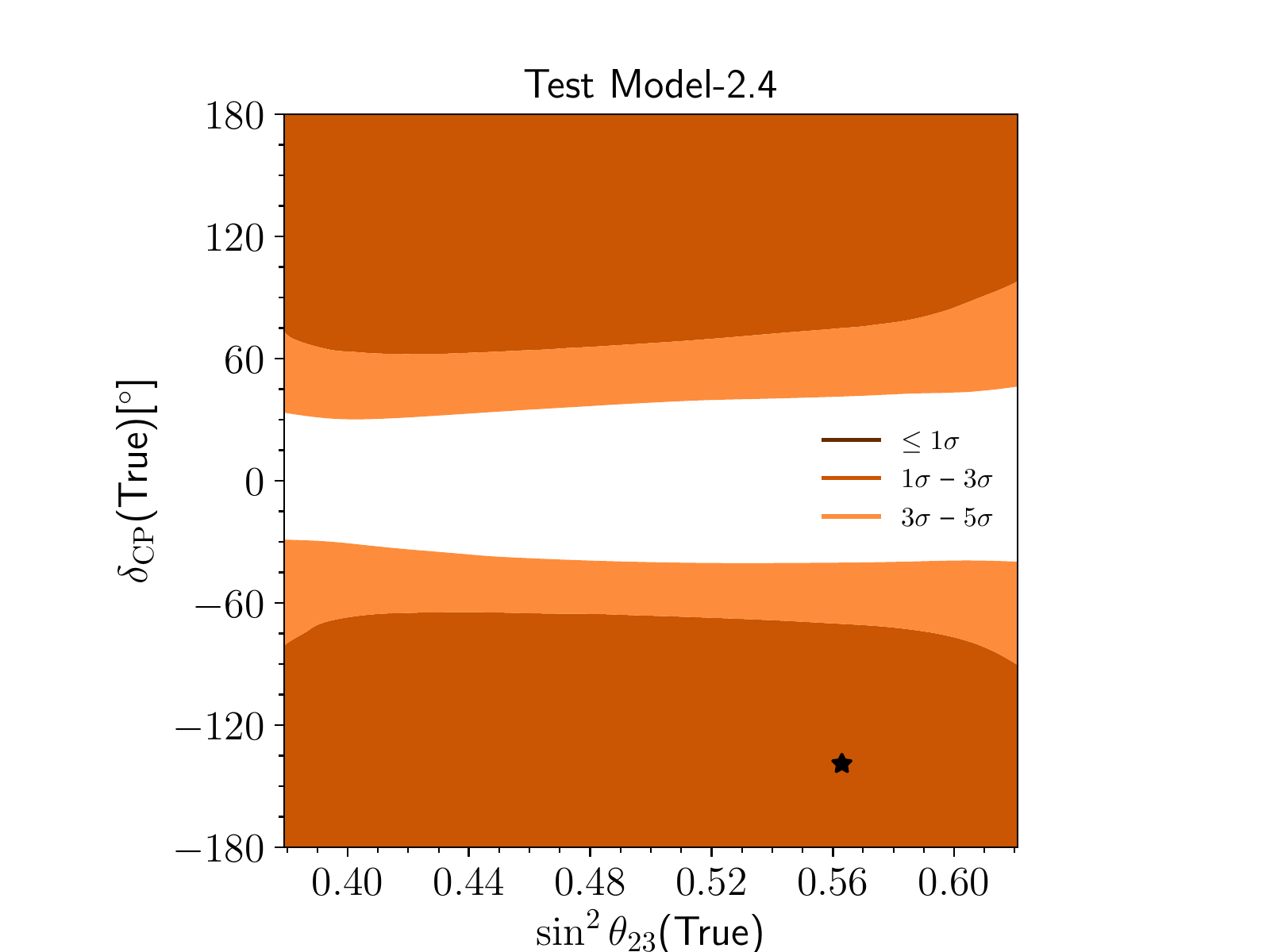} \\[1mm]
\multicolumn{2}{c}{\includegraphics[scale=0.55]{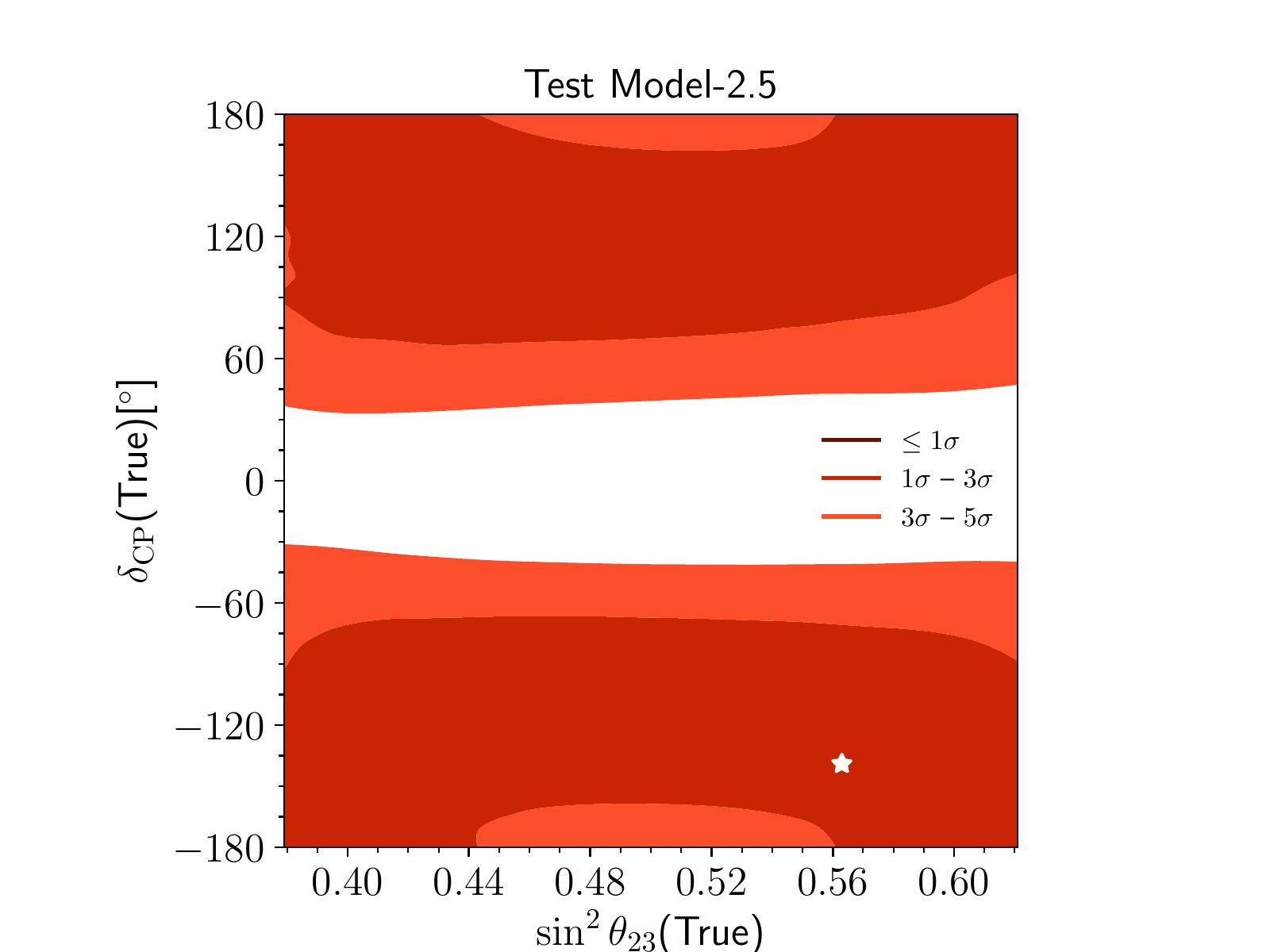}}
\end{tabular}
\caption{Compatibility of the five allowed two-parameter models with any potentially true values of $\sin^2 \theta_{23}$ and $\delta_{\rm CP}$ for ESSnuSB. For the other neutrino oscillation parameters, we assume the following true values $\sin^2 \theta_{12} = 0.310$, $\sin^2 \theta_{13} = 0.02237$, $\Delta m_{21}^2 = 7.39 \cdot 10^{-5} \, {\rm eV}^2$, and $\Delta m_{31}^2 = 2.528 \cdot 10^{-3} \, {\rm eV}^2$. A star (``$\star$'') indicates the present best-fit values of $\sin^2 \theta_{23}$ and $\delta_{\rm CP}$ from global neutrino oscillation data. The contours correspond to the indicated number of sigmas for 2 d.o.f.}
\label{fig:t23delta_twoparamod}
\end{figure}

Finally, in Fig.~\ref{fig:t23delta_twoparamod}, we present the same as in Fig.~\ref{fig:t23delta_oneparamod}, but for the two-parameter models. For the same reason as discussed for the one-parameter models, as we progress from Model~2.1 to Model~2.5, the regions where the models cannot be ruled out become smaller.  We observe that for Models~2.1 and 2.2, there is a region where the models cannot be ruled out even at $1 \sigma$, whereas for Models 2.3--2.5, such regions do not exist. If the true values of $\theta_{23}$ and $\delta_{\rm CP}$ remain close to the current best-fit values, then Models~2.1 and 2.2 would be compatible with the Asimov data at $5 \sigma$, while Models 2.3--2.5 would be compatible at $3 \sigma$. Note that although Model~2.1 leads to a better fit to the current data than Model~2.2, the parameter region where Model~2.1 cannot be ruled out is smaller than that of Model~2.2 (in particular, the $1\sigma$ region). This is due to the fact that the precision of ESSnuSB on $\delta_{\rm CP}$ is best near $0$ and, since Model~2.1 predicts $\delta_{\rm CP} \in (\pm 11.6^\circ,\pm 62.2^\circ)$, whereas Model~2.2 predicts $\delta_{\rm CP} \in (\pm 58.7^\circ,\pm 87.6^\circ)$, we find a smaller region for Model~2.1 than for Model~2.2. It is important to note that although Model~2.1 predicts a very narrow range of $\sin^2\theta_{23}$ around 0.55, the region of $\sin^2\theta_{23}{\rm (true)}$ where it cannot be excluded is rather broad. This is due to the poor $\theta_{23}$ measurement capability of ESSnuSB, which was discussed earlier. This may be mitigated by combining the LBL data with the sample of atmospheric neutrinos that the detector would collect~\cite{Blennow:2019bvl}. 

%===============
\section{Summary and Conclusions}
\label{sec:sc}
%===============

In this work, we have investigated the capability of the proposed long-baseline neutrino oscillation experiment ESSnuSB to discriminate between a class of lepton flavor models along with testing the viability of such models. In the framework of the discrete symmetry approach to lepton flavor, we have reviewed various lepton mixing patterns arising from breaking a flavor symmetry $G_f$ (or its CP version $G_\mathrm{CP}$) to residual symmetries $G_e$ and $G_\nu$ of the charged lepton and neutrino mass matrices. We have classified these patterns according to the number of free parameters entering the predicted form of the leptonic mixing matrix $U_\mathrm{PMNS}$. Further, we have concentrated on one- and two-parameter patterns originating from relatively small (in terms of the number of elements) discrete groups. Namely, we have considered eleven one-parameter models with the free parameter $\theta$, arising from $G_\mathrm{CP} = S_4 \rtimes \text{CP}$~\cite{Feruglio:2012cw} and $A_5 \rtimes \text{CP}$~\cite{Li:2015jxa} (see also Refs.\cite{DiIura:2015kfa,Ballett:2015wia}) broken to $G_e > Z_2$ and $G_\nu = Z_2 \times \text{CP}$, and seven two-parameter models with the free parameters $\theta$ and $\phi$, originating from  $G_f = A_4$, $S_4$, and $A_5$ broken to   $G_e~(G_\nu)= Z_2$ and $G_\nu~(G_e)> Z_2$~\cite{Girardi:2015rwa,Petcov:2018snn}. Since both residual symmetries are non-trivial, but one of them is $Z_2$, these models have a relatively high predictive power, providing at the same time necessary freedom in fitting them to the data.

First, to test the compatibility of the models with the current neutrino oscillation data, we have fitted the models with a simple $\chi^2$ function using the current best-fit values of the three leptonic mixing angles as the true values, see Eqs.~\eqref{eq:chisq1p} and \eqref{eq:chisq2p}. We have listed all eleven one- and seven two-parameter models according to how well they fit the current data. We have also calculated the $3\sigma$ ranges of the free parameter $\theta$ for the one-parameter models as well as $\theta$ and $\phi$ for the two-parameter models, see Table~\ref{tab:fits}. Second, using the $3\sigma$ ranges of the free parameters, we have calculated the predicted values of the leptonic mixing parameters for these models. We have found that among the eleven one-parameter models, six models are already excluded by the current data at more than $3\sigma$, see Table~\ref{tab:fits2} and as partly shown in Fig.~\ref{fig:pred_oneparamod}. However, all seven two-parameter models are consistent with the individual $3\sigma$ ranges of the leptonic mixing angles, see Fig.~\ref{fig:pred_twoparamod}. For our analysis with ESSnuSB, we have selected only those models that are allowed by the current data within $3\sigma$, which are Models~1.1--1.5 and 2.1--2.5. Among the five allowed one-parameter models, Models~1.3 ($S_4 \rtimes \text{CP}$) and 1.5 ($A_5 \rtimes \text{CP}$) predict $|\sin\delta_{\rm CP}| = 1$ and $\theta_{23} = 45^\circ$, whereas Models~1.1 ($A_5 \rtimes \text{CP}$), 1.2 ($A_5 \rtimes \text{CP}$), and 1.4 ($S_4 \rtimes \text{CP}$) predict $\sin\delta_{\rm CP} = 0$ with $\theta_{23}$ lying in the higher octant. For the two-parameter models, all models (i.e., Models~2.1--2.5) predict the higher octant of $\theta_{23}$. Furthermore, the predictions for $\delta_{\rm CP}$ of Models~2.3--2.5 ($A_5$ for all, and in addition, $A_4$ and $S_4$ for 2.5) are compatible with CP conservation, while Model~2.2 ($S_4$) predicts $\delta_{\rm CP}$ close to maximal CP violation. On the other hand, Model~2.1 ($A_5$) is compatible with neither CP conservation nor maximal CP violation.

Next, we have studied the capability of ESSnuSB to distinguish the different one- and two-parameter models. We have found that the one-parameter models, which predict $\sin\delta_{\rm CP} = 0$, i.e., Models~1.1 ($A_5 \rtimes \text{CP}$), 1.2 ($A_5 \rtimes \text{CP}$), and 1.4 ($S_4 \rtimes \text{CP}$), can be separated from the one-parameter models that predict $|\sin\delta_{\rm CP}| = 1$, i.e., Models~1.3 ($S_4 \rtimes \text{CP}$) and 1.5 ($A_5 \rtimes \text{CP}$), by at least $7\sigma$, as can be seen from Fig.~\ref{fig:comp_oneparamod}. However, it is not possible to discriminate among Models~1.1 ($A_5 \rtimes \text{CP}$), 1.2 ($A_5 \rtimes \text{CP}$), and 1.4 ($S_4 \rtimes \text{CP}$) and also not between Models~1.3 ($S_4 \rtimes \text{CP}$) and 1.5 ($A_5 \rtimes \text{CP}$). For the two-parameter models, we have found that ESSnuSB suggests Models~2.1--2.3 ($A_5$ for 2.1 and 2.3 and $S_4$ for 2.2) to be the most preferred models if they are assumed to be the true models for most of the parameter space, see Figs.~\ref{fig:comp_twoparamod_21}--\ref{fig:comp_twoparamod_25}. In general, ESSnuSB is helpful to disfavor the other models. In addition, we have studied the capability of ESSnuSB to discriminate between Models~1.1 ($A_5 \rtimes \text{CP}$) and 2.1 ($A_5$), which are the models that lead to the best fit to the current data for the one- and two-parameter models, respectively. We have found that Models~1.1 ($A_5 \rtimes \text{CP}$) and~2.1 ($A_5$) can be separated by $6 \sigma$ or more (cf.~Figs.~\ref{fig:comp_1_2} and~\ref{fig:comp_2_1}), depending on the true parameter values assumed. Finally, while studying the capability of ESSnuSB to exclude models, we have found that if the best-fit point (after the running of ESSnuSB) remains close to the current best-fit point from global neutrino oscillation data, then Models~1.1 ($A_5 \rtimes \text{CP}$), 1.2 ($A_5 \rtimes \text{CP}$), and 1.4 ($S_4 \rtimes \text{CP}$) are excluded by ESSnuSB at more than $5 \sigma$. However, Models~1.3 ($S_4 \rtimes \text{CP}$), 1.5 ($A_5 \rtimes \text{CP}$), 2.1 ($A_5$), and 2.2 ($S_4$) are consistent with the data at $5 \sigma$ and Models~2.3--2.5 ($A_5$ for all, and in addition, $A_4$ and $S_4$ for 2.5) are consistent with the data at $3 \sigma$, see Figs.~\ref{fig:t23delta_oneparamod} and \ref{fig:t23delta_twoparamod}.

%===============
\section*{Acknowledgments}

We would like to thank Marcos Dracos, Tord Ekel{\"o}f, and Marcus Pernow for useful discussions. We would also like to thank Marie-Laure Schneider for comments on our work. This project is supported by the COST Action CA15139 {\it ``Combining forces for a novel European facility for neutrino-antineutrino symmetry-violation discovery''} (EuroNuNet). It has also received funding from the European Union's Horizon 2020 research and innovation programme under grant agreement No 777419. T.O.~acknowledges support by the Swedish Research Council (Vetenskaps\-r{\aa}det) through Contract No.~2017-03934 and the KTH Royal Institute of Technology for a sabbatical period at the University of Iceland.
%===============

%=======================
%\bibliographystyle{JHEP}
\bibliography{Flavor_Models_at_ESSnuSB_v2}
%=======================

%============
\end{document}